\documentclass[twocolumn]{aastex62}
\UseRawInputEncoding 
\usepackage{graphicx}
\usepackage{grffile}

\usepackage{ulem}
\usepackage{url}
\usepackage{threeparttable}
\usepackage{xcolor}
\usepackage{soul}
\usepackage{booktabs}
\usepackage{natbib}
\usepackage{amsmath}
\usepackage{mathrsfs}
\hypersetup{colorlinks, linkcolor=black, anchorcolor=green, citecolor=blue}
\usepackage{txfonts}

\usepackage{rotating}

\usepackage[]{subfigure,graphicx}

\usepackage{multirow}
\usepackage{graphbox} 

\received{2022 April 4}
\revised{2022 June 3}
\accepted{2022 June 3}
\submitjournal{ApJL}

\shorttitle{A detailed temperature map of the archetypal protostellar shocks in L1157}
\shortauthors{Feng et al.}

\begin{document}

\title{
A detailed temperature map of the archetypal protostellar shocks in L1157
}

\correspondingauthor{Siyi Feng}
\email{syfeng@xmu.edu.cn}

\author[0000-0002-4707-8409]{S. Feng (Chinese Name)}
\affil{Department of Astronomy, Xiamen University, Zengcuo'an West Road, Xiamen, 361005, PeopleÕs Republic of China}

\author[0000-0003-2300-2626]{H. B. Liu}
\affil{Academia Sinica Institute of Astronomy and Astrophysics, No.1, Sec. 4, Roosevelt Rd, Taipei 10617, Taiwan, Republic of China} 

\author[0000-0003-1481-7911]{P. Caselli}
\affil{Max-Planck-Institut f\"ur extraterrestrische Physik, Gie{\ss}enbachstra{\ss}e 1,  D-85748,  Garching bei M\"unchen, Germany}

\author[0000-0003-0799-0927]{A. Burkhardt}
\affil{Wellesley College, 106 Central Street, Wellesley, MA  02481, USA}

\author[0000-0002-7489-0179]{F. Du}
\affil{Purple Mountain Observatory and Key Laboratory of Radio Astronomy, Chinese Academy of Sciences, Nanjing 210033, PeopleÕs Republic of China}
\affil{School of Astronomy and Space Science, University of Science and Technology of China, Hefei, Anhui 230026, PeopleÕs Republic of China}

\author[0000-0002-5331-5386]{R. Bachiller}
\affil{Observatorio Astronomico Nacional (OAN, IGN), Calle Alfonso XII, 3, E-28014 Madrid, Spain}

\author[0000-0003-1514-3074]{C. Codella}
\affil{INAF-Osservatorio Astrofisico di Arcetri, Largo E. Fermi 5, I-50125, Florence, Italy} 
\affil{Univ. Grenoble Alpes, CNRS, IPAG, F-38000 Grenoble, France} 

\author[0000-0001-9664-6292]{C. Ceccarelli}
\affil{Univ. Grenoble Alpes, CNRS, IPAG, F-38000 Grenoble, France}


\begin{abstract} 
{We present sensitive  $\rm NH_3$\,(1,1)--(7,7) line images from the Karl G. Jansky Very Large Array toward successive shocks, which are associated with the blueshifted outflow lobe driven by the {compact protobinary system} L1157.
Within a projection distance of 0.1\,pc, our observations not only trace the quiescent and cold gas in the flattened envelope, but also illustrate the complex physical and chemical processes that take place where the high-velocity jet impinges on its surrounding medium. 
Specifically,  the $\rm NH_3$ ortho-to-para ratio is enhanced  by a factor of 2--2.5 along the jet path, where the velocity offset between the line peak and the blueshifted wing reaches values as high as $\rm 10\,km\,s^{-1}$; it also shows a strong spatial correlation with the $\rm NH_3$ column density, which is enhanced to $\rm >10^{16}\,cm^{-2}$ toward the shock cavities. 
At a linear resolution of 1500\,au, our refined temperature map  from the seven $\rm NH_3$ lines shows a gradient from the warm B0 eastern cavity wall ($\rm >120\,K$) to the cool cavity B1 and the earlier shock B2 ($\rm <80\,K$), indicating  shock heating.
}

\end{abstract}


\section{Introduction}
Protostellar shocks are commonly detected  in the earliest stage of star formation. Generated by the impact of supersonic jets from the central protostars  on their dense natal parental cloud \citep[e.g.,][]{frank14}, the shocks create a dense and warm environment in a short timescale by compressing and heating the  surrounding gas. Such an environment speeds up  chemical processes, which are infeasible in the preshock gas, including, but not limited to, endothermic chemical reactions, ice mantle vaporization, and sputtering \citep[e.g.][]{viti11}.

The target region of this work is an archetypal region containing successive shocks (named as B0, B1, and B2), which are  associated with the blue-shifted outflow lobe from the nearby ($d\sim$352\,pc; \citealp{zucker19}) Class 0 {compact protobinary system} L1157-mm \citep[$\simeq$ 3 $L_{\rm \odot}$;][]
{tobin10,tobin22}. Kinematically, an episodic precessing jet driven by the {protobinary system} hits the shock cavity wall, generating bright knots  \citep[e.g.][]{gueth96,gueth98,podio16}. Within a projection of 0.1\,pc from the {protobinary system} in the plane of the sky, the kinematic age difference  of these shock knots is $\sim$1000\,yr,  so this region provides us with one of the best space laboratories to study the time-dependent shock chemistry \citep[e.g.,][]{lefloch10,codella17}.

Multiwavelength line surveys have intensively observed this region, especially B1, over the decades (e.g., {\it Herschel-CHESS}, \citealt{ceccarelli10}; {\it IRAM 30\,m-ASAI}, \citealt{lefloch18}; {\it NOrthern Extended Millimeter Array (NOEMA)-SOLIS}, \citealt{ceccarelli17}) and revealed chemical complexities from diatomic molecules to complex organics
\citep[e.g., ][]{tafalla95,bachiller01,benedettini07,arce08,codella09,codella10,lefloch12,benedettini13,busquet13,podio14,fontani14c,codella15,codella17,lefloch17,lefloch18,codella20,feng20a,spezzano20}.
It is clear that sufficient observational data exist to carry out a systematic study of shock chemistry, i.e., 
examining the origin, excitation, and  chemical complexity of different species.
However, a crucial input for chemical modeling is the physical structure of this region, which is still missing.

Previous observations attempted to use several inversion lines of $\rm NH_3$ at centimeter wavelengths to provide constraints on the kinetic temperature of this region, given that these lines have been widely used as an interstellar thermometer\footnote{The majority of the $\rm NH_3$ population stays in the metastable levels $J$=$K$, where $J$ is the total angular momentum quantum number of $\rm NH_3$, and $K$ is the projection of $J$ on the rotational axis. The inversion transitions from different rotational ladders of $\rm NH_3$ are coupled only collisionally. Having similar frequencies at 1.3\,cm wavelength, several inversion lines of $\rm NH_3$ can be observed simultaneously, so that the calibration uncertainties in the line ratio measurements are minimized.
}
for molecular gas with number density $n\rm>10^4\,cm^{-3}$ \citep[][]{ho83,walmsley83,crapsi07,rosolowsky08,juvela11,caselli17}. However, hindered by the low spatial/velocity resolutions of single-dish point observations (e.g., \citealp[][]{bachiller93} observed the (1,1)--(4,4) lines; \citealp[][]{umemoto99} observed the (1,1)--(6,6) lines) and the insufficient sensitivity of interferometric observations (\citealp{tafalla95} observed only the (1,1)--(3,3) lines), several assumptions were made in their works,  leading to large uncertainties in their conclusions.

Interferometric observations have been carried out recently, targeting line ladders of $\rm CH_3CN$, $\rm H_2CO$, CS, and molecular ions. Although these works  provide  constraints on the $\rm H_2$ volume density $n$ \citep[e.g., ][]{benedettini13,gomezruiz15} and molecular rotation temperature  \citep[e.g., ][]{codella09,podio14} toward several knots, {these attempts are only toward  B1}.

In this paper, with the newest $\rm NH_3$\,(1,1)--(7,7) observations  at high-spatial and high-velocity resolutions, 
{we provide a detailed temperature map over B0-B1-B2, and discuss the complex physical and chemical processes that take place when the high-velocity jet impinges on its surrounding medium.}

\section{Observations and data reduction} \label{sec:obs}
Using the  Karl G. Jansky Very Large Array (JVLA), we have performed the observations toward L1157 B0-B1-B2 at the K-band in  D-array configurations from September to November in 2018. For all observations, a common phase center at $\rm +20^h39^m10^s.2$, $\rm +68^\circ01^{'}10^{''}.5$ (J2000) and  a systemic velocity $V_{sys} \sim \rm 2.7\,km\,s^{-1}$ were adopted from the {\it Herschel} results on the $o$-$\rm NH_3$\,($\rm 1_0-0_0$) line \citep{codella10}.
Employing the three-bit sampler, our correlator setup  uses 27 independent spectral windows to cover the NH$_{3}$ (1,1)--(7,7) transitions (with $E_u/k_B$ ranges from 24 to 539\,K) and broadband continuum simultaneously.

All epochs of observations used 1331+305 (3C 286), J1642+3948, and J2022+6136 for absolute flux, passband, and complex grain calibrations, respectively.

Following the standard strategy by using the Common Astronomy Software Applications \citep[CASA;][]{mcmullin07} package release 5.4.0, we manually calibrated the JVLA data.
The absolute fluxes of the flux calibrator 3C286 were referenced from the  Perley-Butler 2017 standards \citep{perley17}.
For different epochs, a $\rm 10\%$--$\rm 15\%$ nominal absolute flux calibration accuracy may have to be assumed according to the official documentation\footnote{\href{https://science.nrao.edu/facilities/vla/docs/manuals/oss/performance/fdscale}{https://science.nrao.edu/facilities/vla/docs/manuals/oss/performance/fdscale}}.
Nevertheless, the absolute flux calibration errors are factored out when deriving spectral line ratios.

Using the CASA task \textsc{tclean} in CASA 5.4.6, we performed the 
spectral line imaging {(setting the specmode as `cube') and broadband continuum imaging (setting the specmode as `mfs')} by applying the `multiscale' imaging option with scales values of 0, 5, and 15 times the pixel size.
 The primary beam (pb) and the maximum recoverable angular scales for the single pointing observations are $\sim$119\arcsec and $\sim$61\arcsec at 23.7\,GHz, respectively. 
We test the Briggs robust weighting\footnote{The naturally weighted ($r$=2.0) image has high sensitivity but low spatial resolution, the uniformly weighted ($r$=-2.0) image has low sensitivity but high spatial resolution, and the image with $r$=0.5 is a good trade-off between resolution and sensitivity.} $r$ as 2.0,  -2.0, and 0.5. 
{The continuum shows  $\rm >4\,\sigma$ emission only toward the protobinary system, although this system is not resolved from our observations. The continuum intensity peak, measured at different weightings before and after pb correction, is 5--10\,$\sigma$, with 1\,$\sigma$ rms varying in the range of 0.01--0.04\,$\rm mJy\,beam^{-1}$, which is consistent with \citet{tobin22}. }
For all the targeted $\rm NH_3$ lines, their image properties\footnote{Line image properties have negligible changes before and after continuum subtraction.} are listed in Table~\ref{tab:line}.

\begin{table*}
\small
\centering
\caption{ Spectroscopic parameter of the targeted $\rm NH_3$ lines and their image properties from the JVLA observations
 \label{tab:line}}
 \scalebox{0.8}{
\begin{tabular}{p{1.0cm}llllllllllll}
\hline\hline
Mol. &Freq.  &Transition  &$\rm S\mu^2$$^a$  
&$E_u/k_B$$^b$     &No. hfs$^c$ &$\rm \Delta V$$^d$    &\multicolumn{3}{c}{$\rm \Delta \theta$ (P.A.)$^e$}   &\multicolumn{3}{c}{1 $\sigma$ rms$^f$}\\
         &(GHz) &                                                &($\rm D^2$)   
          &  (K)                        &    &($\rm km\,s^{-1}$)       &\multicolumn{3}{c}{($\rm \arcsec \times \arcsec$,$^\circ$)}                                                           &\multicolumn{3}{c}{($\rm mJy\,beam^{-1}\,ch^{-1}$)}  \\
         & &                                                &      
          &                       &    &       &$r$=2.0 &$r$=0.5 &$r$=-2.0                                                           &$r$=2.0 &$r$=0.5 &$r$=-2.0   \\             
\hline

$p$-$\rm NH_3$    &23.694           &$\rm 1( 1)0a-1( 1)0s$    &6.6 
&24            &18$^g$  &0.049   &$4.\arcsec63\times3.\arcsec77~(17^\circ)$   &$4.\arcsec24\times3.\arcsec19~(26^\circ)$   &$3.\arcsec71\times2.\arcsec63~(31^\circ)$             &6.3  &5.1  &6.1 \\  
$p$-$\rm NH_3$    &23.723        &$\rm 2( 2)0a- 2( 2)0s$   &14.7   
&65            &21$^g$  &0.197 &$4.\arcsec90\times3.\arcsec81~(19^\circ)$     &$4.\arcsec60\times3.\arcsec36~(23^\circ)$  &$4.\arcsec34\times2.\arcsec93~(29^\circ)$             &2.2  &2.5 &2.3\\ 
$o$-$\rm NH_3$    &23.870          &$\rm 3( 3)0a- 3(3)0s$    &46.4 
&124             &26$^h$ &0.196 &$4.\arcsec20\times3.\arcsec60~(-6^\circ)$   &$3.\arcsec67\times3.\arcsec01~(2^\circ)$&$3.\arcsec20\times2.\arcsec46~(10^\circ)$           &2.8  &2.9 &2.8 \\  
$p$-$\rm NH_3$    &24.139        &$\rm 4( 4)0a- 4(4)0s$   &31.8   
&201               &7$^h$   &0.194 &$4.\arcsec12\times3.\arcsec53~(-1^\circ)$  &$3.\arcsec62\times2.\arcsec97~(3^\circ)$ &$3.\arcsec15\times2.\arcsec43~(10^\circ)$            &1.3 &1.3 &2.2\\ 
$p$-$\rm NH_3$    &24.533           &$\rm 5( 5)0a- 5(5)0s$    &40.5 
&296                &7$^i$   &0.191   &$4.\arcsec25\times3.\arcsec65~(0^\circ)$  &$3.\arcsec74\times3.\arcsec05~(10^\circ)$&$3.\arcsec31\times2.\arcsec48~(18^\circ)$              &1.2  &1.3 &2.2     \\  
$o$-$\rm NH_3$    &25.056       &$\rm 6( 6)0a- 6(6)0s$   &98.5 
&409                &7$^i$    &0.187  &$4.\arcsec50\times3.\arcsec65~(20^\circ)$  &$4.\arcsec09\times3.\arcsec10~(26^\circ)$ &$3.\arcsec62\times2.\arcsec51~(28^\circ)$            &1.3   &1.3 &2.3         \\ 
$p$-$\rm NH_3$    &25.715       &$\rm 7( 7)0a- 7( 7)0s$   &56.3  
&539               &7$^i$   &0.182   &$4.\arcsec09\times3.\arcsec46~(2^\circ)$   &$3.\arcsec61\times2.\arcsec90~(8^\circ)$ &$3.\arcsec19\times2.\arcsec34~(16^\circ)$             &1.1 &1.2 &2.2      \\ 


\hline
\hline
\multicolumn{13}{l}{{\bf Note.} {\it a}.  Sum of all hyperfine multiplets.}\\
\multicolumn{13}{l}{~~~~~~~~~~{\it b}. Higher upper energy level of the transition. }\\
\multicolumn{13}{l}{~~~~~~~~~~{\it c}. Number of hyperfine multiplets used for deriving the  optical depth of the main component. }\\
\multicolumn{13}{l}{~~~~~~~~~~{\it d}.  Channel width.}\\
\multicolumn{13}{l}{~~~~~~~~~~{\it e}.  Synthesized beam.}\\
\multicolumn{13}{l}{~~~~~~~~~~{\it f}.  Without pb correction (with uniform noise level over the entire map), without angular resolution smoothing,}\\
\multicolumn{13}{l}{~~~~~~~~~~~~~~~~measured with the referred beam size (``beam") per channel (``ch"). }\\
\multicolumn{13}{l}{~~~~~~~~~~{\it g}. Number of hyperfine multiplets, adopted from \citet{mangum15}.}\\
\multicolumn{13}{l}{~~~~~~~~~~{\it h}. Number of hyperfine multiplets, adopted  from \texttt{PySpecKit}  \citep{ginsburg11}.}\\
\multicolumn{13}{l}{~~~~~~~~~~{\it i}. Number of hyperfine multiplets, adopted  from  TopModel in ``Splatalogue" database (\url{https://www.cv.nrao.edu/php/splat}).}\\
\end{tabular}

}

\end{table*}

\section{Results}\label{sec:obsresult}

\subsection{Molecular spatial distribution}\label{sec:distribution}

The hyperfine multiplets of  $\rm NH_3$\,(1,1)--(7,7) cover a velocity offset from $\rm -30$ to $\rm +30\,km\,s^{-1}$ with respect to the $V_{sys}$ \citep[e.g., ][]{ho83,mangum15}. We detected all the seven lines toward our target region, with their hyperfine multiplets partially resolvable at a velocity resolution of 0.05--0.2\,$\rm km\,s^{-1}$ (Table~\ref{tab:line}). 

\begin{figure*}
\centering
\includegraphics[width=19cm]{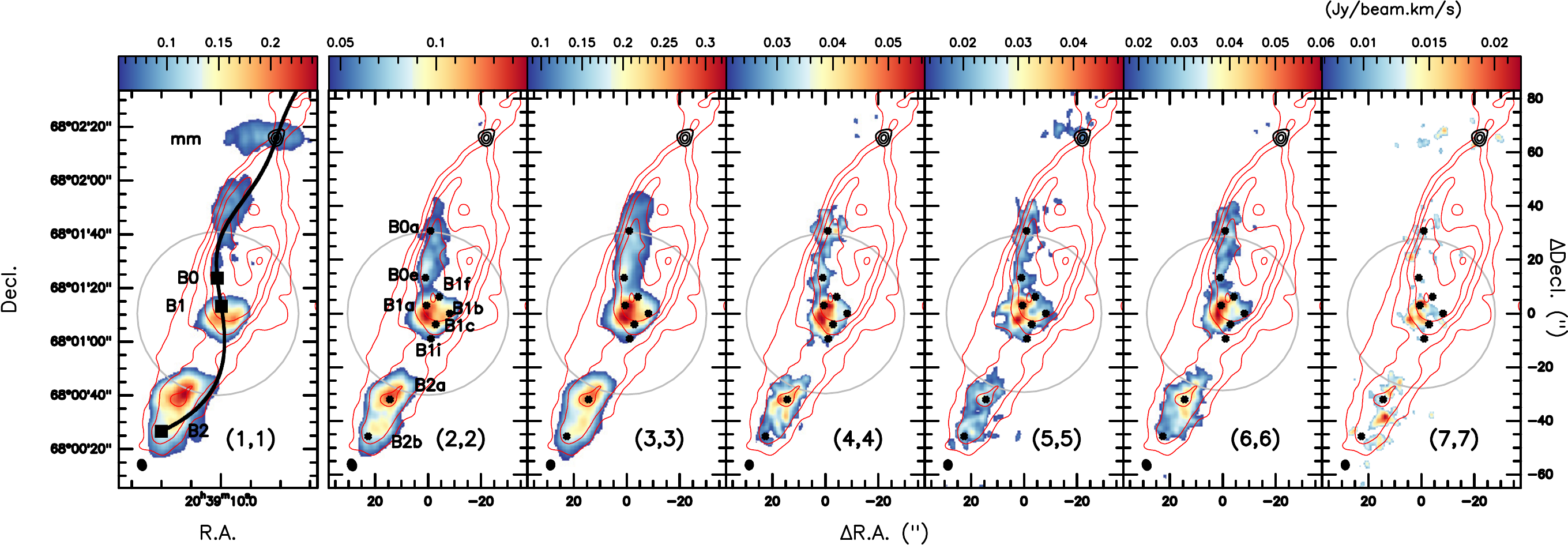}
\caption{
The outline of the southern (blue-shifted) outflow lobe from the Class 0 {compact protobinary system} L1157-mm. Color maps show the naturally weighted ($r$=2.0) intensity maps of $\rm NH_3$\,(1,1)--(7,7) integrated over the velocity range of $\rm -25$ to $\rm +25\,km\,s^{-1}$ (without pb correction).
Line emissions with $\rm <5\sigma$ rms are blank in each panel. The red contours, starting from $\rm 8\sigma$ ($\rm \sigma=0.27\,Jy\,beam^{-1}$) and increasing with a step of  $\rm 8\sigma$, show the CO (1-0) emission, 
obtained from the NOEMA and IRAM-30\,m combination \citep{gueth96}.  
{The black contours, starting from $\rm 4\sigma$ ($\rm \sigma=0.01\,mJy\,beam^{-1}$) and increasing with a step of  $\rm 4\sigma$, show the naturally weighted continuum emission at 1.3\,mm (without pb correction).}
The gray circle in each panel indicates the largest recoverable scale  at the corresponding line rest frequency, and the  angular resolution  is plotted at the bottom-left of each panel.
In the panel of the  (1,1) line, the black curve guidelines the path of the precessing jet from the driving source L1157-mm as modeled by \citet[][]{podio16}. The black squares indicate the shock knots B0, B1, and B2. In the panels of the (2,2)--(7,7) lines, nine clumpy substructures are labeled. 
}
\label{fig:jet}
\end{figure*}

Extracting their naturally weighted spectra  toward different jet knots  from B0 to B2   (Figure ~\ref{fig:velpro}), we found that  the satellite components of each line have a signal-to-noise ratio $\rm S/N>3$ in the velocity range of $\rm -25$ to $\rm +25\,km\,s^{-1}$. {Over this velocity range, the integrated intensities of all seven lines trace an extended structure (Figure~\ref{fig:jet} and Figure~\ref{fig:jetpb}) --from the rim of the B0 eastern cavity wall to two bow-shock cavities associated with B1 and B2} \citep[denoted as C2 and C1, respectively, in ][]{gueth98}.

{Comparing the total fluxes of the (1,1)--(4,4) lines 
from our JVLA observations  with those given by the single-dish observations} \citep[][]{bachiller93,umemoto99}\footnote{The line emission peaks observed with Nobeyama 45\,m  \citep{umemoto99} are in general lower than those observed with Effelsberg 100\,m \citep{bachiller93}  by a factor of 1.5--2, probably due to the beam dilution.}, we found 20\% and 15\% of the flux  missing toward B1 and B2, respectively.
Note that the angular resolution of Effelsberg observations  \citep{bachiller93}  is coarser than that of JVLA by a factor of 10,  so effects such as pointing errors and the attenuation at the edge of the Gaussian-shaped Effelsberg beam, may bring uncertainty in the comparison. Although we cannot precisely recover the flux per pixel from the Effelsberg single-point data, such comparison indicates that  the images  in this work cover the  $\rm NH_3$ extended emissions toward B1 and B2 at a similar level.

Limited by the largest recoverable angular scale with one {pointing} in these observations, only the central region  (i.e., within the gray circle in Figure~\ref{fig:jet}) of the image has high fidelity.
In order to apply the kinetic temperature map derived from $\rm NH_3$ lines to the chemical property measurements (i.e., column density, abundance) of the other molecular lines at a spatial resolution of $\sim$1500\,au, we labeled seven clumpy substructures (B0e, B1a,  B1b,  B1c,  B1f,  B1i, and B2a), which were identified from previous observations \citep[e.g., by using line emissions from $\rm CH_3CN$, HCN, $\rm CH_3OH$, SiO, CS, $\rm H_2CO$, $\rm HCO^+$, HNCO, SO, and $\rm SO_2$ in ][]{benedettini07,codella09,gomezruiz13,burkhardt16,feng20a}. 
Moreover,  the large pb of JVLA allows us to investigate
the complete B0 cavity wall and the B2 shock front together with the phase center B1.
In the  maps  with pb correction (Figure~\ref{fig:jetpb}), we found two clumpy structures  showing stronger emissions  than the rest of B0 and B2. Both structures are along the jet path derived by \citet[][]{podio16} and were previously denoted as B0a and B2b in \citet[][]{burkhardt16}.

{Each substructure  may be traced by more than one molecule, and the absolute coordinate of these intensity peaks (listed in Table~\ref{tab:clump}) are different from tracer to tracer}\footnote{B2a corresponds to the emission peak of CO toward B2 in the present work, which is $\sim 5$\arcsec offset to the SO and $\rm SO_2$ emission peak reported in \citet{feng20a} and is $\sim 2$\arcsec offset to the coordinate reported in \citet{burkhardt16}.}  by 1\arcsec--2\arcsec. 
Nevertheless, these nine clumpy substructures disentangled three chemical layers \citep[e.g.,][]{fontani14,codella17} and marked down the kinematic history from B0 to B1-B2 shocks. For example, B0e, B1a, and B2b indicate the knots where episodic ejection impacts against the cavity wall \citep[][]{gueth98,podio16,spezzano20}, B1a-B1c-B1b indicate the shock front, and B1i indicates the possible magnetic precursor \citep{gueth98}.

The field of view is sufficiently large to cover part of the flattened envelope surrounding the  driving source denoted as mm. In the naturally weighted images without pb correction (Figure~\ref{fig:jet}),  lines have the same noise level over the entire map, only the (1,1) line shows $\rm >3\sigma$ emission toward the flattened envelope. Applying the pb correction,  the integrated intensities of all lines toward the phase center B1 are the same as before, but they increase toward B0, B2a, B2b, and mm by a factor of 1.2, 1.3, 1.7, and 4.2 respectively, and all line emissions reveal {part of} the flattened structure  (Figure~\ref{fig:jetpb}). 
Note that the region with $\rm >5\sigma$ emissions on the (2,2)-(7,7) images is  $\sim\rm10$\arcsec~ north of the flattened structure, {a relatively larger portion of which is} revealed by the (1,1) map. The shift for the higher $\rm NH_3$ transitions may be a temperature effect (e.g., the northern warmer portion may be slightly tilted and closer to the heating source). Because this region is  at the edge  of our pb, it is also likely to be contaminated by the sidelobe fringe pattern, {and its missing flux cannot be estimated}. The uniformly weighted images should have less sidelobe effect toward the pb edge, but the worse sensitivity there makes this structure quite diffuse (see Figure~\ref{fig:jetpb}). In the following analysis, we use the naturally weighted images for better sensitivity.

\subsection{Velocity structure}\label{sec:kin}
The integrated intensity map projects the molecular line distribution  in the 2D plane of the sky,  but it misses the velocity information in the line of sight.
Our source contains two successive shocks, as well as an entrained processing jet, which is associated with an outflow. Characterizing the velocity structure resulting in such complicated kinematics is not straightforward. 
Pixel by pixel, {we measure two characteristic velocities:  $V_p$, where the line intensity peaks, and $V_b$, where the line intensity of the blueshifted wing goes down to zero.} 

Traditionally, the peak velocity  $V_p$ toward each pixel can be shown as the ``first moment" (the intensity-weighted average velocity) map or the centroid velocity map provided by the hyperfine multiplets fit to a particular line. Because of the shocks, all spectra observed toward our target show deviations from a Gaussian profile (Figure~\ref{fig:velpro}). Therefore,  neither map can provide a reliable gradient of  $V_p$  over the entire region. {Instead, we  apply the ``ninth moment"  algorithm implemented by CASA and obtain a $V_p$  map
smoother than the ``first moment" map and the centroid velocity map. Similar image quality is also achieved when we test the \texttt{bettermoments} code\footnote{\url{https://github.com/richteague/bettermoments}} \citep[][]{teague18} to fit a quadratic model to the pixel of maximum intensity  and its two neighboring pixels.}

For the spectrum of each pixel, starting from the intensity peak, we modify the ``eleventh moment" algorithm\footnote{https://casa.nrao.edu/docs/casaref/image.moments.html}  implemented by CASA, search blueward to find the first velocity channel (within $\rm -10\,km\,s^{-1}$ to $\rm 5\,km\,s^{-1}$) $V_b$, where the intensity becomes equal to or less than zero.

\begin{figure*}
\centering
\begin{tabular}{lp{4.5cm}p{4.1cm}p{4.cm}p{1cm}p{3cm}}
\multirow{3}{*} {\begin{sideways}Decl. offset (\arcsec)\end{sideways}}
&\includegraphics[clip, trim=0.0cm 1.4cm 0.0cm 0.0cm, height=8.1cm]{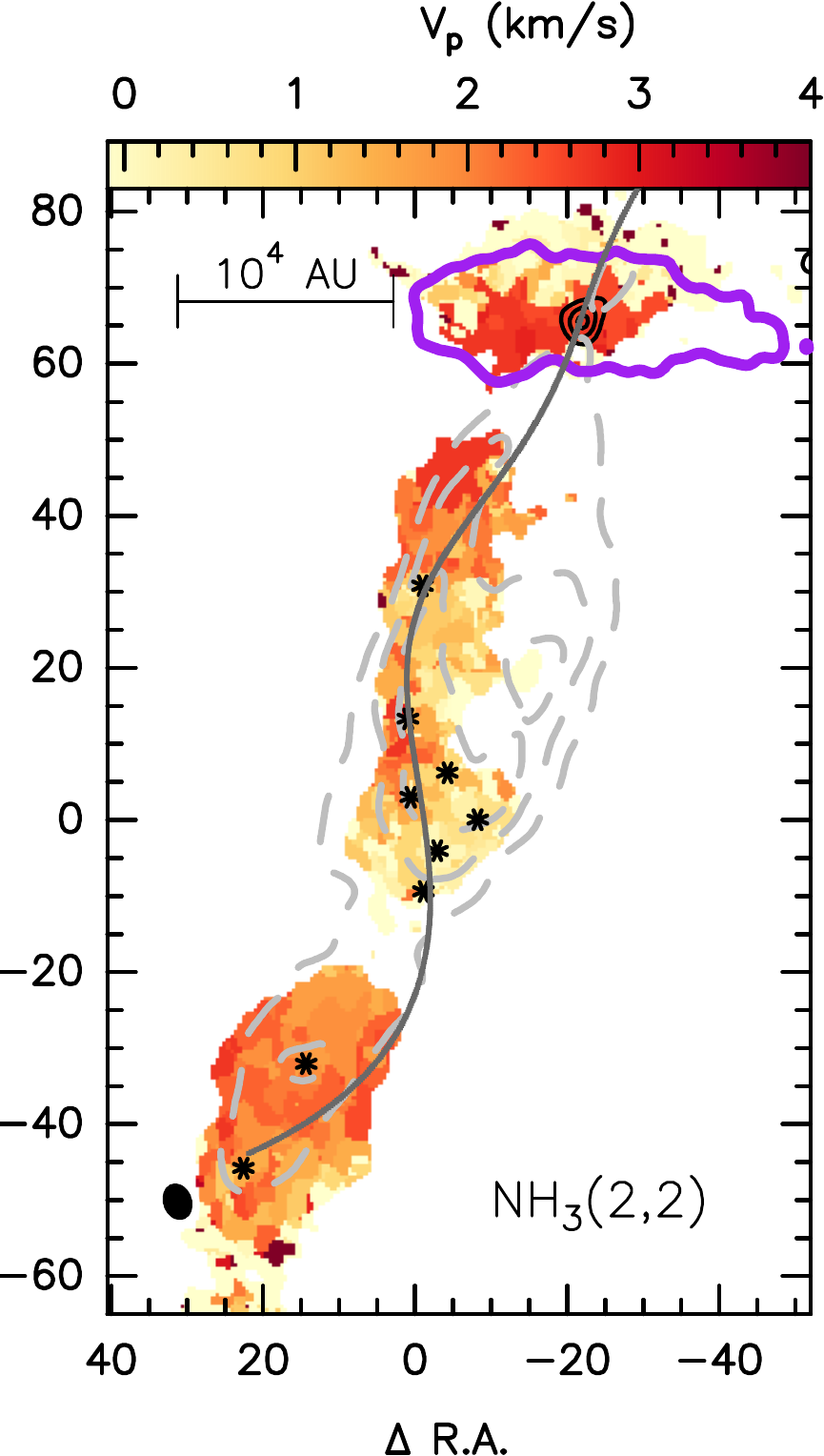}
&\includegraphics[clip, trim=0.9cm 1.4cm 0.0cm 0.0cm, height=8.1cm]{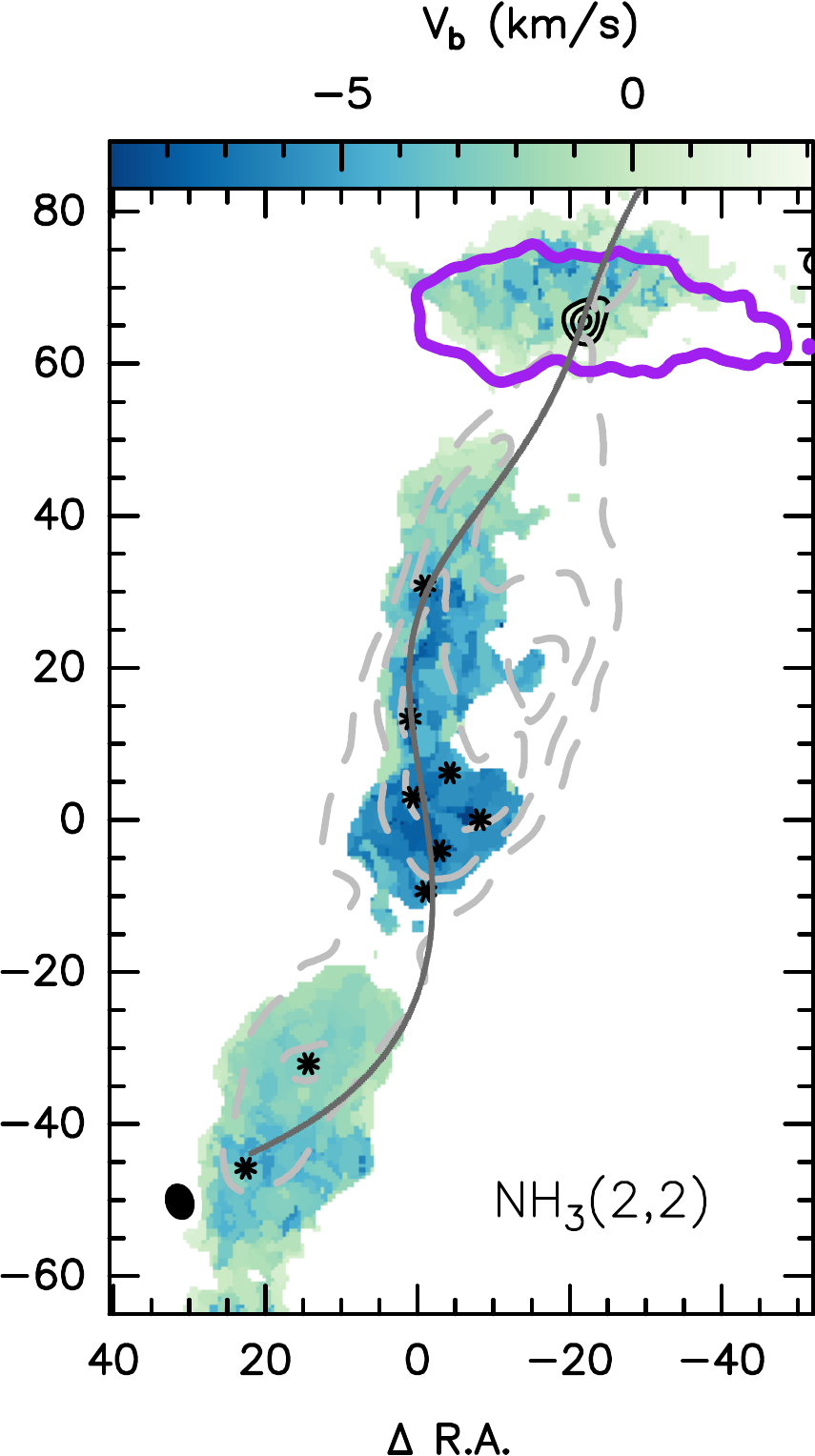}
&\includegraphics[clip, trim=0.9cm 1.4cm 0.0cm 0.0cm, height=8.1cm]{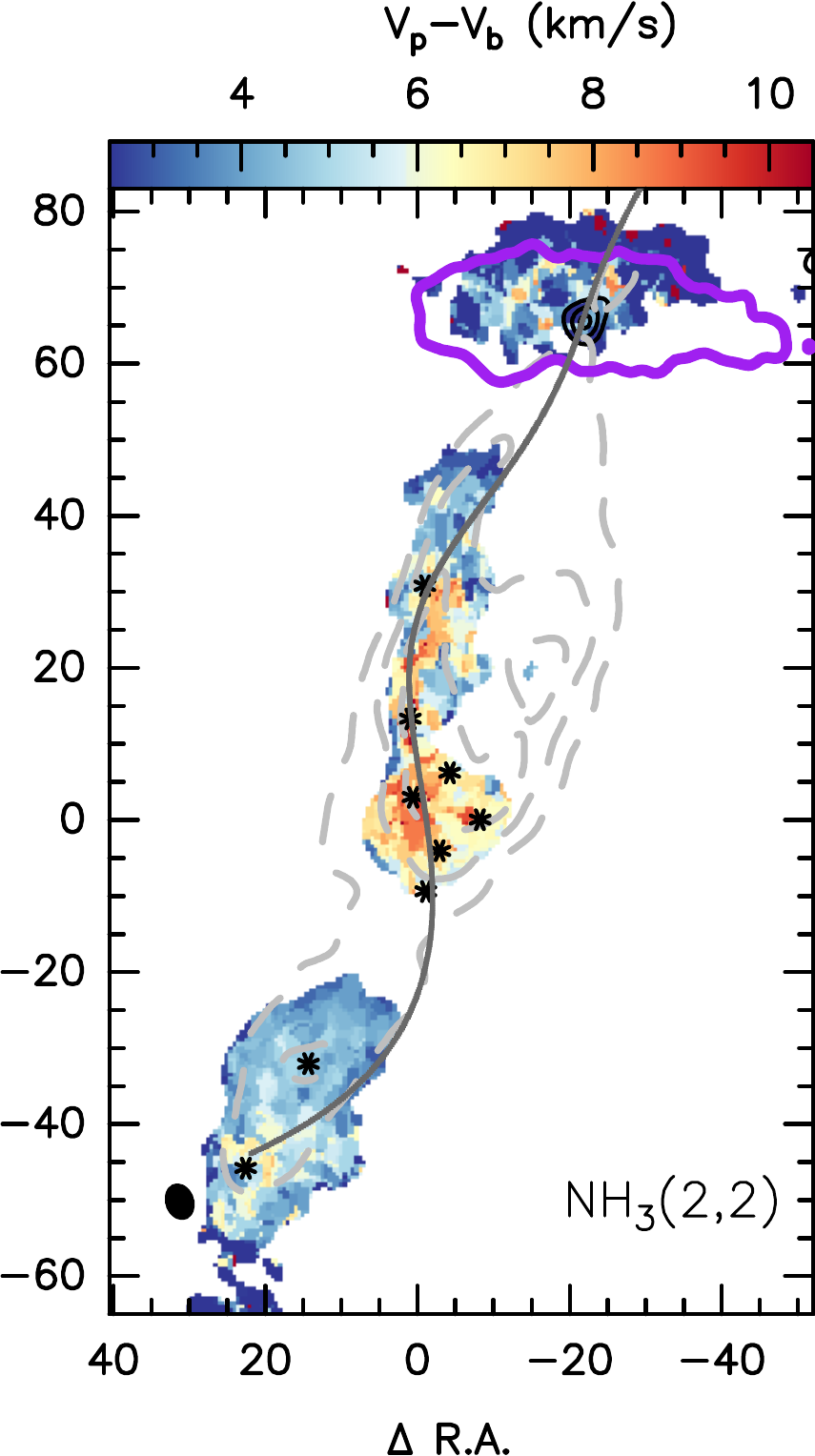}

&~~~~~~~~\multirow{3}{*} {\begin{sideways}$\rm T_B$ (K)\end{sideways}}
&\includegraphics[height=8.3cm]{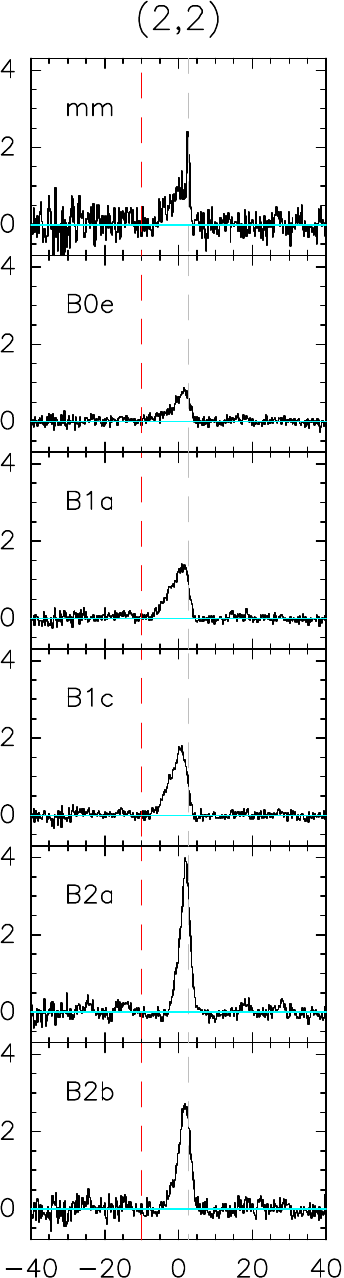}\\

&\includegraphics[clip, trim=0.0cm 0.5cm 0.0cm 0.7cm, height=8.2cm]{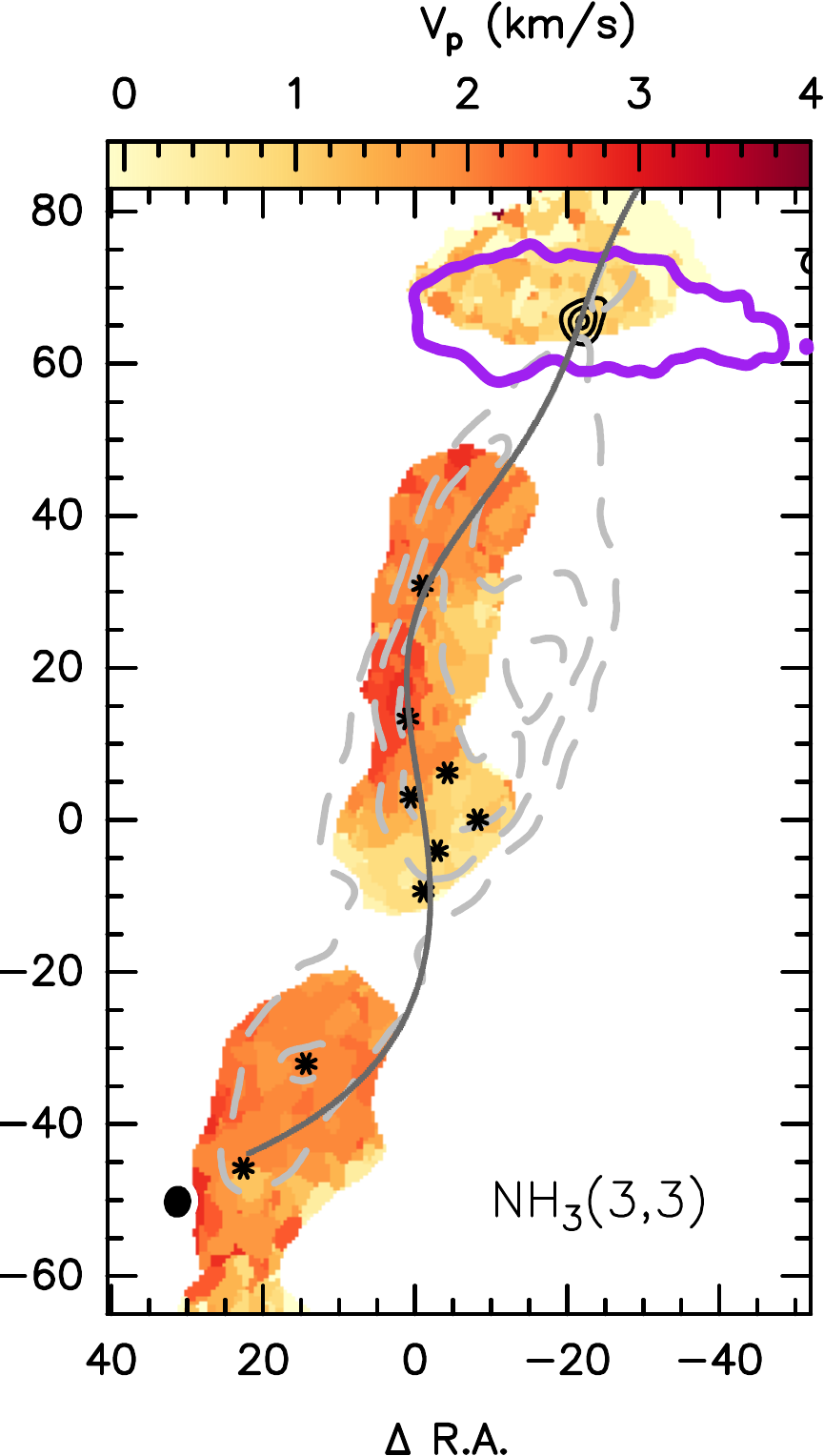}
&\includegraphics[clip, trim=0.9cm 0.5cm 0.0cm 0.7cm, height=8.2cm]{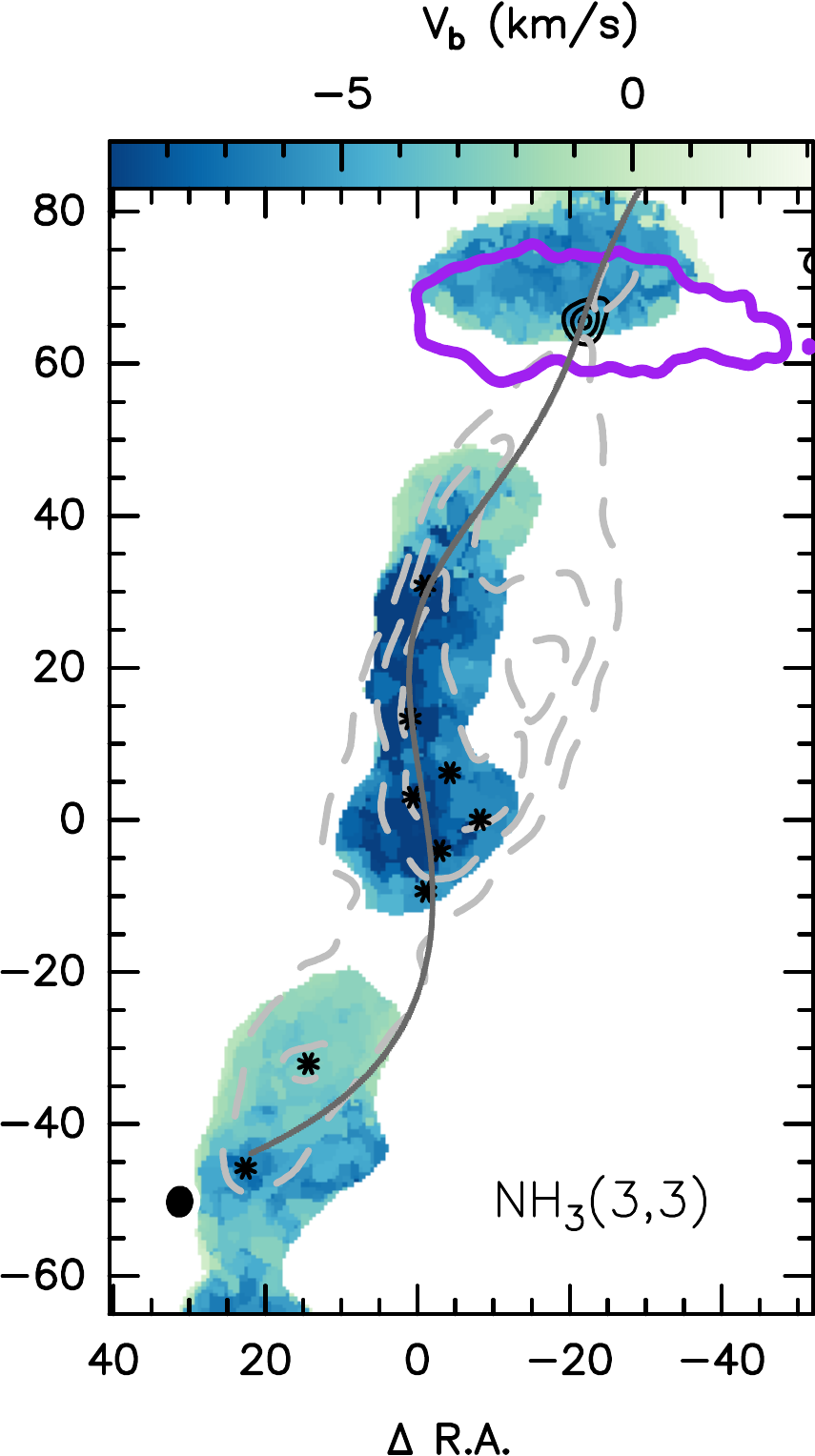}
&\includegraphics[clip, trim=0.9cm 0.5cm 0.0cm 0.7cm, height=8.2cm]{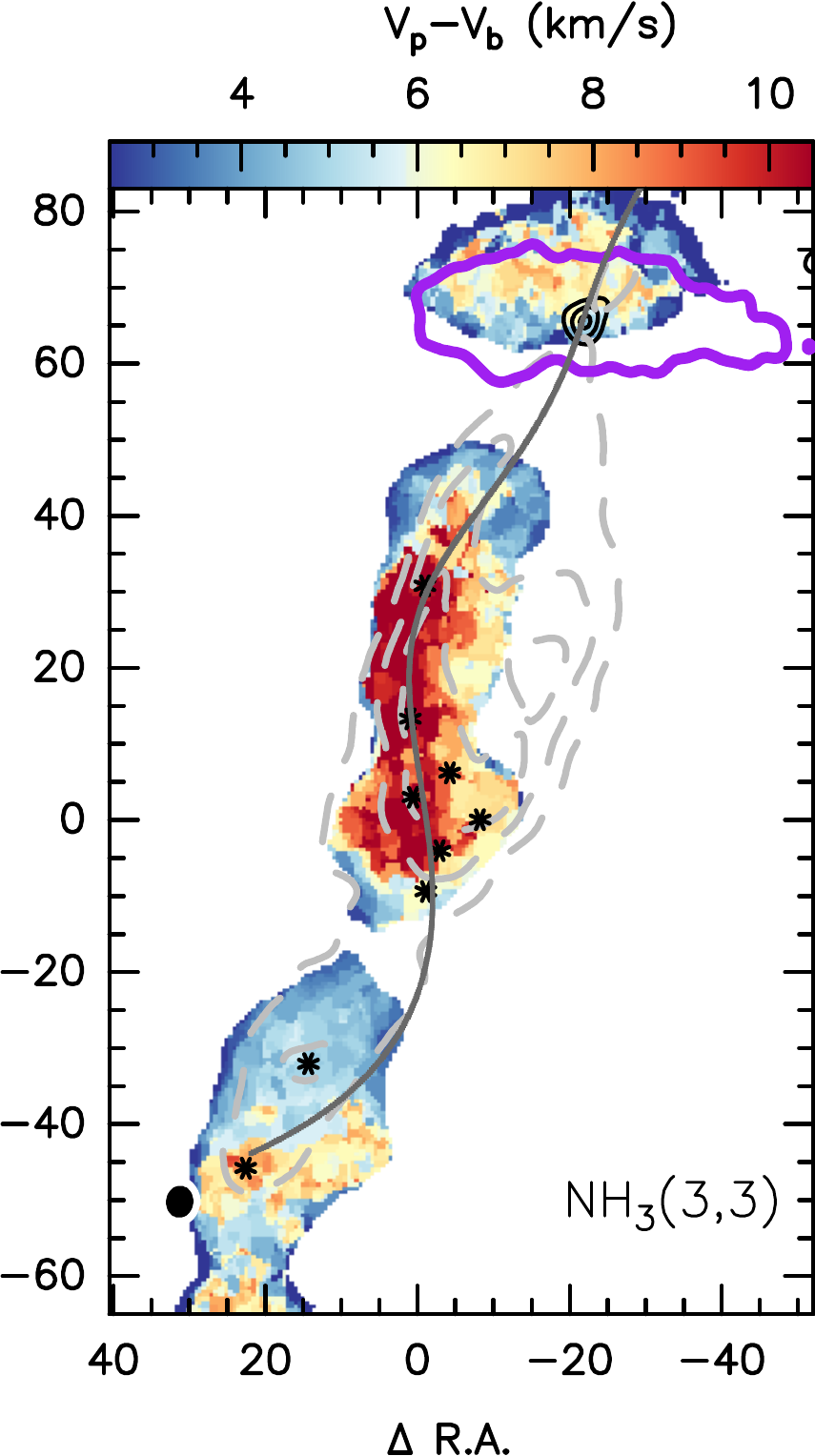}
&
&\includegraphics[height=8.3cm]{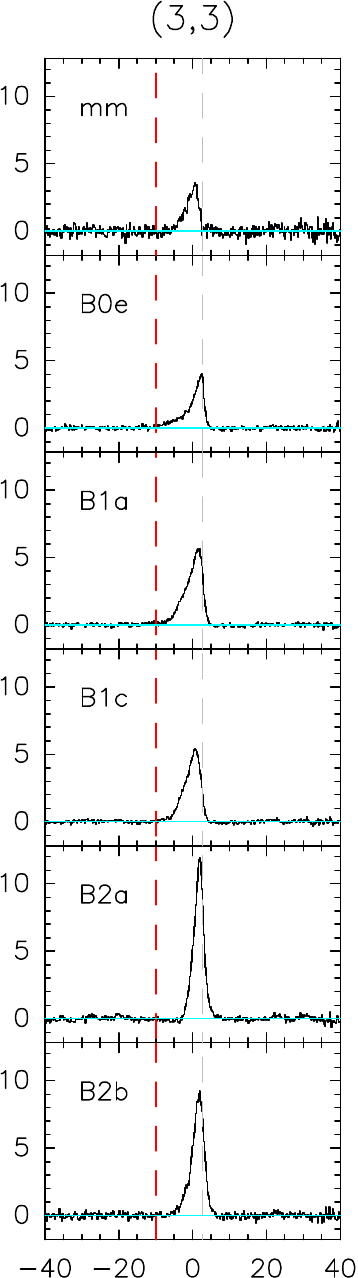}\\

&\multicolumn{3}{c}{R.A. offset (\arcsec)} &&~~~~~{$\rm V_{lsr}$ ($\rm km\,s^{-1}$)}

\end{tabular}
\caption{The velocity structure from the cavity wall B0 to shocks B1 and B2 (the same for both cases with or without pb correction). 
The three left columns show the velocity structure maps of the (2,2) and (3,3)  lines.
The background in each panel shows the velocity map $V_p$, where the line intensity peaks  (the first column),  $V_b$, where the blueshifted wing  extends to  (the second column), and the peak-to-bluest velocity offset $V_p-V_b$ (the third column).
All maps are naturally weighted.
{The black contours are the same continuum emissions as those in Figure~\ref{fig:jet}.}
The purple contour indicates the $\rm 5\sigma$ emission from the (1,1) integrated intensity toward the flattened envelope around the protobinary system, which is perpendicular to the outflow direction. 
The gray dashed contours and the labeled clumpy substructures are the same as shown in Figure~\ref{fig:jet}, and the  synthesized beam is plotted in the bottom-left corner.
 {The pb-corrected} line emissions with $\rm <5\sigma$ rms are blank in each panel. The black curve guidelines the path of the precessing jet from L1157-mm as modeled by \citet[][]{podio16}.  
 The last column shows the (2,2) and (3,3)  line profiles  extracted from clumpy substructures along the jet path. The gray dashed vertical line indicates {the system velocity of the cloud} $V_{sys} \rm = 2.7\,km\,s^{-1}$, and the red dashed vertical line indicates {the velocity where the main component of each line blueward} $V_{lsr} \rm = -10\,km\,s^{-1}$. The line intensity $\rm Jy\,beam^{-1}$ is converted to the brightness temperature $\rm T_{B}$, and the horizontal cyan line indicates the baseline ($\rm T_B$ = 0\,K).
{A narrow feature appears toward the protobinary system mm on the spectrum, probably tracing the quiescent and cold  gas in the flattened envelope.}     
\label{fig:gradient}}
\end{figure*}

We present the $V_p$ (``ninth moment") and $V_b$ (``eleventh moment")  maps of one para ($p$-, $K\neq3n$) $\rm NH_3$ line (2,2) and one ortho ($o$-, $K=3n$) $\rm NH_3$ line (3,3)  throughout B0-B1-B2  in  Figure~\ref{fig:gradient}. These lines were selected because they are observed  at a relatively high-velocity resolution ($\rm 0.197\,km\,s^{-1}$) and show higher S/Ns than the other lines toward the entire region.
{From the $V_p$ maps of both lines,  their intensities peak at around the $V_{sys}$ (2--3\,$\rm km\,s^{-1}$) toward the cavity wall along B0a-B0e-B1a-B2a, while the peaks {are blueshifted} to 0--1\,$\rm km\,s^{-1}$  toward the west of the bow shock B1f-B1b-B1i.
The $V_b$ maps indicate that the blueshifted wings of both lines extend to  -3\,$\rm km\,s^{-1}$ toward B2a, -5\,$\rm km\,s^{-1}$ toward B2b, and further down to -9\,$\rm km\,s^{-1}$ toward B0 and B1.
A lower radial velocity downstream may be a deceleration effect of  the underlying jet or wind after traveling large distances \citep{zhang00}.
In each shock, the pattern of blueshifted wings becoming broader with the distance from the outflow source indicates the ``Hubble-law", which is modelled as a consequence of the bow shock \citep[e.g., ][]{smith97,downes99}.}

When comparing the peak-to-bluest velocity offset ($V_p-V_b$) in the line of sight, we note that a high offset of 10\,$\rm km\,s^{-1}$ is present along the curved jet path.

\section{Analysis and discussion}\label{sec:analysis}
{To determine the  temperature and density profile over a region in the ISM,  the rotational diagram (RD) method was previously widely used with two  assumptions: (1) All lines are in  the local thermal equilibrium (LTE) condition; (2) the lines are optically thin or the optical depth is known to allow correction. However, our test of the RD method shows that  significant blue-shifted emissions on the line profiles toward this shocked region lead to large parameter uncertainties (see Appendix \ref{sec:tau} for details).}

Instead, with  a proper assumption of the source geometry, the large velocity gradient (LVG)  escape probability approximation is preferred to constrain the gas properties  toward a particular location, when the  collision rates of a particular species are known \citep{sobolev60,
goldreich76}.

\subsection{LVG approximation}\label{sec:lvg}
Using the non-LTE statistical equilibrium radiative transfer code RADEX \citep{vandertak07} and a related solver (Fujun Du's \texttt{myRadex})\footnote{https://github.com/fjdu/myRadex.}, we 
apply the MultiNest Algorithm \citep{feroz07,feroz08,feroz13} and derive the probability density function (PDF) of the kinetic temperature $T_{kin}$, the $\rm H_2$ volume density $n$, and the column density $N_T$ for $p$-$\rm NH_3$ and $o$-$\rm NH_3$ toward all pixels. 

Radiative and nonreactive collisional transitions in the gas phase do not change the molecular spin orientations. Therefore, we treat  $o$/$p$-$\rm NH_3$ 
as distinct species and assume that the transitions between them are forbidden (interconversion processes between $o$- and $p$- transitions are ignored).
In  this work, the collision rates are relative to the sum of all the hyperfine components and between  $\rm NH_3$\,(1,1)--(6,6) and $p$-$\rm H_2$, which are provided by the Leiden Atomic and Molecular Database \citep[LAMDA, ][]{schoier05}\footnote{No data are provided for the collision rate between $\rm NH_3$\,(7,7) and $p$-$\rm H_2$ in LAMDA.}.

All the pb-corrected  images are smoothed to the same angular resolution (i.e., $\rm 4.90\arcsec\times3.81\arcsec$) and the same pixel size  (i.e., $\rm 0.5\arcsec$).
The intensities of the lines are integrated over the velocity range of -25 to $\rm +25\,km\,s^{-1}$
 by assuming a single velocity component.
The geometry is set as ``LVG" (``Slab" was tested and  no appreciable differences were shown). The full width at half maximum (FWHM) line widths of all lines are adopted as $\rm 4\,km\,s^{-1}$, based on the median from all spectrum fittings (see Table~\ref{tab:hfsline}). {In the case that multivelocity components are smeared within the VLA's synthesized beam, an FWHM line width  of 9\,$\rm km\,s^{-1}$ is adopted from \citet[][]{bachiller93} and \citet[][]{lefloch12}  for the test.
Two different values have been considered for the beam-filling factor: (i) unity for all lines; and (ii) 0.5 or 0.1 for transitions higher than (3,3).}
The best-fit parameters are searched over relatively large ranges for the $\rm H_2$ number density $n$ ($\rm 10^3$--$\rm 10^7\,cm^{-3}$), $T_{kin}$ (5--$\rm 10^{4}$\,K), and $N_T$ ($\rm 10^{14}$--$\rm 10^{17}\,cm^{-2}$).

An example of the PDF from the LVG approximation  is shown toward B1a in Figure~\ref{fig:lvgeg}, and examples of parameter maps are given in Figure~\ref{fig:lvgtestmap}. With different line width and beam-filling factor combinations, the best fits  toward all nine clumps are listed in Table~\ref{tab:lvgfit}.
Although degeneracy among the above three parameters is inevitable, we found that  the $T_{kin}$ and the $N_T$ for $o$- and $p$-$\rm NH_3$ are well constrained toward these representative positions, with uncertainties less than 30\%. 

From Figure~\ref{fig:lvgeg} we also note that the $n$ of each clump cannot be constrained from the LVG fittings, {and thus, varying its value in the aforementioned range does not change the best-fit result of   $T_{kin}$ and the $N_T$.}
In this warm shocked environment, the critical densities of all seven  $\rm NH_3$ lines {are in the range of $\rm 5\times10^3$--$\rm 3\times10^4\,cm^{-3}$ (estimated by using the Einstein coefficients and the collision rates provided by LAMDA in the temperature range of 50--300\,K, which are consistent with the effective critical density given by, e.g., \citealp{shirley15}). }
Previous CO and CS observations from \citet[][]{lefloch12,benedettini13,gomezruiz15}  indicate  $n\rm >10^4\,cm^{-3}$ at an angular resolution in the range of 3\arcsec--20\arcsec, which should validate the LTE assumption for the $\rm NH_3$ lines. 
{Because this number density cannot be confirmed from our LVG fittings, the RD method would need to be applied with caution.}

{Our tests show that a smaller beam-filling factor for higher transitions result in better consistency of the  $T_{kin}$ derived from $o$-/$p$-$\rm NH_3$  separately (Table~\ref{tab:lvgfit}). The collisional coefficients, the line width and the integrated intensity of each line adopted in the fittings change the best-fit absolute value of $T_{kin}$ and  $N_T$ toward individual pixels, though they do not change the contrast  between pixels for both parameter maps (see Appendix~\ref{app:lvg} for details). }

\subsection{The gas kinetic temperature map}\label{sec:tkin}

{Assuming a line width of $\rm 4\,km\,s^{-1}$, a unity beam-filling factor for (1,1)--(3,3) and  0.1 for (4,4)--(7,7), Figure~\ref{fig:lvgmap} shows the maps of gas kinetic temperature, the relative abundance ratio between $o$- and $p$-$\rm NH_3$ and their total column density, derived from the above-mentioned LVG best fit.}   

In a 0.1\,pc scale field of view, good agreement on $T_{kin}$ derived from $o$/$p$-$\rm NH_3$ separately (relative ratio $\sim$1, shown in yellow in the first column of Figure~\ref{fig:lvgmap}, with $\rm <10$\% uncertainty) indicates the appropriateness of these assumptions.

Throughout the entire southern outflow shocks, the mean $T_{kin}$ map reveals an intrinsic gradient (second column  of Figure~\ref{fig:lvgmap}):  
{warm components  ($\rm >120K$) appear toward the spots where the jet impinges on the eastern cavity walls (B0a-B0e-B1a), cool components ($\rm <80K$) toward  the cavity B1b-B1c and the older shock B2.  }

{
Although the temperature gradient from B0 to B2 has been roughly indicated by  previous observations  \citep[e.g, ][]{umemoto99,tafalla95} and was proposed by the model \citep[e.g., ][suggested a slightly larger temperature gradient from B0 with $>200$\, K  to B1 with $\sim$60\,K  and B2 with $\sim$20\,K]{lefloch12,podio14} }, it has never been confirmed by a detailed map until this work. At a linear resolution of $\sim$1500\,au, our refined results indicate that the older components have experienced more postshock cooling.

\subsection{The column density map and the ortho-to-para ratio  map of $\rm NH_3$}\label{sec:opr}

Given that  no continuum {at millimeter and centimeter} wavelengths is detected toward B0-B1-B2 from the existing data and that the molecular abundance with respect to $\rm H_2$ measured from different observations has a large uncertainty especially in the shocked region, the column density and ortho-to-para ratio (OPR) of $\rm NH_3$ are more reliable physical indicators than its abundance with respect to $\rm H_2$  in this work.

\begin{figure*}[tbh]
\centering
\begin{tabular}{lp{4.5cm}p{3.95cm}p{3.95cm}p{3.95cm}}

\multirow{1}{*} {\begin{sideways}Decl. offset (\arcsec)\end{sideways}}
&\includegraphics[clip, trim=0cm 1.4cm 0.0cm 1.0cm, height=7.75cm]{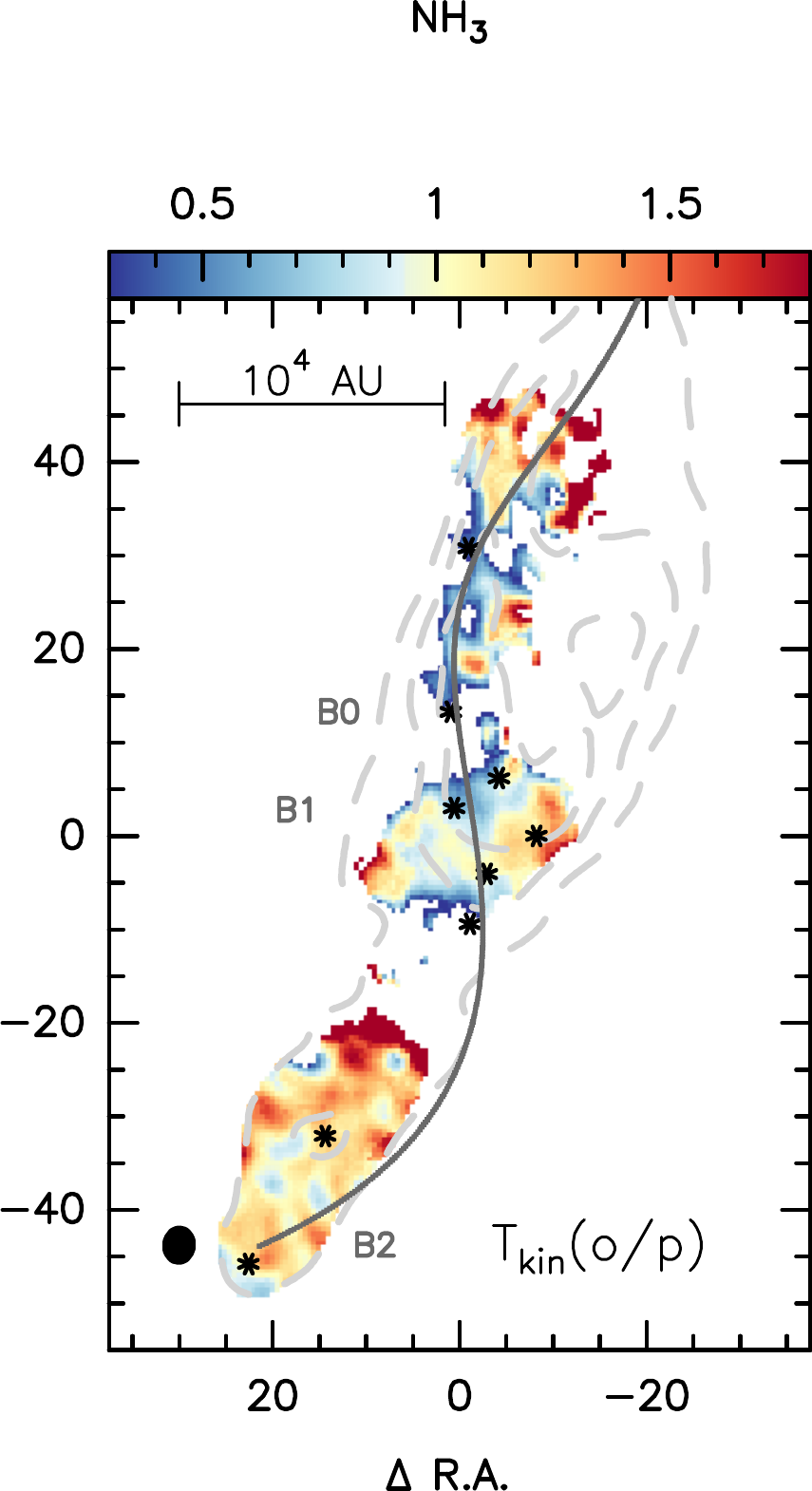}
&\includegraphics[clip, trim=1.cm 1.4cm 0.0cm 1.0cm, height=7.75cm]{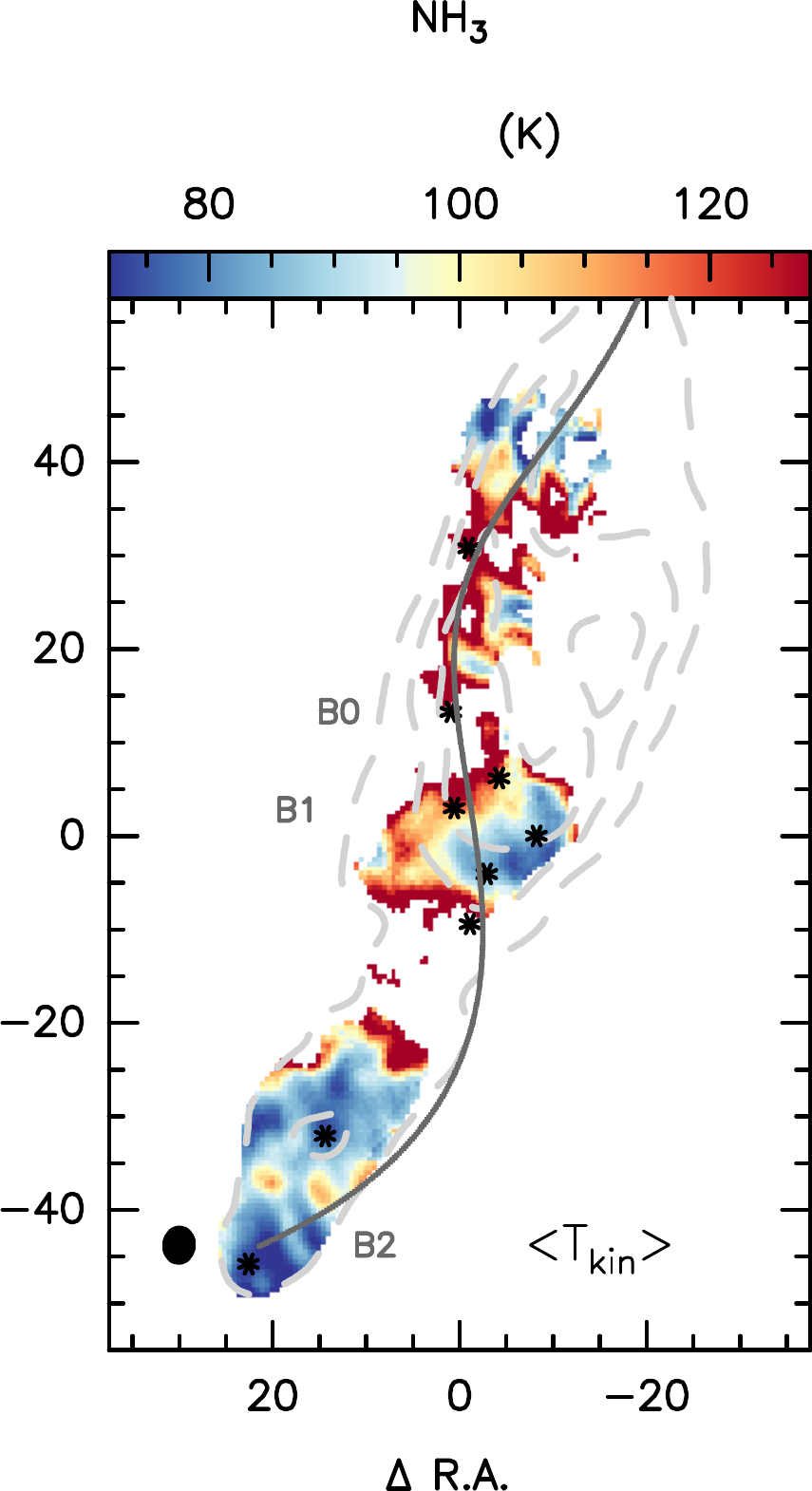}
&\includegraphics[clip, trim=1.cm 1.4cm 0.0cm 1.0cm, height=7.75cm]{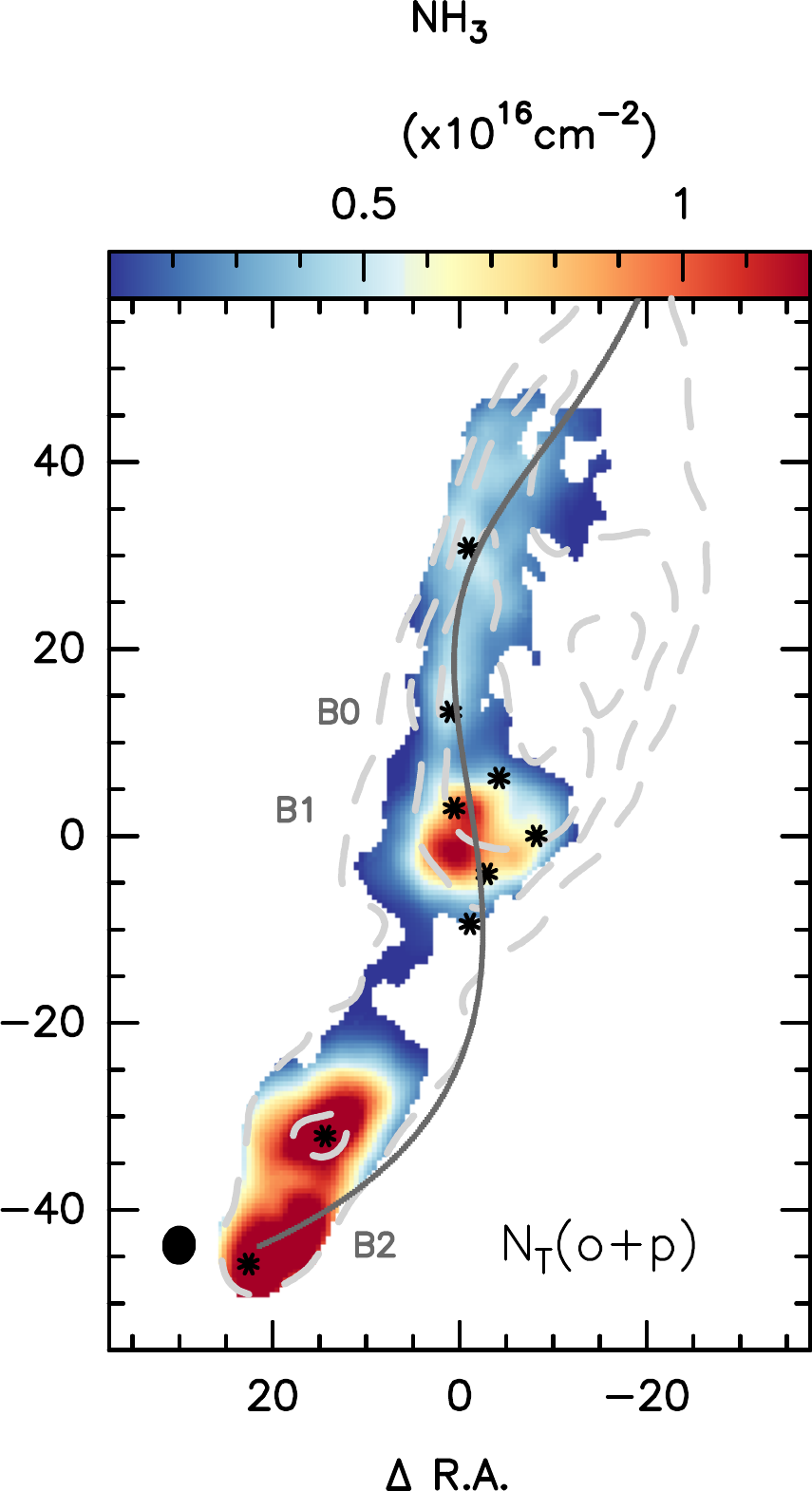}
&\includegraphics[clip, trim=1.cm 1.4cm 0.0cm 1.0cm, height=7.75cm]{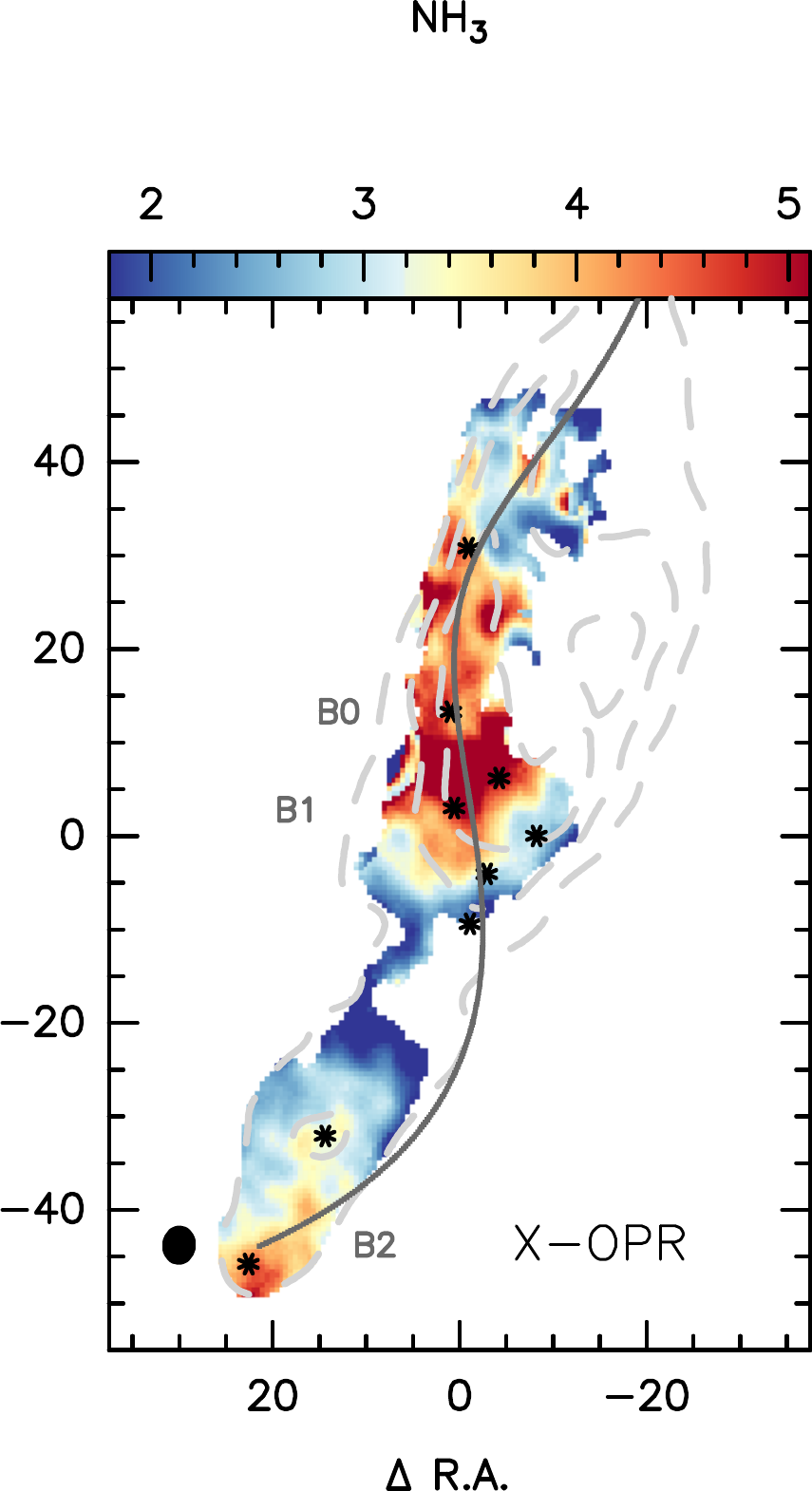}\\

&\includegraphics[clip, trim=0cm 0.5cm 0.0cm 1.5cm, height=7.95cm]{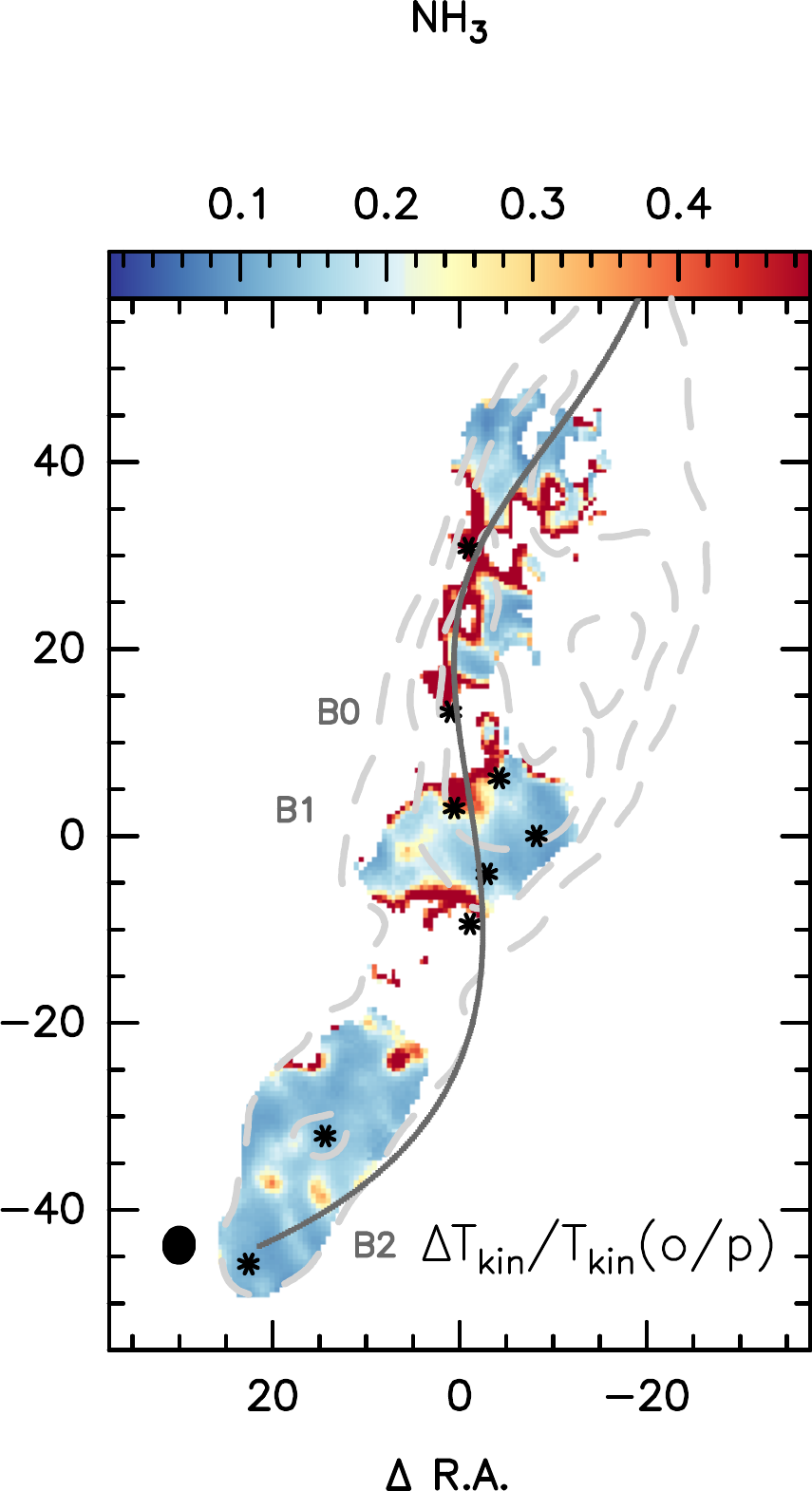}
&\includegraphics[clip, trim=1.cm 0.5cm 0.0cm 1.5cm, height=7.95cm]{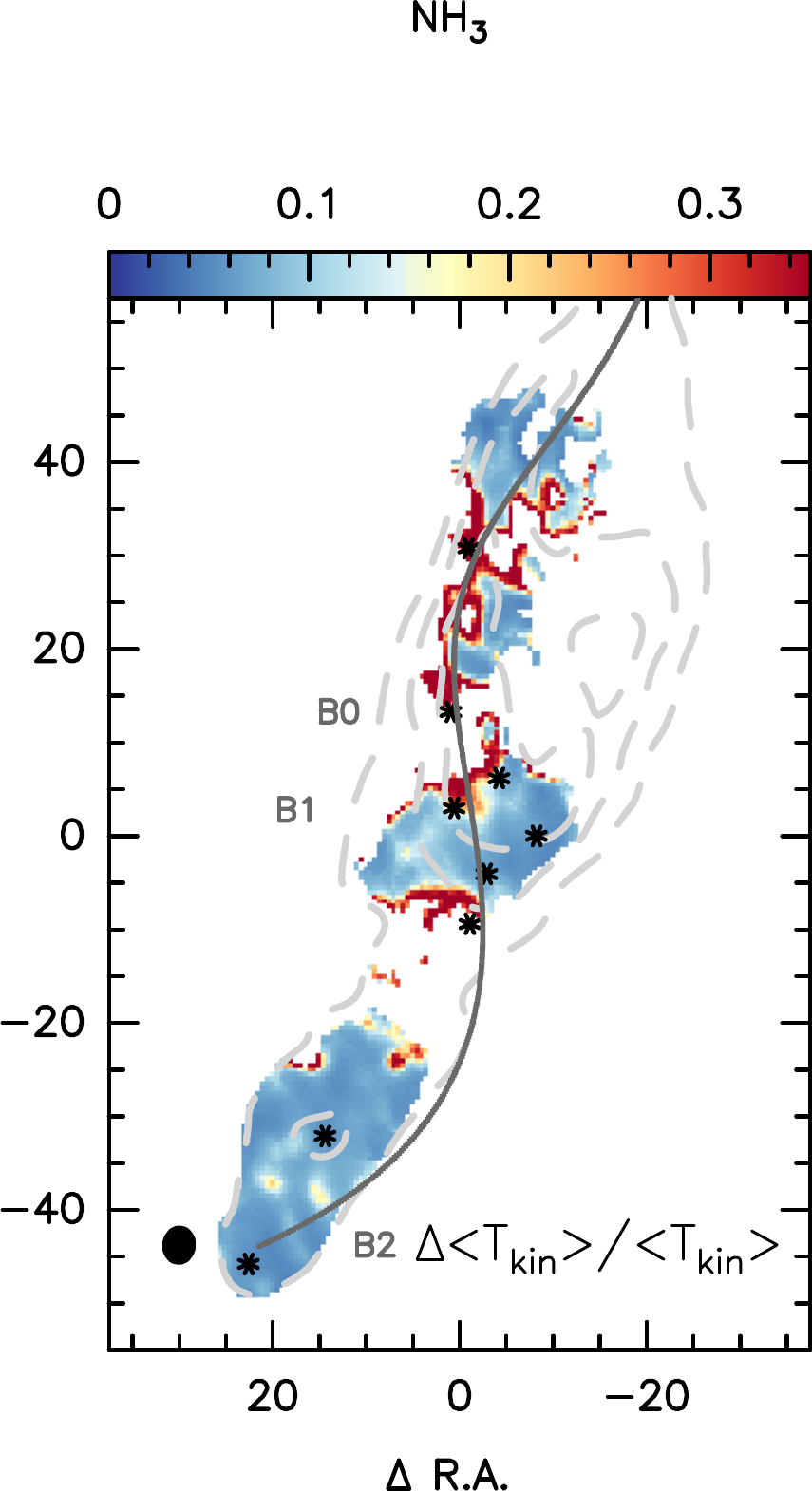}
&\includegraphics[clip, trim=1.cm 0.5cm 0.0cm 1.5cm, height=7.95cm]{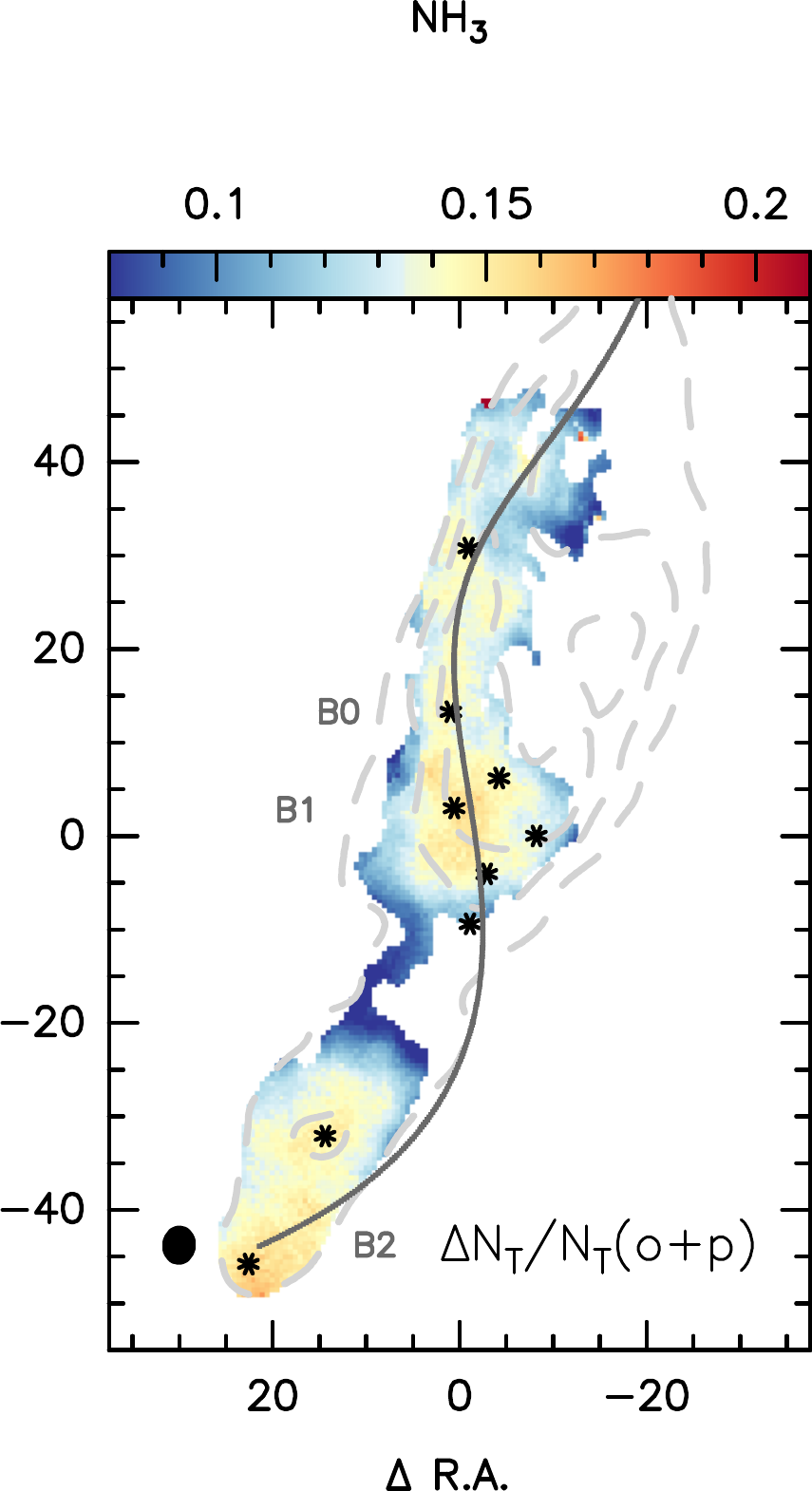}
&\includegraphics[clip, trim=1.cm 0.5cm 0.0cm 1.5cm, height=7.95cm]{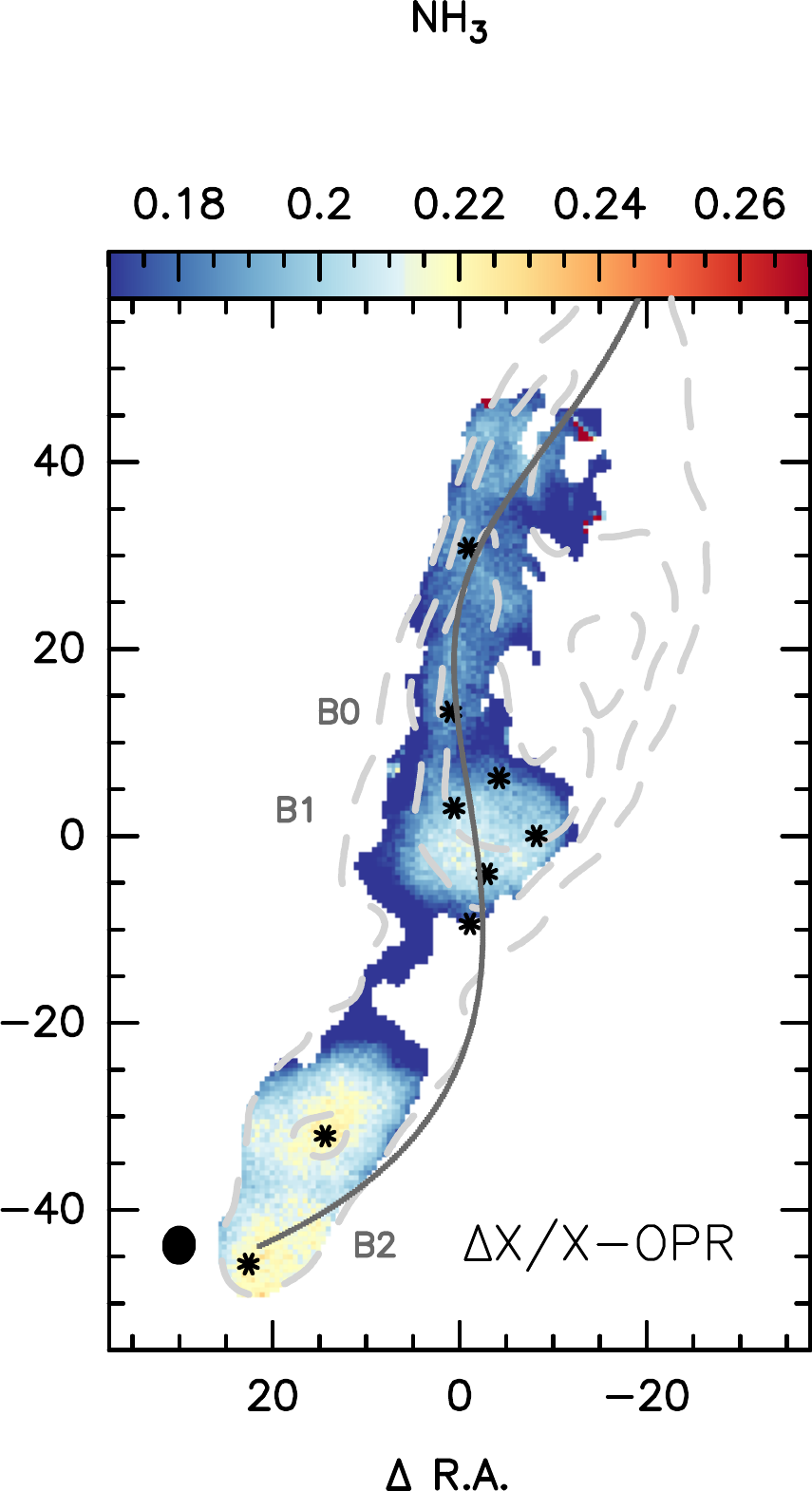}\\

&\multicolumn{4}{c}{R.A. offset (\arcsec)}
\end{tabular}
\caption{Kinetic temperature maps and column density maps for $o$- and $p$-$\rm NH_3$, derived from LVG MultiNest best fitting, by assuming a single component (with an FWHM line width of $\rm 4\,km\,s^{-1}$) 
{and two filling factors (1 for (1,1)--(3,3) lines and  0.1 for (4,4)--(7,7) lines) toward each position.}
Panels from left to right in the upper row show  the relative ratio  and the mean of $T_{kin}$ from $o$-/$p$-$\rm NH_3$, as well as the total column density and the OPR of $\rm NH_3$.
Panels in the lower row show the uncertainty with respect to the value of the above parameters.
The pixels where  {the pb-corrected} $\rm NH_3$ (5,5) and (6,6) lines $\rm <5\sigma$ emission and CO\,(1--0) $\rm <8\sigma$ emission are blank.  
The gray dashed contours and the labeled clumpy substructures are the same as shown in Figure~\ref{fig:jet}.
The black curve guidelines the path of the precessing jet from L1157-mm as modeled by \citet[][]{podio16}. 
These maps are derived by smoothing all lines to the same pixel size and angular resolution, i.e., $\rm 4.90\arcsec\times3.81\arcsec$,  
shown as the synthesized beam in the bottom left of each panel.
\label{fig:lvgmap}}
\end{figure*}

A U-shape structure appears on both  $o$- and $p$-$\rm NH_3$ column density map toward B1, connecting the spots B1a-B1c-B1b, with the total value reaching $\rm >1\times10^{16}\,cm^{-2}$ to the eastern wall.  Two U-shape structures  also appear toward B2a and B2b, with the column density peak twisting toward the western wall (third column  of Figure~\ref{fig:lvgmap}).  {These U-shapes are significant on the integrated intensity maps, especially for the lines observed with larger pb. 

Measuring the Spearman's rank correlation\footnote{The SpearmanÕs rank correlation coefficient $\rho$ is a nonparametric measure of statistical dependence between two variables. This coefficient can assess how well a monotonic function (no matter whether linear or not) can describe the relationship between two variables. The coefficient $\rho$ is in the range from -1 (decreasing monotonic relation) to 1 (increasing monotonic relation), with zero indicating no correlation.} coefficient $\rho$ \citep{cohen88}, the total column density shows moderate (0.4--0.5) spatial correlation with the velocity offset  in the line of sight ($V_p-V_b$)  toward B0 and B2 locally, as well as strong  (0.63)  correlation toward B1 locally (Figure~\ref{fig:correlation}). However, the correlation is weak (0.18) if treating the successively shocks as one entity.
This could imply that the sputtering of ammonia off the surface of ice grains is not very sensitive to shock {at a velocity up to $\sim\rm 10\,km\,s^{-1}$}, possibly indicating that $\rm NH_3$ is efficiently returning in the gas phase already at the lowest values of $V_p-V_b$ (which may then be related to the shock velocity threshold for ice sputtering). 
}

 {The OPR seems more enhanced along the curved jet path (B0a-B0e-B1a-B1c-B2b) compared to the rest by a factor of 2--2.5   (the fourth column  of Figure~\ref{fig:lvgmap}). Interestingly, } these locations where the OPR reaches a maximum (as high as 4--5 or 2.8--3.5 when adopting a line width of $\rm 4\,km\,s^{-1}$ or $\rm 9\,km\,s^{-1}$, respectively) are spatially coincident with the largest velocity offset  in the line of sight ($V_p-V_b\rm \sim 10\,km\,s^{-1}$)  in Figure~\ref{fig:gradient}, showing a strong Spearman's rank correlation  ($\rho\rm >0.5$)  toward  the entire mapping region (Figure~\ref{fig:correlation}). {Although the locations of the maximum OPR and  the total column density peak are not identical, the Spearman's rank correlation between these parameters is strong ($\rho\rm >0.5$)  toward  B0, B1, and B2 locally. Such spatial correlation has also been reported toward other protostellar shocked regions  \citep[e.g., shocks associated with Orion-KL,][]{goddi11b}. }

Considering that the uncertainties from observation, data reduction, and LVG estimations ($\rm <15\%$) are uniform for the entire map, we believe
the enhancement of column density toward the cavity, the enhancement of OPR toward the jet path,  {as well as its spatial correlation with the velocity offset are intrinsic dynamic effects}.

The OPR gradient revealed by our observations at high angular resolution improves the single-dish result from  \citet[][]{umemoto99}\footnote{\citet[][]{umemoto99} assumed all lines are optically thin and reported the OPR toward B1 to be  $\sim1.7_{0.3}^{+0.2}$, while toward B2 to be $\sim1.7\pm0.2$.}. {Enlightened by the explanation from the comprehensive gas-grain chemical network in \citet{harju17}, 
surface and gas-phase processes may have different contributions along the cavity walls and within the cavity.
}
{According to  \citet[][]{faure13}, OPR $<$ 1 can be reproduced if $\rm H_2$ is mainly in para form. Therefore, a high OPR (2--5) given from our study may be the result of  $o$-$\rm H_2$, which is not included in our LVG fitting but whose abundance may indeed increase as expected in shocked gas (see discussion in Appendix~\ref{app:lvg})}.

\subsection{The flattened envelope}\label{sec:envelope}
Previous studies have revealed that
the flattened structure surrounding L1157-mm is a  mixture of outflow components with the inner envelope on the scale of thousands of astronomical units \citep[e.g., ][]{looney07,chiang10,tobin10,lee12}. Although  the protobinary system mm is at the edge of our pb where the noise level is high from the pb-corrected images (Figures~\ref{fig:velpro}--\ref{fig:jetpb}),  line emissions especially those from (1,1) and (2,2) have $\rm S/N> 3$. 

\begin{figure*}
\centering

\begin{tabular}{lp{4.3cm}p{1cm}p{4.3cm}}
\multirow{2}{*}{\begin{sideways}Decl. offset (\arcsec)\end{sideways}}
&\multicolumn{1}{c}{\includegraphics[clip, trim=0.0cm 0.9cm 0.0cm 0.0cm, width=9cm]{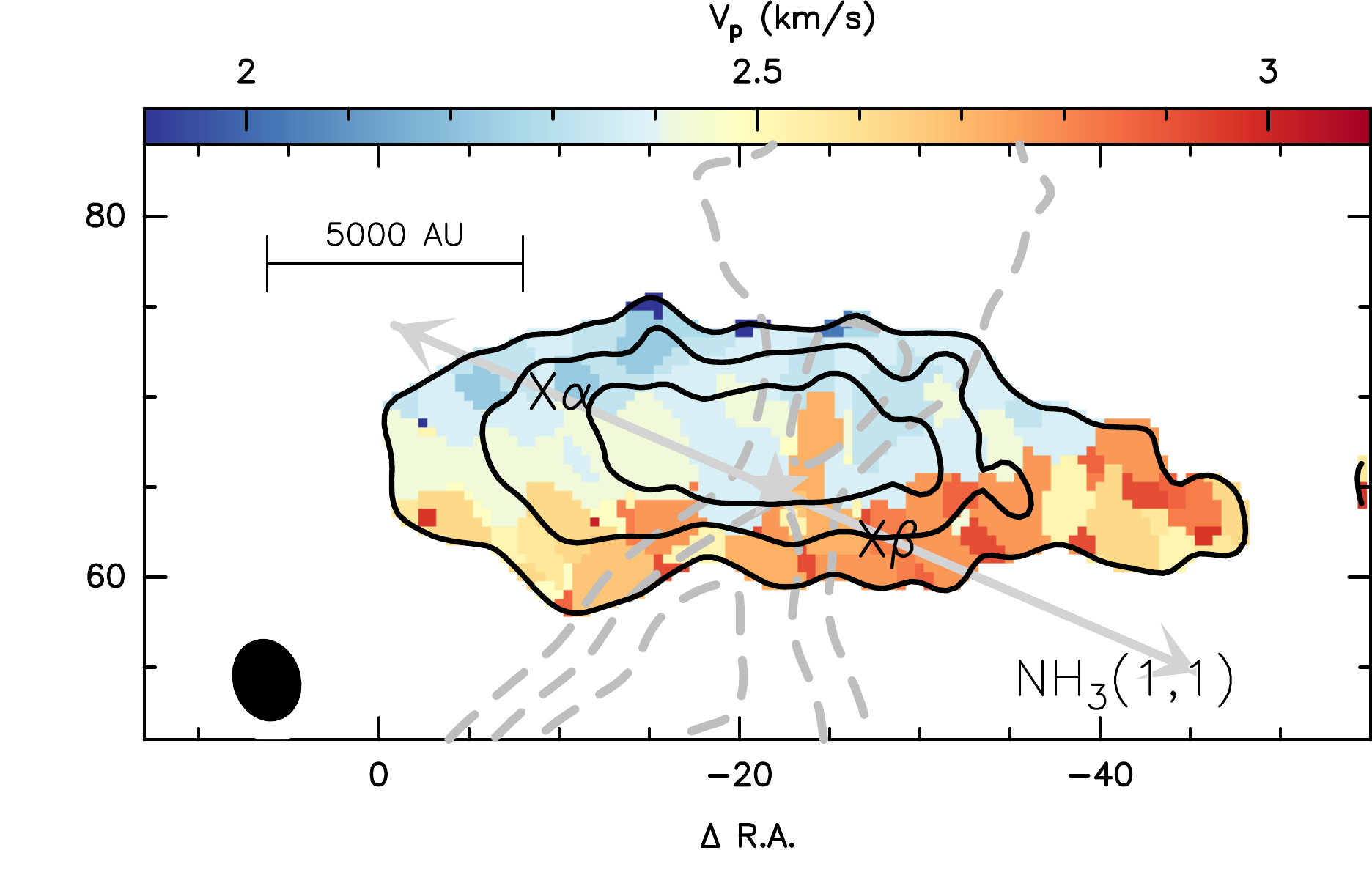}}
&~~~~~~~~\multirow{2}{*}{\begin{sideways}$T_B$ (K)\end{sideways}}
&\multicolumn{1}{c}{\includegraphics[clip, trim=0.0cm 0.0cm 0.0cm 0.0cm, height=5cm]{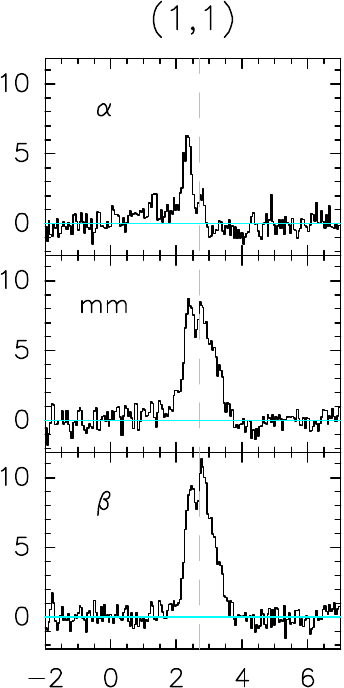}}\\

&\multicolumn{1}{c}{\includegraphics[clip, trim=0.0cm 0.9cm 0.0cm 0.5cm, width=9cm]{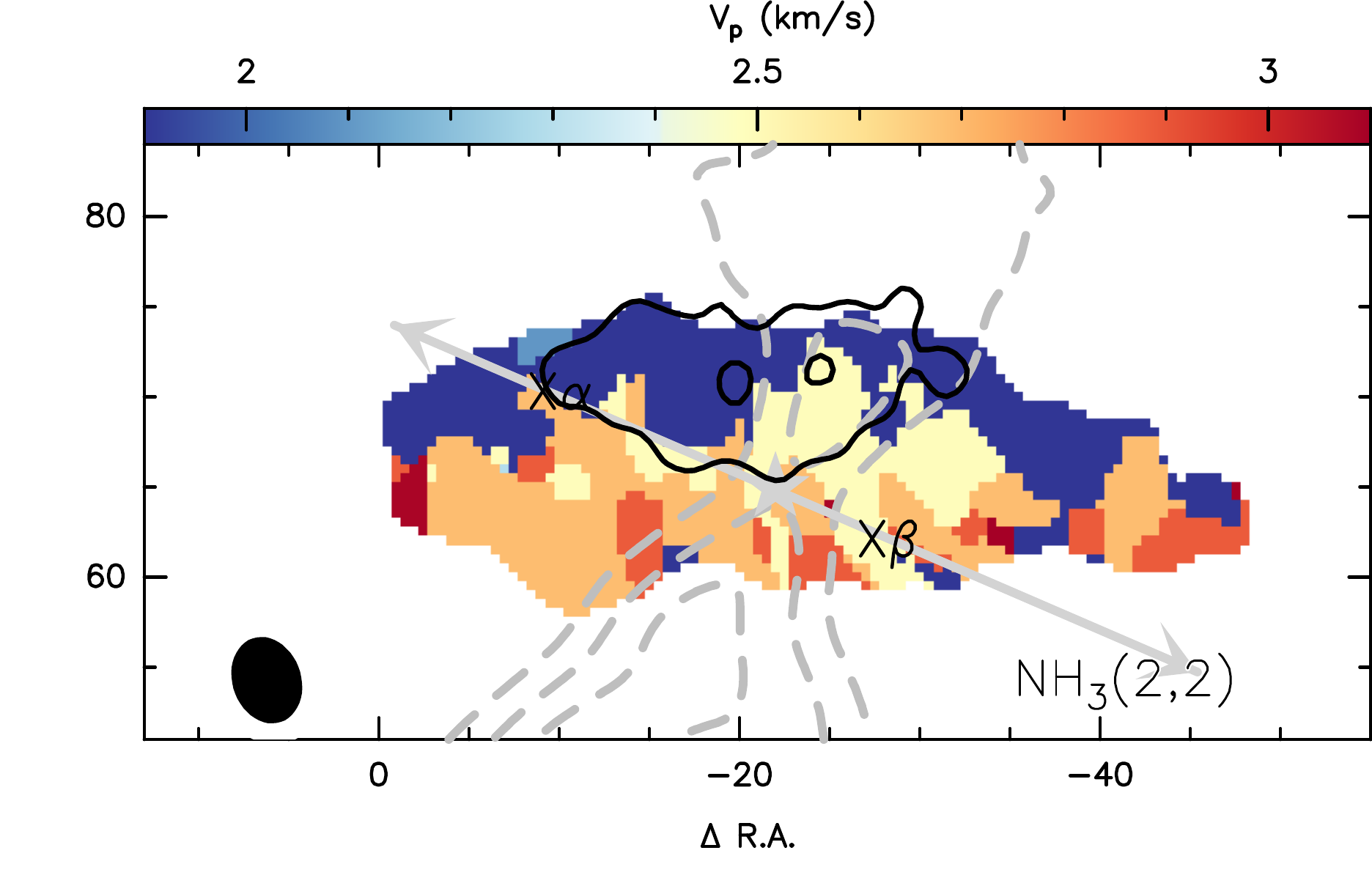}}
&
&\multicolumn{1}{c}{\includegraphics[clip, trim=0.0cm 0.0cm 0.0cm 0.0cm, height=5cm]{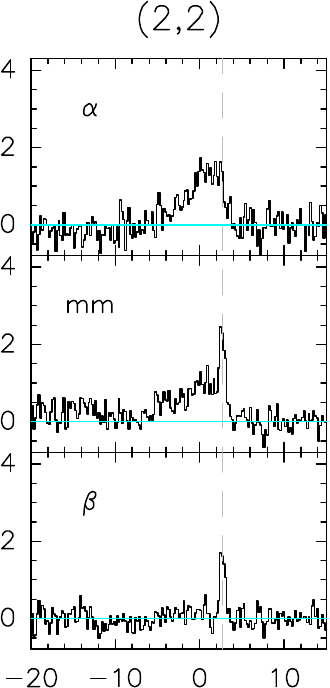}}\\
&\multicolumn{1}{c}{R.A. offset (\arcsec)} 
&&\multicolumn{1}{c}{$V_{lsr}$ ($\rm km\,s^{-1}$)}
\end{tabular}

\caption{Left column: The $V_p$ map of the main hyperfine component of the (1,1) and (2,2) lines  toward the flattened envelope around the marked protobinary system, which should elongate in the direction shown as the gray two-headed arrow  \citep{chiang10}, perpendicular  to the outflow.
In each panel, the black contours indicate the integrated intensity  (with pb correction), starting from $\rm 5\sigma$  and increasing by $\rm 3\sigma$. 
The {gray dashed contours}, starting from $\rm 4\sigma$ ($\sigma \rm = 0.27\,Jy\,beam^{-1}$) and increasing with the step of  $\rm 5\sigma$, show the CO (1-0) emission \citep{gueth96}, and the  synthesized beam is plotted in the bottom-left corner.
 {The pb-corrected} (1,1) line emissions with $\rm <5\sigma$ rms are blank in each panel.  
Right column: beam-averaged line profiles  extracted from the protobinary system and two positions perpendicular to the outflow direction in the flattened structure (labeled as $\alpha$ and $\beta$ on the $V_p$ map). The gray dashed vertical line indicates the $V_{sys} \rm = 2.7\,km\,s^{-1}$ of the cloud. The line intensity $\rm Jy\,beam^{-1}$ is converted to the brightness temperature $\rm T_{B}$, and the horizontal cyan line indicates the baseline ($\rm T_B$ = 0\,K).
\label{fig:flatten}}
\end{figure*}

{To study the velocity gradient across the flattened envelope, we present the $V_p$ maps of both lines within the region where the (1,1) line shows  $\rm >5\sigma$ emission after pb correction (Figure~\ref{fig:flatten}).}
Comparing the beam-averaged line profiles from the protobinary system mm, a northeast position $\alpha$, and  a southwest position $\beta$ perpendicular to the outflow, we find that {a $\rm 1\,km\,s^{-1}$ width narrow feature appears toward mm on the spectra of the (2, 2) line. Such feature appears on the (3,3) line as well (Figure~\ref{fig:gradient}), probably tracing the quiescent and cold  gas in the flattened envelope. 
The blueshifted wing 
may trace the warmer material partially belonging to the outflow.}

A velocity gradient of the $\rm N_2H^+$\,(1--0) line is reported by \citet{chiang10} and \citet{tobin10}, which is normal to the outflow elongation and indicates the presence of rotation.
{In contrast, our $\rm NH_3$ lines present a velocity gradient from the northeast ($\sim \rm 2\,km\,s^{-1}$) to the southwest ($\sim \rm 3\,km\,s^{-1}$).
Nevertheless, this gradient} is the same as what was shown on the JVLA $\rm NH_3$\, (1,1) map  by \citet{tobin10}, covering $\rm \pm30\arcsec$~ east and west offset to the protobinary system mm.
Another consistency with \citet{tobin10} is the red-shifted  ($\sim \rm 2.7\,km\,s^{-1}$)  gas of $\rm NH_3$, which seems to be curved down toward the B0.
Therefore, instead of an effect of sidelobe contamination (e.g., in dark blue on the (2,2) map of Figure~\ref{fig:flatten}), we believe the northeast to southwest velocity gradient of the $\rm NH_3$ lines indicates the interaction between the outflow and the envelope.

\section{Conclusions}\label{sec:conclusion}
Our newest JVLA observations provide high-sensitivity images of $\rm NH_3$\,(1,1)--(7,7)  lines toward the archetypal protostellar shock, which is associated with  the chemically rich blueshifted outflow in L1157. In the 0.1\,pc scale field of view,  we draw a detailed kinetic temperature {and $\rm NH_3$ column density} maps of this successively shocked region  for the first time  at a linear resolution of 1500\,au.  Our conclusions are as follows:\\

\begin{enumerate}
\item The emissions of all seven lines highlight the curved precessing jet path, from the eastern cavity wall of B0 and B1 to the cavity B2.  In the line of sight, the map of the peak-to-bluest velocity offset 
reaches as high as 10\,$\rm km\,s^{-1}$ on this path.

\item Treating $o$- and $p$-$\rm NH_3$ as distinct species, the kinetic temperature maps from LVG  fittings show good agreement.  {The high-precision ($\rm <10\%$ uncertainty) temperature map  reveals an intrinsic gradient  from  the warm B0 eastern cavity wall ($\rm >$120\,K)  to the cool cavity B1 and the earlier shock B2 ($\rm <$80\,K).}

\item Both $o$- and $p$-$\rm NH_3$ show a column density enhancement toward  {three U-shape cavities in B1 and B2, reaching as high as $\rm >1\times10^{16}\,cm^{-2}$}. The OPR is enhanced  by a factor of 2--2.5 along the curved jet path compared to the rest, {showing a strong spatial correlation with the peak-to-bluest velocity offset and with the total column density}. All of these chemical gradients may be linked to the shocks. 

\item A flattened envelope surrounding the protobinary system appears at the edge of our pb. {We find a $\rm 1\,km\,s^{-1}$ width narrow feature  on the line spectrum, probably tracing the quiescent and cold gas in the flattened envelope.  The velocity map for the line intensity peak velocity shows a gradient from the northeast ($\sim \rm 2\,km\,s^{-1}$) to the southwest ($\sim \rm 3\,km\,s^{-1}$) of this structure, with the  red-shifted  ($\sim \rm 2.7\,km\,s^{-1}$)  gas extended toward B0,  indicating the interaction between the outflow and the envelope.}

\end{enumerate}

\begin{acknowledgements}

{We gratefully acknowledge the useful and complimentary comments of the  anonymous referee.}

H.B.L. is supported by the Ministry of Science and
Technology (MoST) of Taiwan (grant Nos. 108-2112-M-001-002-MY3).

F.D. is supported by the National Natural Science Foundation of China (NSFC) through grant Nos. 11873094 and 12041305.

C.Co. acknowledges the PRIN-INAF 2016 The Cradle of Life - GENESIS-SKA (General Conditions in Early Planetary Systems for the rise of life with SKA);

C.C. acknowledges receiving funds from the European UnionÕs Horizon 2020 research and innovation programme from the European Research Council (ERC) for the project ÒThe Dawn of Organic ChemistryÓ (DOC), grant agreement No 741002, and from the Marie Sklodowska-Curie for the project ÓAstro-Chemical OriginsÓ (ACO), grant agreement No 811312.

This research made use of NASA's Astrophysics Data System.

 \end{acknowledgements}

\software{
          astropy \citep{astropy13},  
          Numpy \citep{vanderwalt11}, 
          CASA \citep[v5.4.0; v5.6.1;][]{mcmullin07},
          \texttt{bettermoments} \citep[v1.0.0; ][]{teague18},
          \texttt{HfS} \citep[][]{estalella17}
          }
          
\bibliographystyle{apj}
\bibliography{l1157-NH3-proof.bbl}

\newpage
\setcounter{section}{0}
\renewcommand{\thetable}{A\arabic{section}}
\setcounter{table}{0}
\renewcommand{\thetable}{A\arabic{table}}
\setcounter{figure}{0}
\renewcommand{\thefigure}{A\arabic{figure}}
\appendix

\section{Line profile}\label{sec:lineprof}
\begin{figure*}[tbh]
\begin{center}
\includegraphics[angle=-90,scale=0.8]{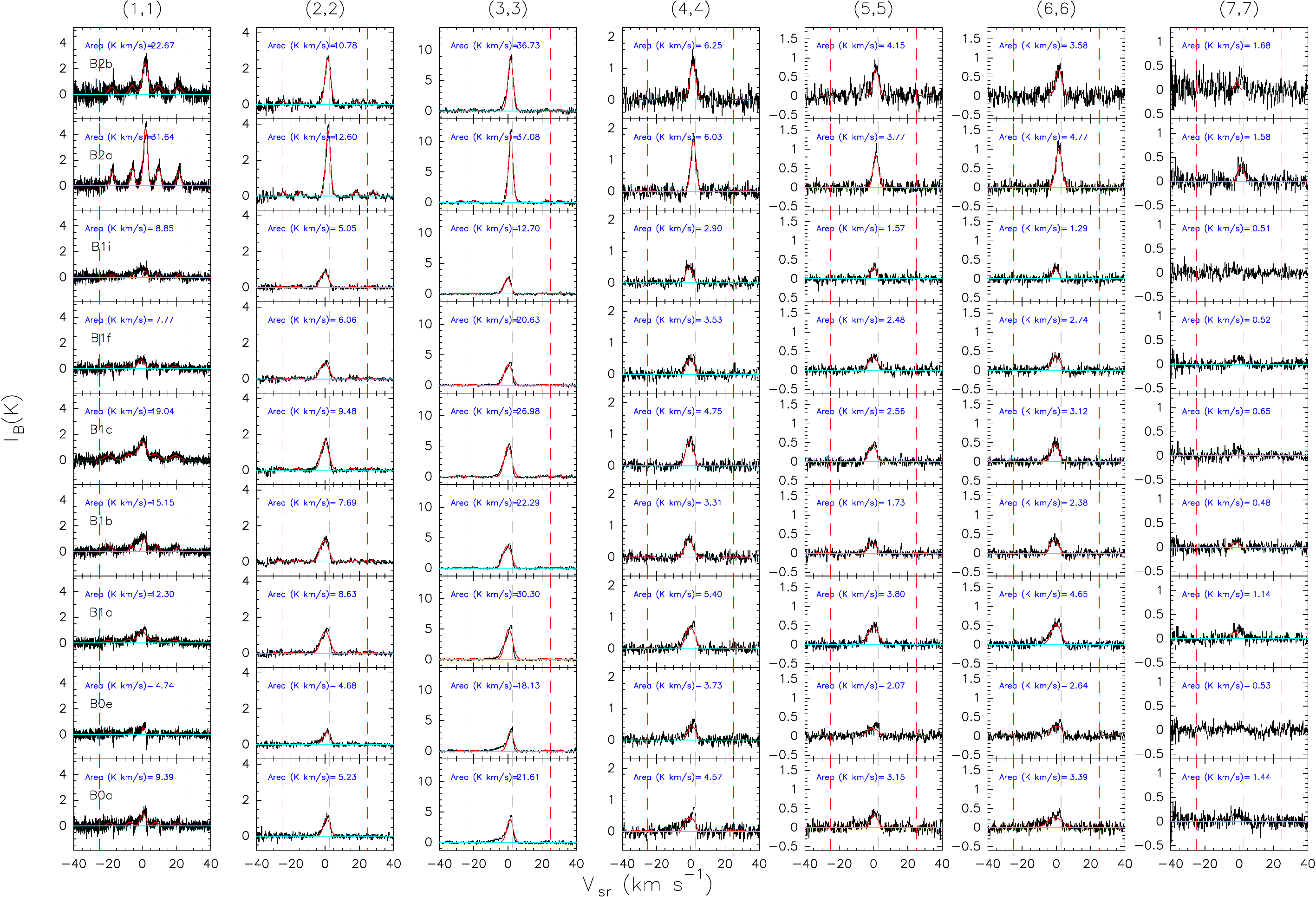}
\end{center}
\caption{Beam-averaged line profiles (the line intensity $\rm Jy\,beam^{-1}$ is converted to the brightness temperature $\rm T_{B}$), with the center toward nine clumpy substructures   (Table~ \ref{tab:clump}) in the plane of the sky. The \texttt{HfS}  fit is shown in red overlying on each spectrum.
All lines are extracted from images smoothed to the same pixel size (0.5\arcsec), and their velocity resolutions are listed in Table~\ref{tab:line}. In each panel, two red vertical lines at -25 and $\rm +25\,km\,s^{-1}$ indicate the velocity range within which we integrate the intensity; the gray dashed vertical line indicates the $V_{sys} \rm = 2.7\,km\,s^{-1}$ of the cloud. The horizontal cyan line indicates the baseline ($\rm T_{B}=0\,K$).
}\label{fig:velpro}
\end{figure*}

Along the jet paths,  the protobinary system L1157-mm and nine representative clumpy substructures from B0 to B1 and B2  are denoted, with their coordinates listed in Table~\ref{tab:clump}. The beam-averaged line profiles of $\rm NH_3$\,(1,1)--(7,7) are extracted from  these nine positions, shown in Figure~\ref{fig:velpro}.

Applying the Monte Carlo fitting tool \texttt{HfS}  developed by \citet[][]{estalella17}, we fit the hyperfine multiplets of each line with a Gaussian shape model by using the ``\url{hfs_fit}" procedure and assuming one velocity component. The best-fit results of the line parameters are listed in  Table~\ref{tab:hfsline}, and shown as a red curve overlying the observation spectrum in Figure~\ref{fig:velpro}.

Testing the other mainstream fitting packages, such as CLASS/GILDAS \citep{pety05} and \texttt{PySpecKit}  \citep{ginsburg11}, we found  similar fitting results.

Given that the following assumptions were made in these mainstream fitting packages, some caveats  need to be noted:

(1) The mainstream fitting packages take the Gaussian profile as a model, but our target lines show significant blue-shifted emissions in this successive shocked region. 
{Therefore, the centroid velocity of the main component is not strictly identical to the velocity of the peak intensity, and the FWHM line width from fitting is slightly narrower than the observed line wing, especially for the (3,3)--(6,6) lines (Figure~\ref{fig:velpro}). }

(2) The mainstream fitting packages provide the optical depth of the main component  $\tau_m$. This optical depth is close to that of the entire line $\tau$, but only when the satellites have $\rm <30\%$ contribution to the line-integrated intensity, i.e., for the (2,2)--(7,7) lines. As to the (1,1) line where satellites may also be optically thick, $\tau_m$ may be significantly smaller than $\tau$.

(3) As a compromise of the fit to hyperfine multiplets, one velocity component is assumed in the fitting. This may bring in an overestimation of the  FWHM line width.

\begin{table}[tbh]
\small
\caption{Positions corresponding to the protostellar object and the clumpy substructures}\label{tab:clump}%
\centering
 \scalebox{1}{
\begin{tabular}{cp{2.5cm}p{2.3cm}}
\hline
\hline
      &R.A. (J2000) &Decl. (J2000)\\

\hline
      mm                         &$\rm 20^h39^m06^s.170$    &$\rm 68^{\circ}02\arcmin15.07s\arcsec$    \\
      B0a      		   &$\rm 20^h39^m10^s.026$    &$\rm 68^{\circ}01\arcmin41.30s\arcsec$    \\
      B0e      		   &$\rm 20^h39^m10^s.365$    &$\rm 68^{\circ}01\arcmin23.80s\arcsec$    \\
      B1a                        &$\rm 20^h39^m10^s.301$    &$\rm 68^{\circ}01\arcmin13.51s\arcsec$    \\
      B1b      		   &$\rm 20^h39^m08^s.734$    &$\rm 68^{\circ}01\arcmin10.54s\arcsec$    \\
      B1c      		   &$\rm 20^h39^m09^s.677$    &$\rm 68^{\circ}01\arcmin06.45s\arcsec$    \\
      B1f      		   &$\rm 20^h39^m09^s.444$    &$\rm 68^{\circ}01\arcmin16.70s\arcsec$    \\
      B1i      		   &$\rm 20^h39^m10^s.000$    &$\rm 68^{\circ}01\arcmin01.11s\arcsec$    \\
      B2a      		   &$\rm 20^h39^m12^s.755$    &$\rm 68^{\circ}00\arcmin38.44s\arcsec$    \\
      B2b      		   &$\rm 20^h39^m14^s.217$    &$\rm 68^{\circ}00\arcmin24.73s\arcsec$    \\
\hline
\hline

\end{tabular}
}
\end{table}

\begin{table*}[tbh]
\caption{The best-fit parameters for spectra in Figure~\ref{fig:velpro}
}\label{tab:hfsline}
\scalebox{0.9}{
\centering
\begin{tabular}{c|c|p{1.8cm}p{1.8cm}p{1.8cm}p{1.8cm}p{1.8cm}p{1.8cm}p{1.8cm}}
\hline\hline

Position &Best-Fit parameter$^{a}$  &(1,1)   &(2,2) &(3,3)  &(4,4) &(5,5) &(6,6)  &(7,7)\\

\hline

 \multirow{5}{*}{B0a}
&$\rm  \delta V\,(km\,s^{-1})$$^{b}$    &$\rm 3.59\pm0.56$     &$\rm 4.47\pm0.50$    &$\rm 4.34\pm0.21$    &$\rm 4.64\pm1.75$    &$\rm 4.80\pm0.65$     &$\rm 9.60\pm2.79$ 	&...$^h$\\ 
&$\rm   V_{lsr}\,(km\,s^{-1}) $$^{c}$    &$\rm 0.56\pm0.20$     &$\rm 1.02\pm0.19$    &$\rm 1.19\pm0.08$    &$\rm -0.61\pm0.89$    &$\rm 0.38\pm0.44$     &$\rm -0.93\pm1.18$ 	&...$^h$\\
&$\rm  A (1-e^{-\tau_m}) (K) $$^{d,e}$   &$\rm 1.04\pm0.11$     &$\rm 1.07\pm0.10$    &$\rm 3.60\pm0.13$    &$\rm 0.46\pm0.11$    &$\rm 0.37\pm0.05$     &$\rm 0.31\pm0.07$ 	&...$^h$\\
&$\rm  \tau_m^*=1-e^{-\tau_m}  $$^{}$ &$\rm 0.26\pm0.43$  &$\rm (4.0\pm4.0)E-4$   &$\rm (2.0\pm2.0)E-4$    &$\rm 1.00\pm0.07$$^g$     &$\rm 0.72\pm0.46$$^g$     &$\rm (4.0\pm4.0)E-4$ 	&...$^h$\\
&$ I \rm(K\,km\,s^{-1})   $$^{f}$           &$\rm 9.54_{-3.64}^{+0.36}$     &$\rm 7.00_{-1.40}^{+0.33}$     &$\rm 23.44_{-2.02}^{+0.49}$      &$\rm 5.28_{-1.09}^{+0.27}$   &$\rm 2.94_{-0.26}^{+0.07}$   &$\rm 3.66 _{-0.78}^{+0.12}$ 	&$\rm 1.17_{-0.08}^{+0.17}$\\ 

\hline
 \multirow{5}{*}{B0e}
&$\rm  \delta V\,(km\,s^{-1}) $$^{b}$          &$\rm 1.57\pm0.53$  &$\rm 4.94\pm1.67$  &$\rm 4.11\pm0.25$   &$\rm 4.31\pm0.45$  &$\rm 5.29\pm4.98$   &$\rm 3.82\pm0.63$ 	&...$^h$\\ 
&$\rm   V_{lsr}\,(km\,s^{-1}) $$^{c}$          &$\rm 0.77\pm0.01$    & $\rm 0.88\pm0.24$    &$\rm 1.49\pm0.09$      &$\rm 0.66\pm0.33$    &$\rm -0.53\pm0.98$   &$\rm 0.20\pm0.34$ 	&...$^h$\\
&$\rm  A (1-e^{-\tau_m}) (K) $$^{d,e}$         &$\rm 0.63\pm0.14$     &$\rm 0.75\pm0.08$     &$\rm 3.15\pm0.14$      &$\rm 0.49\pm0.05$    &$\rm 0.22\pm0.07$   &$\rm 0.29\pm0.03$ 	&...$^h$\\
&$\rm  \tau_m^*=1-e^{-\tau_m}  $$^{}$   &$\rm 0.80\pm0.16$     &$\rm 0.45\pm1.49$$^g$    &$\rm (5.0\pm5.0)E-4$      &$\rm 0.90\pm0.14$$^g$     &$\rm 0.99\pm0.44$$^g$     &$\rm 1.00\pm0.02$$^g$ 	&...$^h$\\
&$ I \rm(K\,km\,s^{-1})   $$^{f}$         &$\rm  4.94_{-1.52}^{+0.43}$		&$\rm  6.46_{-1.65}^{+0.42}$   	&$\rm  19.88_{-1.66}^{+0.44}$	&$\rm  3.86_{-0.42}^{+0.16}$		&$\rm  2.23_{-0.19}^{+0.13}$   	&$\rm  2.37_{-0.09}^{+0.07}$		&$\rm  0.57_{-0.07}^{+0.16}$\\ 

\hline
 \multirow{5}{*}{B1a}
&$\rm  \delta V\,(km\,s^{-1}) $$^{b}$         &$\rm 5.46\pm0.39$     &$\rm 5.57\pm0.49$     &$\rm 5.34\pm0.17$     &$\rm 5.36\pm0.33$     &$\rm 6.87\pm0.65$     &$\rm 6.40\pm0.39$          &$\rm 1.78\pm4.53$\\ 
&$\rm   V_{lsr}\,(km\,s^{-1}) $$^{c}$         &$\rm -0.22\pm0.18$    &$\rm 0.13\pm0.14$     &$\rm 0.58\pm0.07$     &$\rm -0.21\pm0.22$     &$\rm -0.17\pm0.27$     &$\rm -0.21\pm0.23$          &$\rm -2.37\pm0.08$\\
&$\rm  A (1-e^{-\tau_m}) (K) $$^{d,e}$        &$\rm 1.00\pm0.07$     &$\rm 1.28\pm0.06$     &$\rm 4.67\pm0.12$     &$\rm 0.71\pm0.04$     &$\rm 0.50\pm0.04$     &$\rm 0.57\pm0.03$          &$\rm 0.18\pm0.04$\\
&$\rm  \tau_m^*=1-e^{-\tau_m}  $$^{}$  &$\rm 0.27\pm0.28$     &$\rm 0.50\pm0.33$     &$\rm (3.1\pm3.1)E-3$     &$\rm 0.83\pm0.10$     &$\rm 0.01\pm0.01$     &$\rm 0.53\pm0.22$          &$\rm 1.00\pm0.01$\\
&$ I \rm(K\,km\,s^{-1})   $$^{f}$                &$\rm 12.46 _{-4.57}^{+0.14}$	      &$\rm 10.68 _{-2.16}^{+0.42}$	 &$\rm 32.76 _{-2.35}^{+1.04}$    &$\rm  5.35_{-0.17}^{+0.11}$		&$\rm 4.01_{-0.35}^{+0.06}$	&$\rm 4.58 _{-0.12}^{+0.08}$	&$\rm  1.53_{-0.18}^{+0.02}$\\ 

\hline
 \multirow{5}{*}{B1b}
&$\rm  \delta V\,(km\,s^{-1}) $$^{b}$          &$\rm 2.72\pm4.35$          &$\rm 5.30\pm0.44$          &$\rm 4.83\pm0.28$          &$\rm 5.72\pm0.64$          &$\rm 3.62\pm0.73$          &$\rm 4.46\pm0.39$ &$\rm 3.56\pm1.26$\\ 
&$\rm   V_{lsr}\,(km\,s^{-1}) $$^{c}$          &$\rm 1.41\pm0.01$        &$\rm -0.41\pm0.13$         &$\rm -0.37\pm0.07$          &$\rm -1.07\pm0.28$          &$\rm -0.97\pm0.50$          &$\rm -1.24\pm0.30$   &$\rm -1.68\pm0.62$ \\
&$\rm  A (1-e^{-\tau_m}) (K) $$^{d,e}$         &$\rm 1.11\pm0.14$         &$\rm 1.24\pm0.06$          &$\rm 3.50\pm0.08$           &$\rm 0.60\pm0.06$         &$\rm 0.28\pm0.04$           &$\rm 0.35\pm0.03$    &$\rm 0.14\pm0.04$\\
&$\rm  \tau_m^*=1-e^{-\tau_m}  $$^{}$   &$\rm 0.84\pm0.13$         &$\rm 0.55\pm0.28$          &$\rm 0.76\pm0.13$          &$\rm (1.8\pm1.8)E-3$          &$\rm 1.00\pm0.03$$^g$          &$\rm 0.95\pm0.08$$^g$   &$\rm 0.95\pm0.72$$^g$\\
&$ I \rm(K\,km\,s^{-1})   $$^{f}$                 &$\rm 15.15 _{-5.97}^{+0.04}$	&$\rm 9.00_{-1.24}^{+0.41}$  	&$\rm 23.90_{-0.93}^{+0.49}$		 &$\rm 3.42_{-0.25}^{+0.20}$		&$\rm 2.59_{-0.57}^{+0.22}$		&$\rm 2.59_{-0.21}^{+0.05}$		&$\rm 0.84_{-0.11}^{+0.15}$\\ 

\hline
 \multirow{5}{*}{B1c}
&$\rm  \delta V\,(km\,s^{-1}) $$^{b}$          &$\rm 5.30\pm0.25$          &$\rm 5.37\pm0.22$          &$\rm 4.97\pm0.13$          &$\rm 4.64\pm0.28$          &$\rm 4.15\pm0.31$         &$\rm 5.75\pm0.71$ 	&...$^h$ \\ 
&$\rm   V_{lsr}\,(km\,s^{-1}) $$^{c}$          &$\rm 0.00\pm0.12$          &$\rm 0.00\pm0.09$          &$\rm 0.11\pm0.05$          &$\rm -0.60\pm0.19$         &$\rm -0.53\pm0.24$           &$\rm -0.79\pm0.29$ 	&...$^h$\\
&$\rm  A (1-e^{-\tau_m}) (K) $$^{d,e}$         &$\rm 1.51\pm0.07$          &$\rm 1.65\pm0.06$          &$\rm 4.67\pm0.10$          &$\rm 0.82\pm0.05$          &$\rm 0.44\pm0.03$          &$\rm 0.50\pm0.05$ 	&...$^h$\\
&$\rm  \tau_m^*=1-e^{-\tau_m}  $$^{}$   &$\rm 0.32\pm0.17$          &$\rm 0.03\pm0.03$          &$\rm 0.02\pm0.02$         &$\rm 0.76\pm0.13$           &$\rm 0.96\pm0.05$$^g$          &$\rm (1.0\pm1.0)E-4$ 	&...$^h$\\
&$ I \rm(K\,km\,s^{-1})   $$^{f}$                 &$\rm 19.08 _{-7.36}^{+0.09}$		&$\rm  11.20_{-1.55}^{+0.70}$		&$\rm  29.59_{-1.61}^{+0.59}$	&$\rm  5.38_{-0.37}^{+0.03}$		&$\rm  3.22_{-0.44}^{+0.14}$		&$\rm  3.20_{-0.04}^{+0.01}$		&$\rm  1.10_{-0.18}^{+0.09}$\\

\hline
 \multirow{5}{*}{B1f}
&$\rm  \delta V\,(km\,s^{-1}) $$^{b}$          &$\rm 3.88\pm0.36$         &$\rm 5.32\pm0.53$          &$\rm 5.00\pm0.52$         &$\rm 4.02\pm0.33$        &$\rm 6.37\pm0.84$        &$\rm 4.15\pm0.32$ 	&...$^h$\\ 
&$\rm   V_{lsr}\,(km\,s^{-1}) $$^{c}$          &$\rm -0.66\pm0.01$        &$\rm -0.27\pm0.16$         &$\rm 0.24\pm0.09$         &$\rm -0.37\pm0.27$      &$\rm -0.11\pm0.35$        &$\rm -0.38\pm0.28$ &...$^h$\\
&$\rm  A (1-e^{-\tau_m}) (K) $$^{d,e}$         &$\rm 0.76\pm0.08$         &$\rm 0.92\pm0.05$          &$\rm 3.26\pm0.09$          &$\rm 0.50\pm0.04$        &$\rm 0.37\pm0.04$        &$\rm 0.37\pm0.03$ 	&...$^h$\\
&$\rm  \tau_m^*=1-e^{-\tau_m}  $$^{}$   &$\rm 0.58\pm0.28$         &$\rm 0.65\pm0.27$$^g$          &$\rm 0.44\pm0.43$          &$\rm 0.97\pm0.04$$^g$        &$\rm 0.09\pm0.09$        &$\rm 0.99\pm0.02$$^g$ 	&...$^h$\\
&$ I \rm(K\,km\,s^{-1})   $$^{f}$    &$\rm  7.44_{-2.69}^{+0.03}$		&$\rm  7.44_{-1.36}^{+0.25}$		&$\rm 22.54 _{-1.74}^{+0.35}$	 &$\rm  3.22_{-0.14}^{+0.35}$		&$\rm 2.59 _{-0.12}^{+0.08}$		&$\rm  2.95_{-0.18}^{+0.03}$		&$\rm  0.51_{-0.19}^{+0.03}$\\

\hline
 \multirow{5}{*}{B1i}
&$\rm  \delta V\,(km\,s^{-1}) $$^{b}$         &$\rm 1.52\pm5.84$         &$\rm 5.18\pm0.34$         &$\rm 4.18\pm0.38$         &$\rm 3.64\pm0.34$          &$\rm 4.26\pm0.80$          &$\rm 3.96\pm0.82$ 	&...$^h$ \\ 
&$\rm   V_{lsr}\,(km\,s^{-1}) $$^{c}$         &$\rm 1.86\pm0.01$         &$\rm -0.26\pm0.15$        &$\rm -0.21\pm0.07$         &$\rm -0.84\pm0.26$          &$\rm 0.11\pm0.34$          &$\rm -0.45\pm0.34$ 	&...$^h$\\
&$\rm  A (1-e^{-\tau_m}) (K) $$^{d,e}$        &$\rm 0.56\pm0.07$         &$\rm 0.93\pm0.05$         &$\rm 2.44\pm0.07$         &$\rm 0.53\pm0.05$          &$\rm 0.33\pm0.05$           &$\rm 0.32\pm0.05$ 	&...$^h$\\
&$\rm  \tau_m^*=1-e^{-\tau_m}  $$^{}$   &$\rm 0.99\pm0.02$$^g$        &$\rm (4.0\pm4.0)E-4$         &$\rm 0.39\pm0.39$          &$\rm 0.95\pm0.07$$^g$          &$\rm (3.0\pm3.0)E-4$           &$\rm (5.4\pm5.4)E-3$ 	&...$^h$\\
&$ I \rm(K\,km\,s^{-1})   $$^{f}$   &$\rm  9.61_{-3.68}^{+0.04}$		&$\rm  5.73_{-0.68}^{+0.24}$		&$\rm  13.96_{-0.93}^{+0.37}$	&$\rm  3.73_{-0.71}^{+0.35}$		&$\rm  1.56_{-0.14}^{+0.11}$		&$\rm  1.38_{-0.01}^{+0.05}$		&$\rm 0.61 _{-0.04}^{+0.07}$\\ 

\hline
 \multirow{5}{*}{B2a}
&$\rm  \delta V\,(km\,s^{-1}) $$^{b}$           &$\rm 2.70\pm0.10$          &$\rm 3.21\pm0.11$          &$\rm 3.15\pm0.05$          &$\rm 3.68\pm0.26$          &$\rm 3.82\pm0.37$          &$\rm 4.38\pm0.33$          &$\rm 3.43\pm3.09$\\ 
&$\rm   V_{lsr}\,(km\,s^{-1}) $$^{c}$          &$\rm 1.99\pm0.03$           &$\rm 1.81\pm0.04$          &$\rm 1.75\pm0.02$          &$\rm 1.63\pm0.10$          &$\rm 1.47\pm0.15$          &$\rm 1.52\pm0.12$          &$\rm 1.09\pm0.15$\\
&$\rm  A (1-e^{-\tau_m}) (K) $$^{d,e}$         &$\rm 4.62\pm0.13$          &$\rm 3.76\pm0.11$          &$\rm 10.29\pm0.15$          &$\rm 1.67\pm0.10$          &$\rm 0.90\pm0.07$          &$\rm 1.05\pm0.06$          &$\rm 0.34\pm0.12$\\
&$\rm  \tau_m^*=1-e^{-\tau_m}  $$^{}$    &$\rm 0.31\pm0.11$          &$\rm (1.0\pm1.0)E-3$          &$\rm (2.8\pm2.8)E-3$          &$\rm (1.2\pm1.2)E-3$          &$\rm (7.0\pm7.0)E-4$          &$\rm (2.0\pm2.0)E-4$          &$\rm 0.97\pm0.86$$^g$\\
&$ I \rm(K\,km\,s^{-1})   $$^{f}$   &$\rm  30.82_{-11.73}^{+0.18}$		&$\rm  15.80_{-2.93}^{+1.12}$		&$\rm  41.63_{-3.51}^{+1.37}$	&$\rm  7.14_{-0.94}^{+0.06}$		&$\rm  3.64_{-0.03}^{+0.16}$		&$\rm  5.54_{-0.58}^{+0.02}$		&$\rm 1.62 _{-0.02}^{+0.04}$\\ 

 \hline
 \multirow{5}{*}{B2b}
&$\rm  \delta V\,(km\,s^{-1}) $$^{b}$           &$\rm 3.86\pm0.25$          &$\rm 3.90\pm0.22$          &$\rm 3.81\pm0.10$          &$\rm 4.24\pm0.54$          &$\rm 3.95\pm0.68$          &$\rm 4.31\pm0.69$          &$\rm 2.74\pm1.20$\\ 
&$\rm   V_{lsr}\,(km\,s^{-1}) $$^{c}$           &$\rm 1.81\pm0.16$          &$\rm1.65\pm0.08$          &$\rm 1.60\pm0.04$          &$\rm 1.33\pm0.21$          &$\rm 1.48\pm0.26$          &$\rm 1.41\pm0.28$          &$\rm 1.44\pm0.54$\\
&$\rm  A (1-e^{-\tau_m}) (K) $$^{d,e}$          &$\rm 2.57\pm0.16$          &$\rm 2.68\pm0.12$          &$\rm 7.88\pm0.18$          &$\rm 1.23\pm0.13$          &$\rm 0.78\pm0.11$          &$\rm 0.76\pm0.10$          &$\rm 0.25\pm0.08$\\
&$\rm  \tau_m^*=1-e^{-\tau_m}  $$^{}$    &$\rm 0.11\pm0.12$          &$\rm (1.3\pm1.3)E-3$          &$\rm (3.0\pm3.0)E-4$          &$\rm (1.0\pm1.0)E-4$          &$\rm (3.3\pm3.3)E-3$          &$\rm 0.02\pm0.02$          &$\rm 0.97\pm0.93$$^g$\\
&$ I \rm(K\,km\,s^{-1})   $$^{f}$  &$\rm  22.30_{-8.67}^{+0.10}$		&$\rm  14.57_{-3.07}^{+0.99}$		&$\rm  39.51_{-2.39}^{+0.65}$	&$\rm  5.88_{-0.11}^{+0.59}$		&$\rm  4.41_{-0.58}^{+0.17}$		&$\rm  3.60_{-0.07}^{+0.12}$		&$\rm  1.88_{-0.72}^{+0.1}$\\

\hline\hline
  \multicolumn{9}{l}{{ Note.} $a$. Given by using the  ``\url{hfs_fit}" procedure in the \texttt{HfS}  package, fit by assuming a Gaussian line profile; }\\
  \multicolumn{9}{l}{~~~~~~~~~~~~~~ all beam-averaged lines are extracted from images with the same pixel size;}\\
  \multicolumn{9}{l}{~~~~~~~~~~$b$. Intrinsic FWHM line width by taking into account the opacity;}\\
  \multicolumn{9}{l}{~~~~~~~~~~$c$. Centroid velocity of the main component;}\\
  \multicolumn{9}{l}{~~~~~~~~~~$d$. Peak intensity of the main component (for lines broader than channel width);}\\
  \multicolumn{9}{l}{~~~~~~~~~~$e$. Optical depth of the main component $\tau_m$, provided by the \texttt{HfS} as $\tau_m^*$, $\tau_m=-ln(1-\tau_m^*)$;}\\
  \multicolumn{9}{l}{~~~~~~~~~~$f$. Integrated intensity over the velocity range from -25 to $\rm +25km\,s^{-1}$, the uncertainty is given by the difference between this range}\\
  \multicolumn{9}{l}{~~~~~~~~~~~~~~ and the velocity range of -10$\rm \sim+5km\,s^{-1}$ (the main component of the hyperfine multiplets) and -27$\rm \sim+33km\,s^{-1}$ (line intensity down to zero);}\\
  \multicolumn{9}{l}{~~~~~~~~~~{$g$. Fitting results need to be taken cautiously for $\tau_m\rm >1$, and with an uncertainty larger than the value;}}\\
      \multicolumn{9}{l}{~~~~~~~~~~$h$. Line emission is less than $\rm 3\sigma$.}\\
\end{tabular}
}
\end{table*}

\section{Images with primary beam  correction}\label{sec:pbcor}
\begin{figure*}[tbh]
\centering
$r$=2.0\\
\includegraphics[width=19cm]{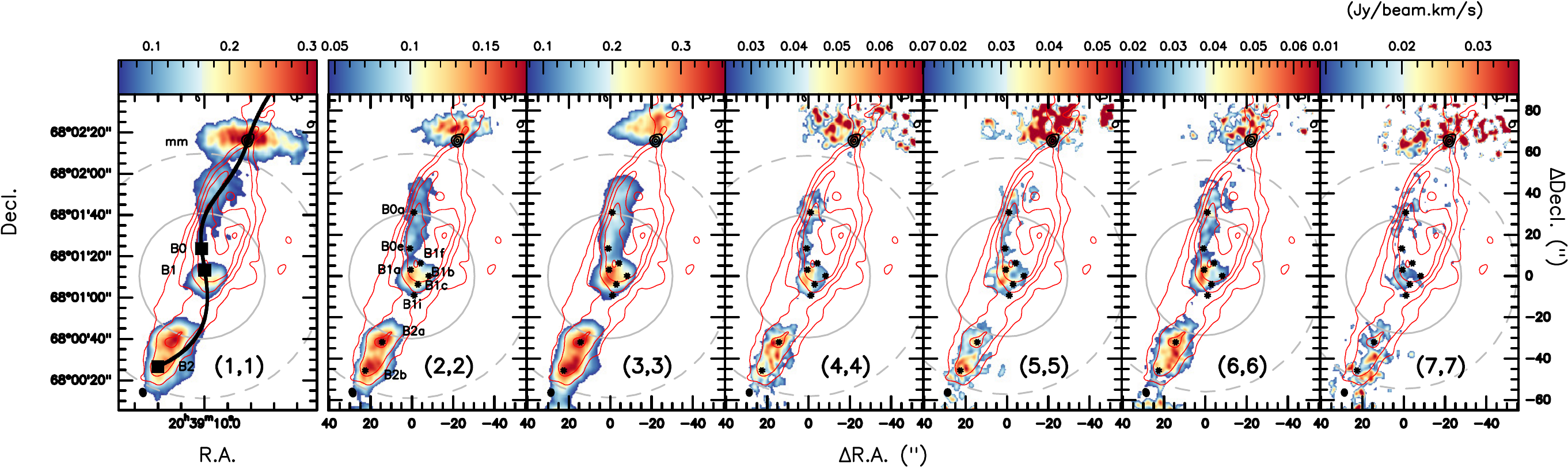}
$r$=0.5\\
\includegraphics[width=19cm]{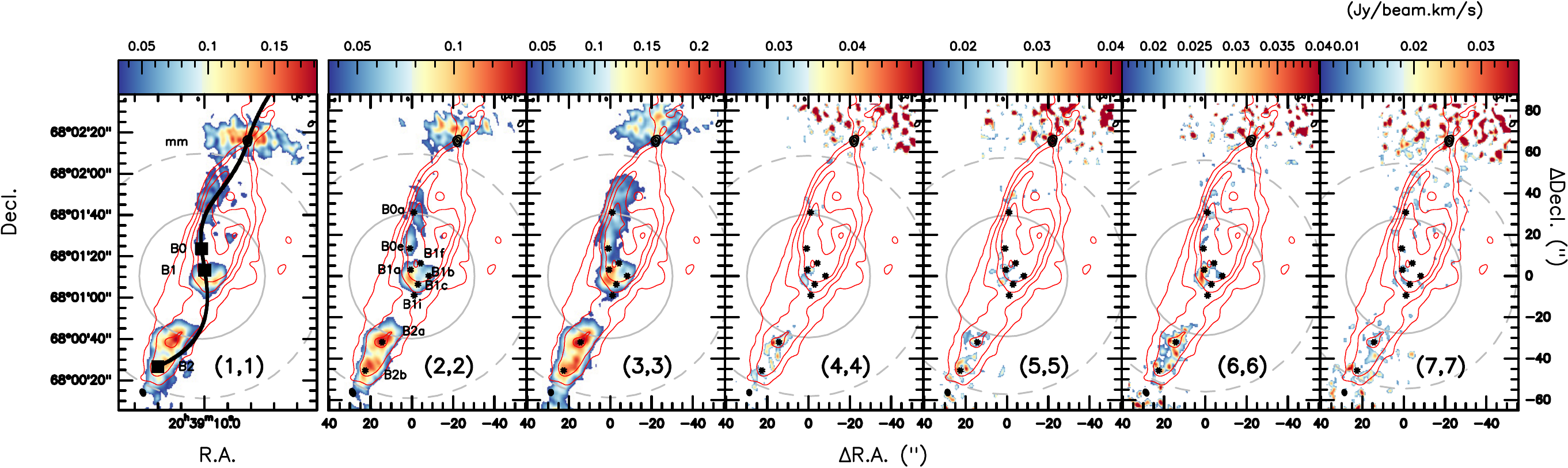}
$r$=-2.0\\
\includegraphics[width=19cm]{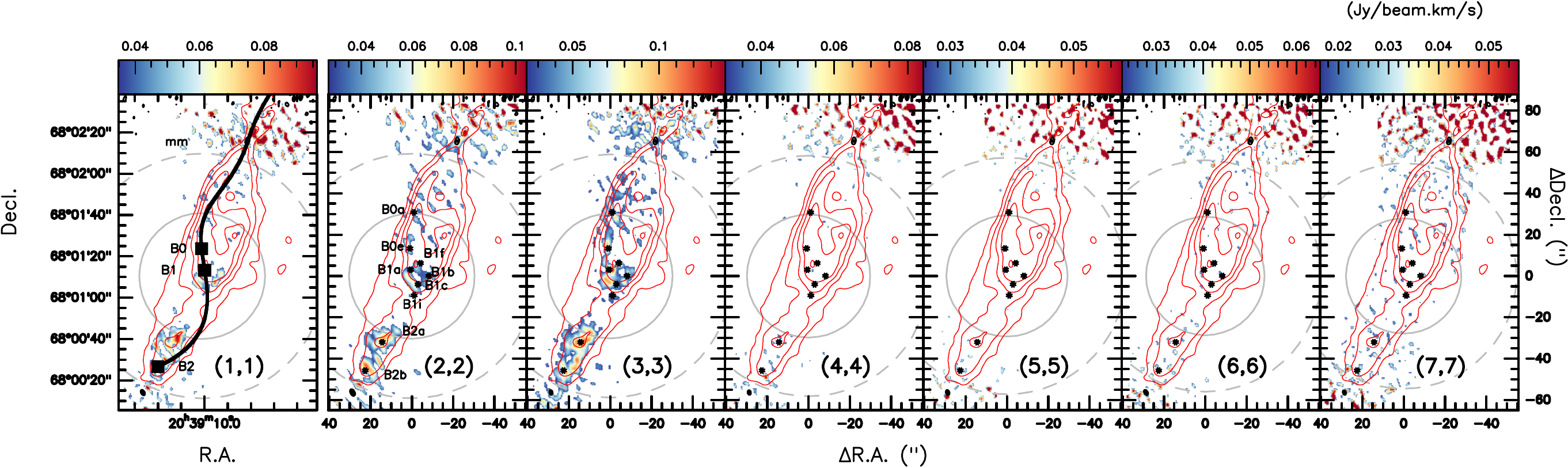}
\caption{
The outline of the southern (blue-shifted) outflow lobe from the Class 0 {compact protobinary system} L1157-mm ({images with different weightings are given  in different  rows}). Color maps show the intensity maps of $\rm NH_3$\,(1,1)--(7,7) integrated over the velocity range of $\rm -25$ to $\rm +25\,km\,s^{-1}$ (with pb correction).
{The black contours, starting from $\rm 4\sigma$ ($\rm \sigma=0.03\,mJy\,beam^{-1}$) and increasing with a step of  $\rm 4\sigma$, show the continuum emission with the same weighting  as the line emissions (with pb correction).}
The rest of the contours and all the labels are the same as in Figure~\ref{fig:jet}. The solid gray circle indicates the largest recoverable scale and the dashed gray circle indicates the primary beam.  }
\label{fig:jetpb}
\end{figure*}

Given that the beam responses are different for each antenna in a synthesis array, the flux density at the beam edge is biased low. 
Dividing the image by the primary beam (pb), we apply the pb correction to each line datacube and provide the integrated intensity maps over the velocity range of $\rm -25$ to $\rm +25\,km\,s^{-1}$ in Figure~\ref{fig:jetpb}. Compared it with Figure~\ref{fig:jet}, line emission toward the edge of the pb (e.g., mm, B2b) are recovered. However, noise levels are increasing toward these positions (see example spectrum in Figure~\ref{fig:gradient}). 

\section{Line optical depth correction}\label{sec:tau}

{Even though the mainstream fitting packages bring in a centroid velocity map a bit different from the peak velocity map  (Section~\ref{sec:lineprof}), they can constrain  
the lower limit of the optical depth for a particular line $\tau$ for each pixel. 
Therein, to correct the molecular column density which is underestimated, 
we can multiply the  line-integrated intensity  from observations by a factor of $\tau/(1-e^{-\tau})$.}

Applying the ``\url{hfs_cube_mp}" procedure in the \texttt{HfS} fitting tool, we focus on deriving the optical depth $\tau_m$ map and its uncertainty map for each line. Because the parameter directly given by the fitting program is $\rm  \tau_m^*=1-e^{-\tau_m}$, the error propagation algorithm enhances the uncertainty of $\tau_m$ due to the fitting caveats listed in Section~\ref{sec:lineprof}.  Therefore, we set a threshold for the pixels with a high fidelity, i.e., when the uncertainty with respect to the value is less than {1\,dex}.

\begin{figure*}[tbh]
\centering
\begin{tabular}{p{0.2cm}p{2.9cm}p{2.5cm}p{2.5cm}p{2.5cm}p{2.5cm}p{2.5cm}}
&\multicolumn{6}{c}{$\tau_m$}\\
{\begin{sideways}~~~~~~~~~~~~~~~~~~~$\Delta$ Decl.  (\arcsec)\end{sideways}}
&\includegraphics[clip, trim=0cm 0.5cm 0.0cm 1.0cm, height=5.6cm]{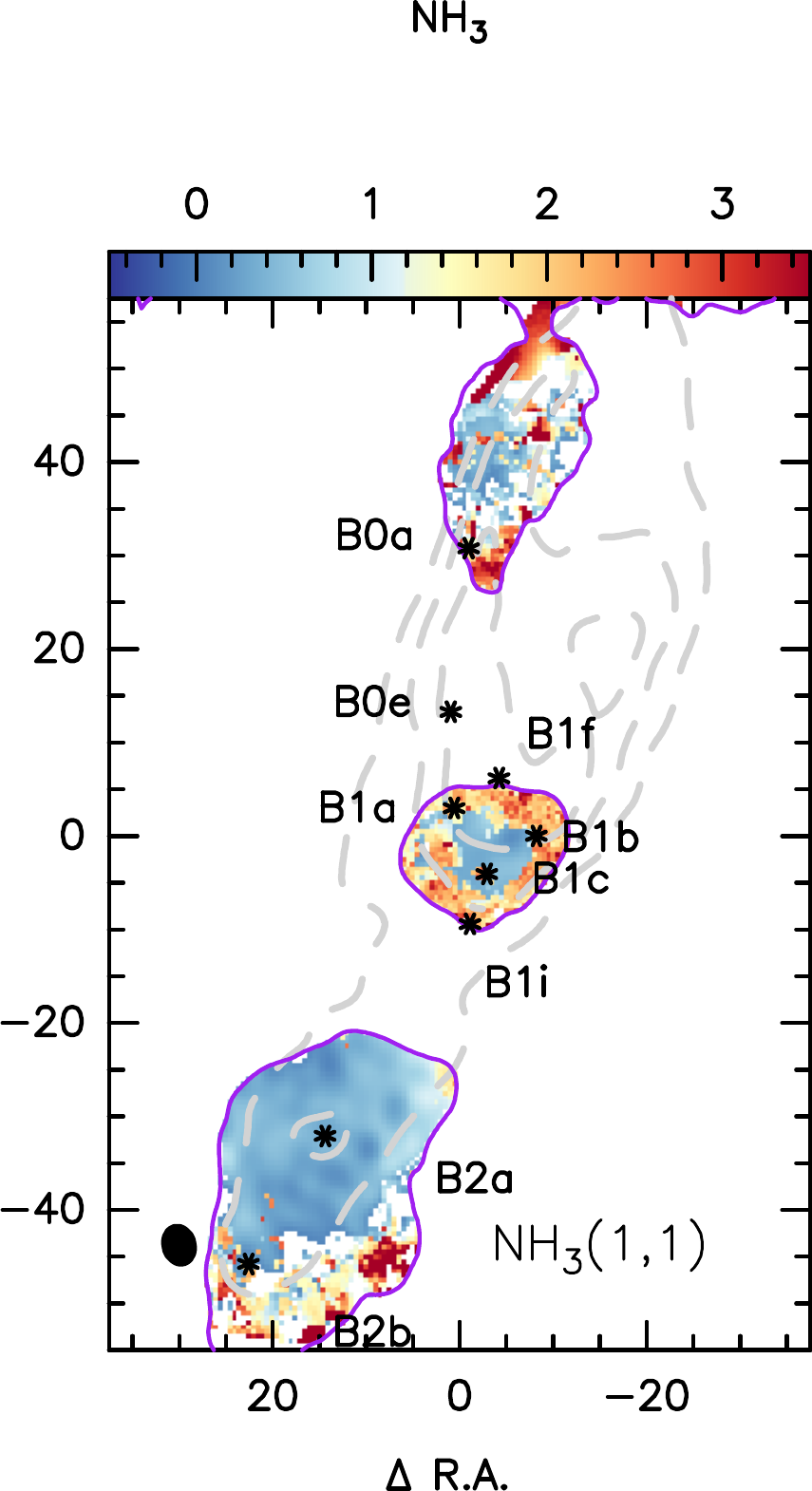}
&\includegraphics[clip, trim=1.0cm 0.5cm 0.0cm 1.0cm, height=5.6cm]{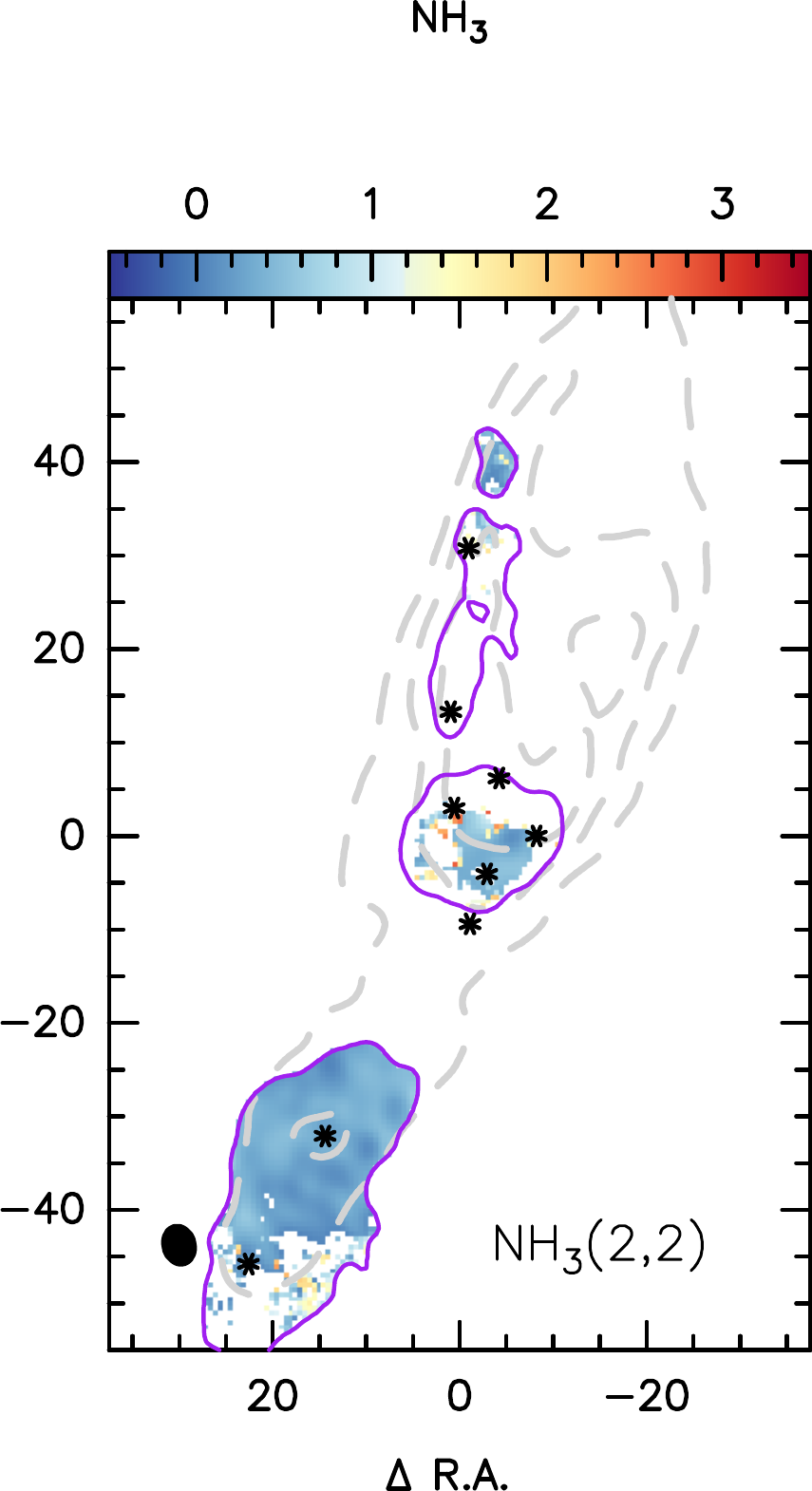}
&\includegraphics[clip, trim=1.0cm 0.5cm 0.0cm 1.0cm, height=5.6cm]{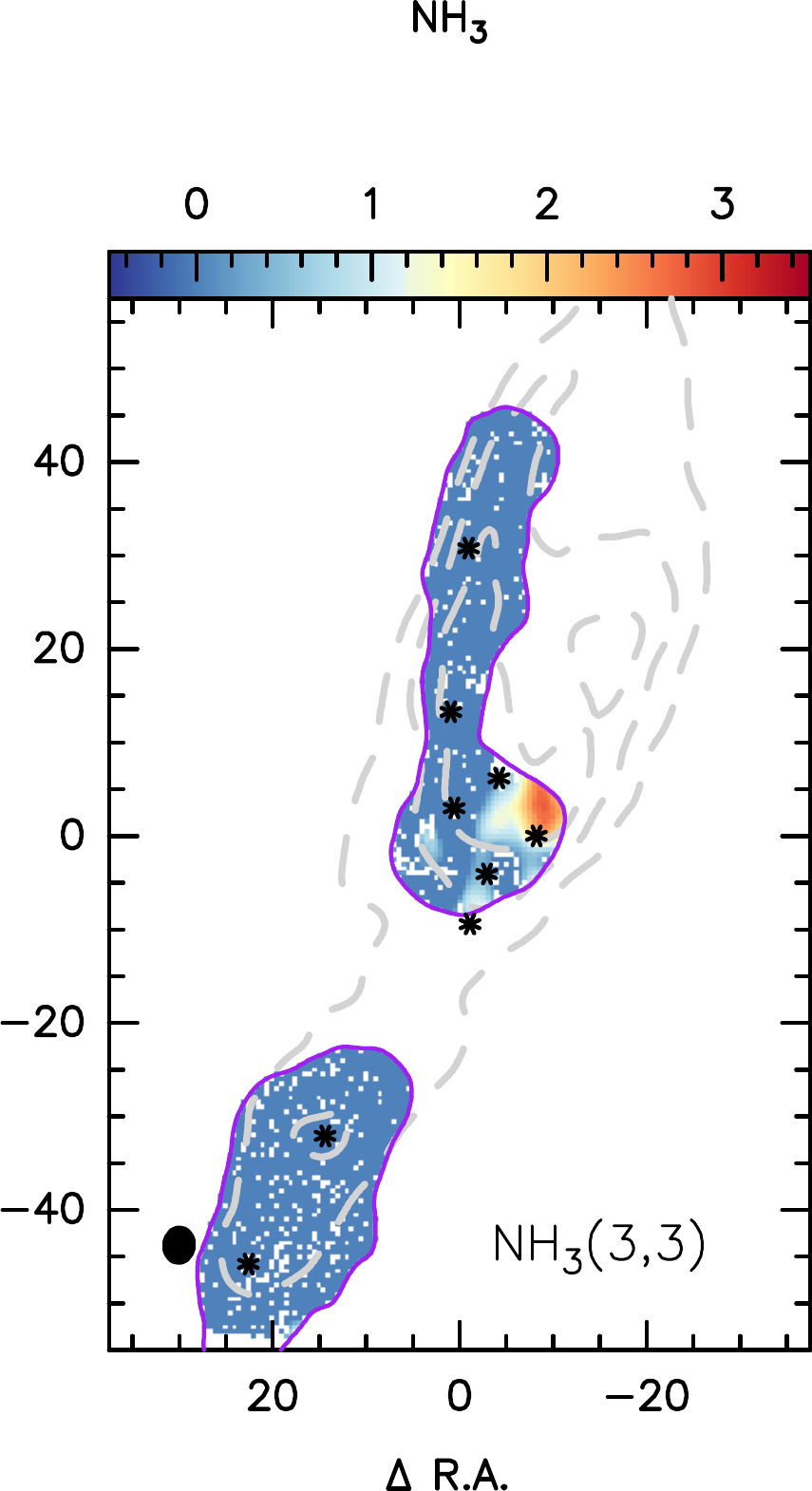}
&\includegraphics[clip, trim=1.0cm 0.5cm 0.0cm 1.0cm, height=5.6cm]{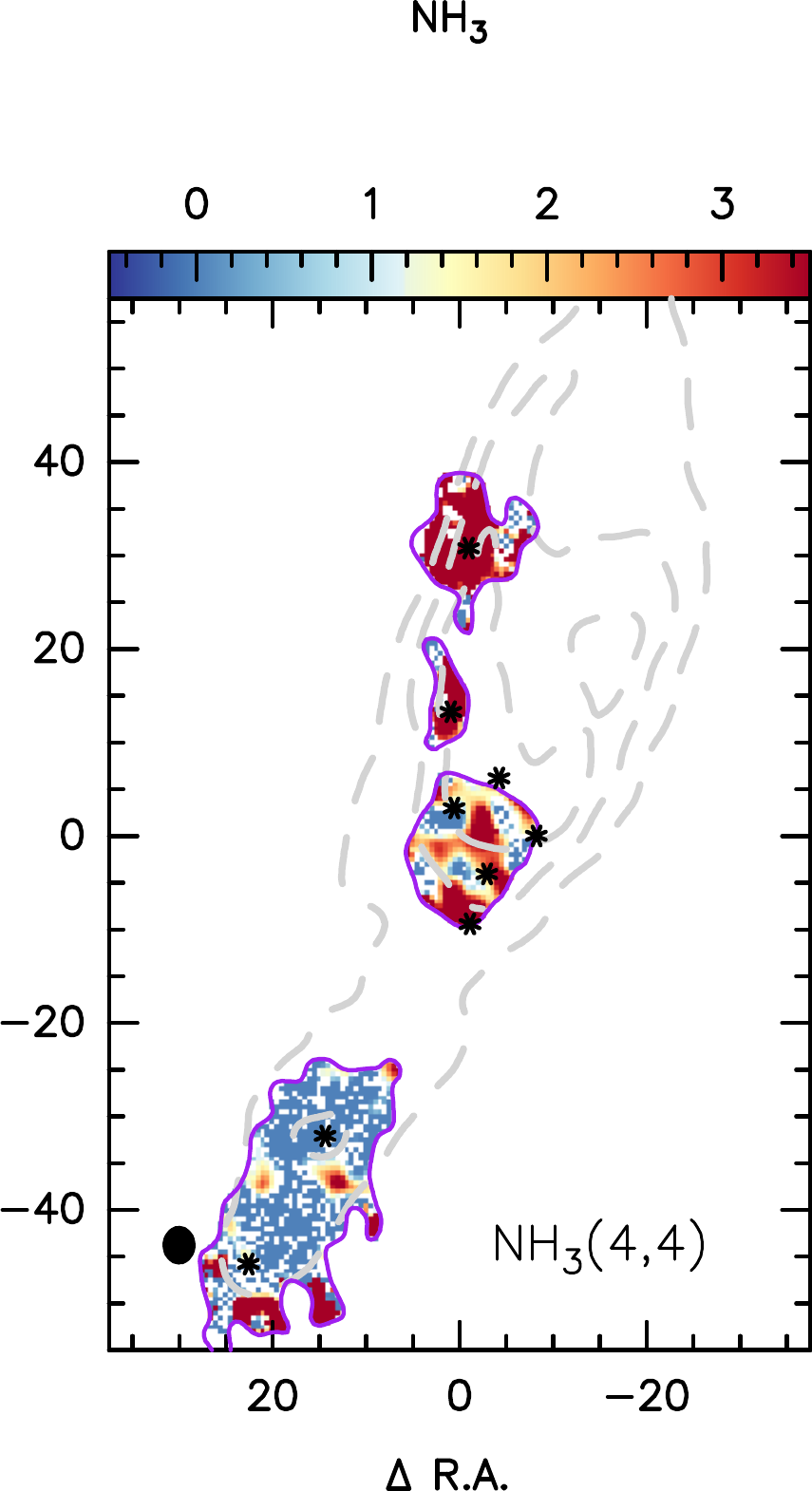}
&\includegraphics[clip, trim=1.0cm 0.5cm 0.0cm 1.0cm, height=5.6cm]{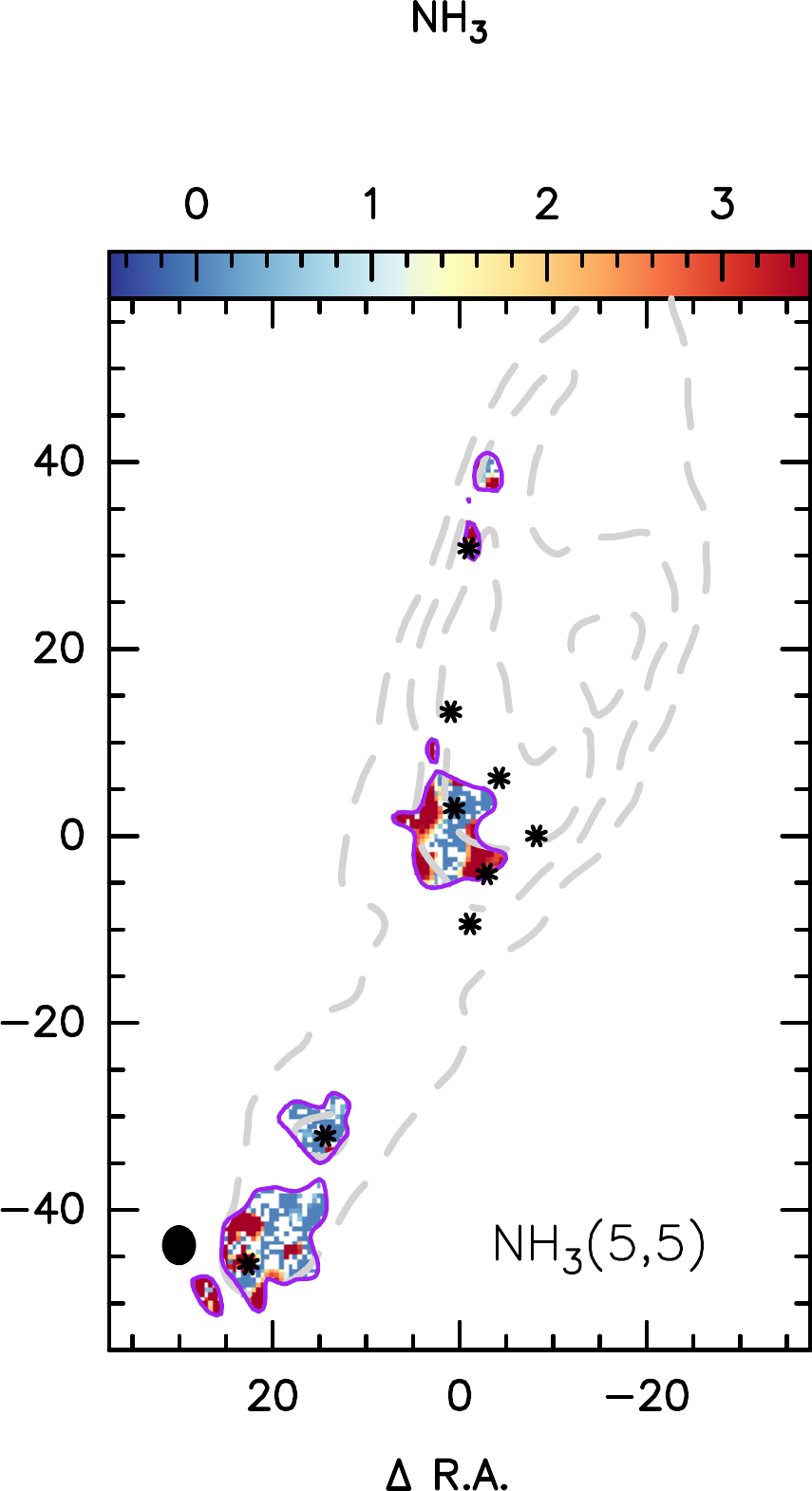}
&\includegraphics[clip, trim=1.0cm 0.5cm 0.0cm 1.0cm, height=5.6cm]{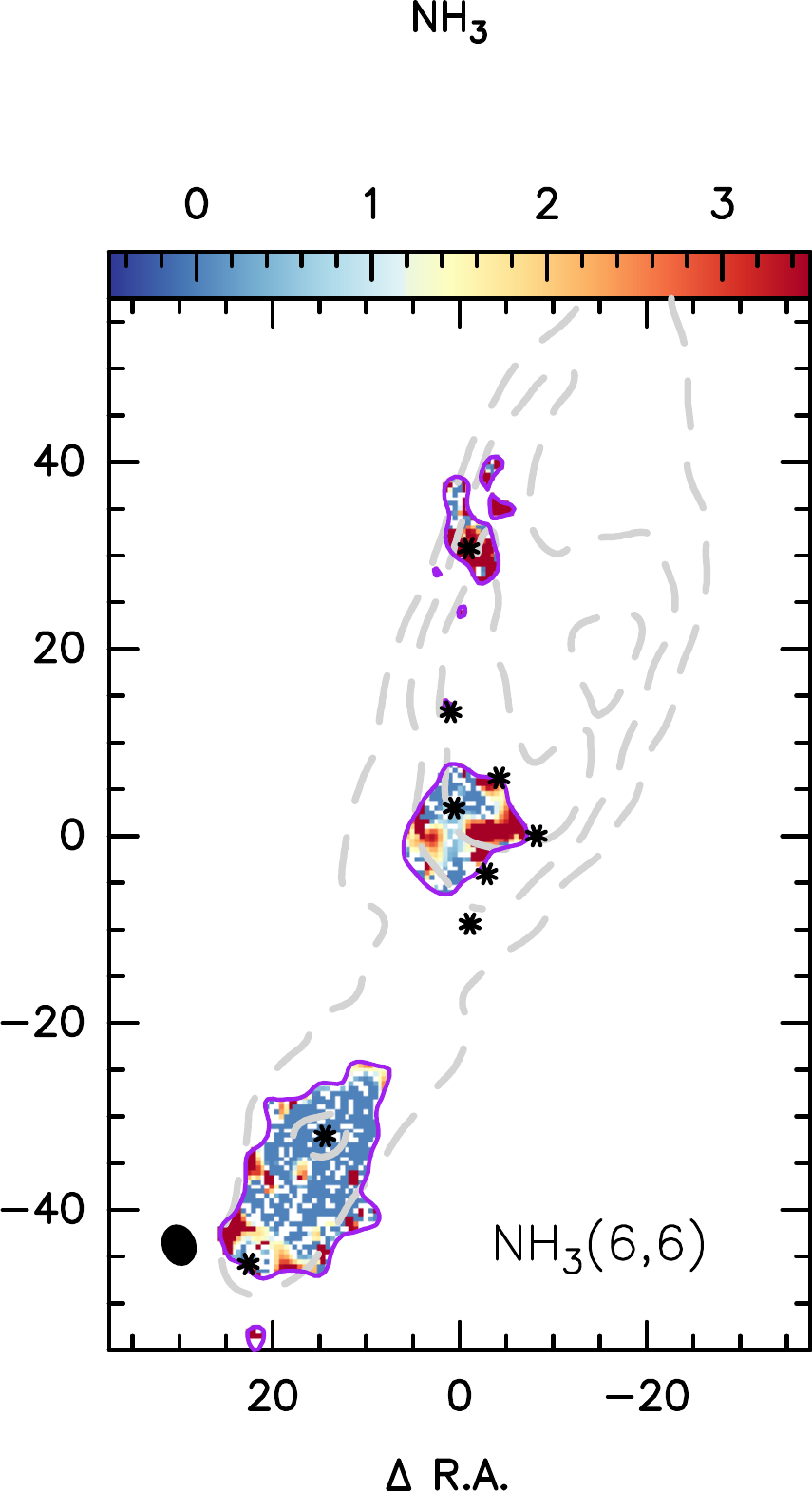}\\
&\multicolumn{6}{c}{$\Delta$ R. A. (\arcsec)} 

\end{tabular}
\caption{
Optical depth $\tau_m$ maps of the (1,1)--(6,6) lines. The gray dashed contours and the labeled clumpy substructures are the same as shown in Figure~\ref{fig:jet}. The integrated intensity of line within the red solid contour shows $\rm >4\sigma$ emissions.  The pixels where lines shows  $\rm <4\sigma$ emissions or uncertainty 1dex higher than the value of $\tau_m$ are blank. 
\label{fig:tauline}
}
\end{figure*}

As shown in  Figure~\ref{fig:tauline}, the optical depths of all seven lines are negligible  ($\rm \tau_m<1$)  toward B2, while those toward the B1 cavity wall (B1a, B1f, B1i, B1b) are larger than 3, except for the (3,3) line. {The $\tau_m$ of (4,4)--(6,6) may be underestimated toward some pixels, where the uncertainties are higher than the value by a factor of more than 5. }

\section{LVG fitting example and caveats}\label{app:lvg}

The optical depth of each line is significant, uncertain, and not uniform throughout the entire region. Even correcting each line-integrated intensity with the optical depth $\tau_m$, the rotation diagram (RD) method is not an optimized approach to deriving the temperature and density structure of our source. Instead, the LVG approximation is suitable under such a circumstance, though it is computationally expensive.

\begin{figure*}[tbh]
\centering
\begin{tabular}{cc}
\includegraphics[height=8cm]{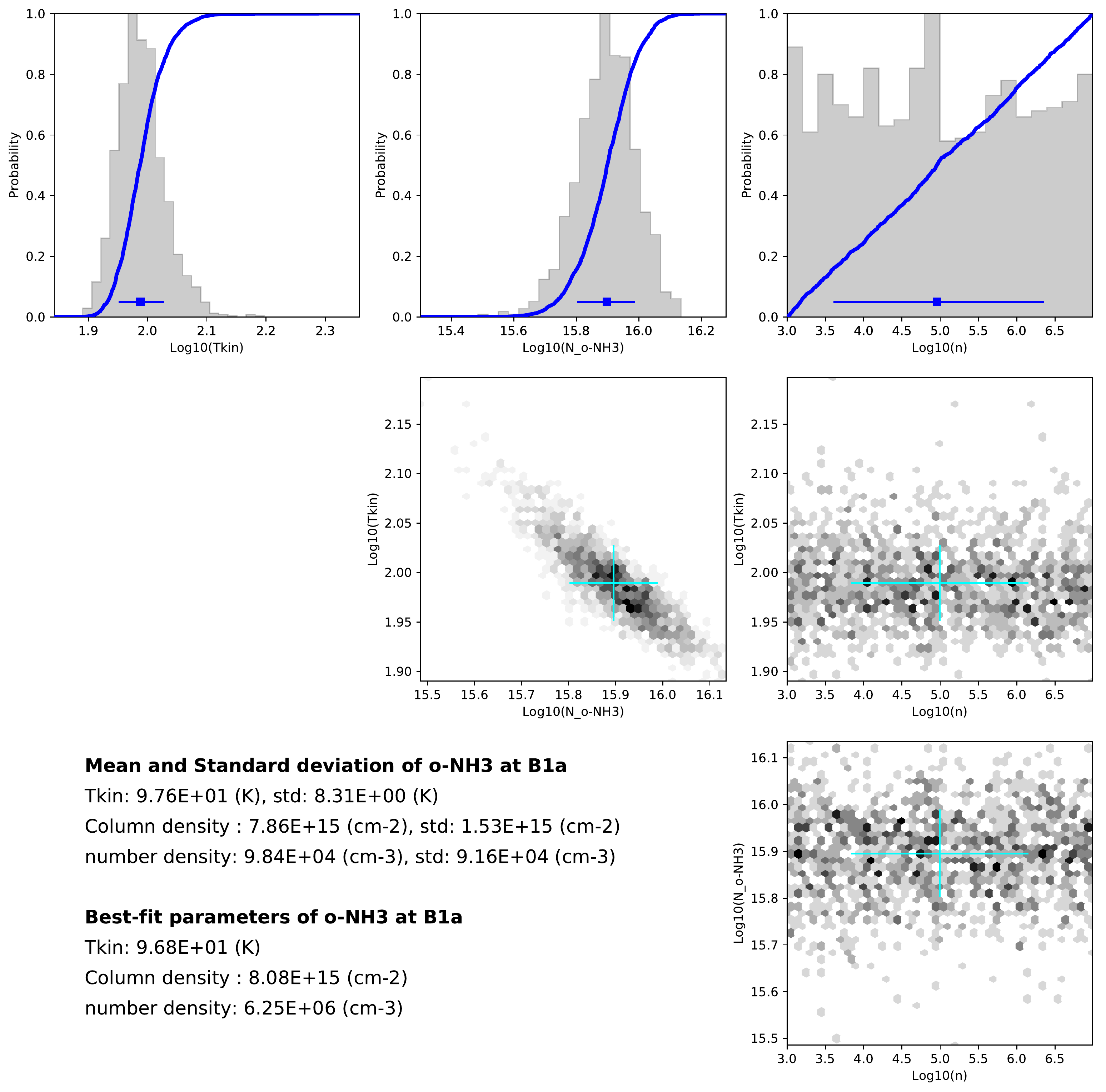}
&\includegraphics[height=8cm]{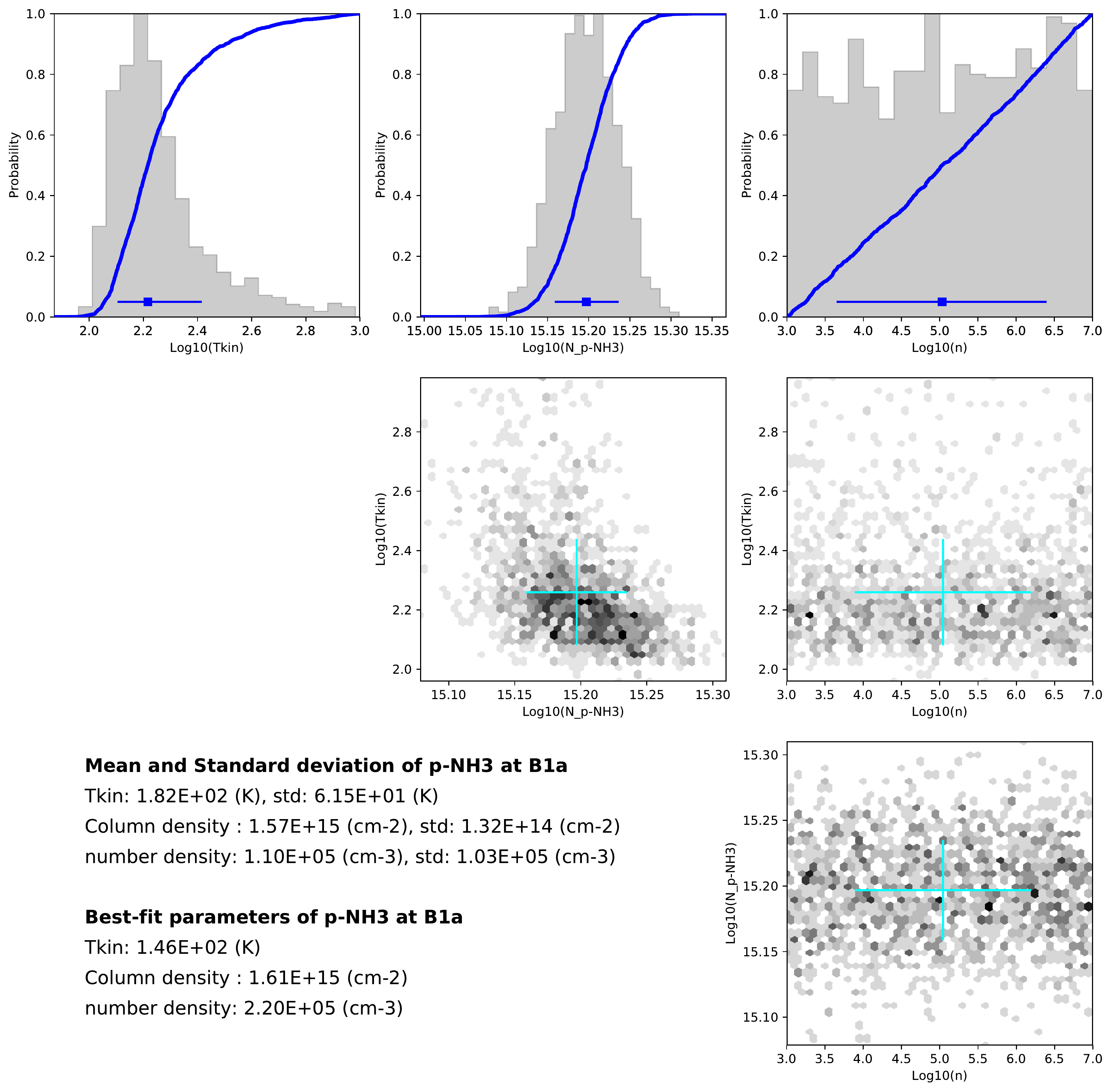}
\end{tabular}
\caption{The PDF of $o$/$p$-$\rm NH_3$ parameters toward B1a, derived by using RADEX and the MultiNest algorithm. A line width of $\rm 4\,km\,s^{-1}$ is adopted for all lines, and the beam-filling factor is assumed to be  unity for (1,1)--(3,3) lines while to be 0.1 for (4,4)--(7,7) lines.  The fitting parameters, including the $\rm H_2$ volume density ($n$), $o$/$p$-$\rm NH_3$ column density N($o$/$p$-$\rm NH_3$) and the gas kinetic temperature ($\rm T_{kin}$), are listed.
\label{fig:lvgeg}}
\end{figure*}

\begin{figure*}[tbh]
\centering
\begin{tabular}{lp{4.5cm}p{3.95cm}p{3.95cm}p{3.95cm}}
&$\rm \Delta V=4\,km\,s^{-1}$, ff=1\\
\multirow{1}{*} {\begin{sideways}Decl. offset (\arcsec)\end{sideways}}
&\includegraphics[clip, trim=0cm 1.4cm 0.0cm 1.0cm, height=7.75cm]{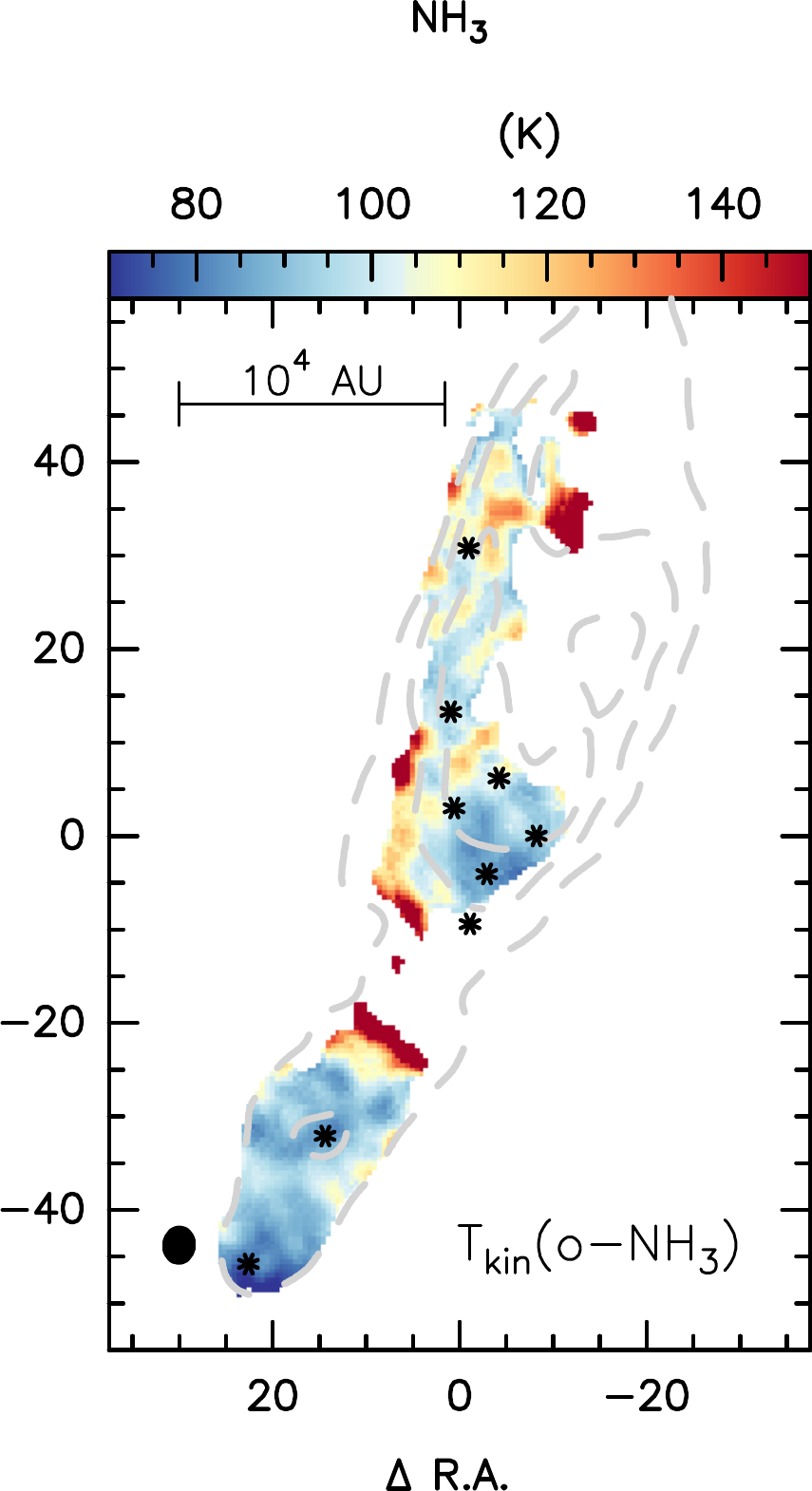}
&\includegraphics[clip, trim=1.cm 1.4cm 0.0cm 1.0cm, height=7.75cm]{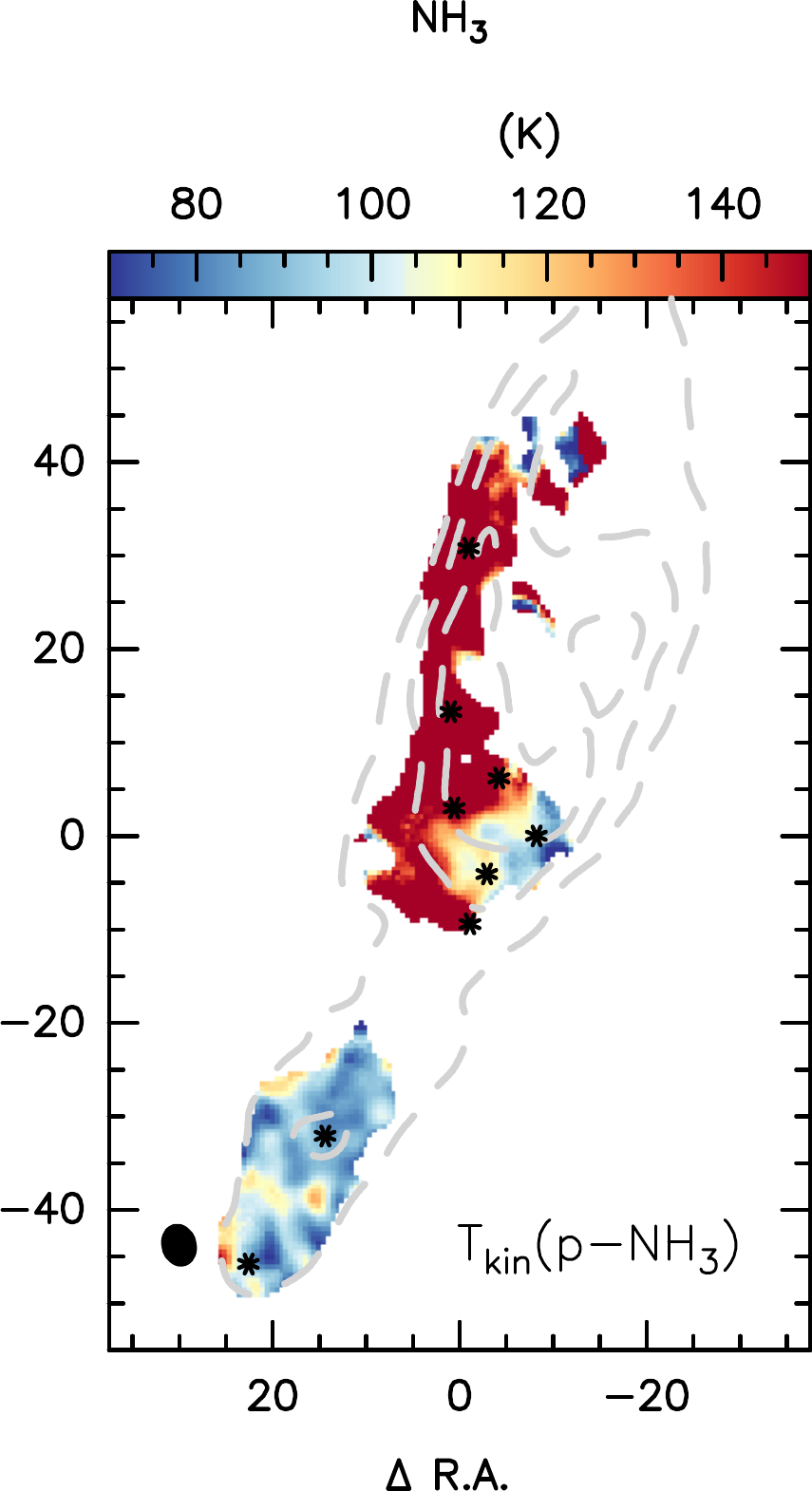}
&\includegraphics[clip, trim=1.cm 1.4cm 0.0cm 1.0cm, height=7.75cm]{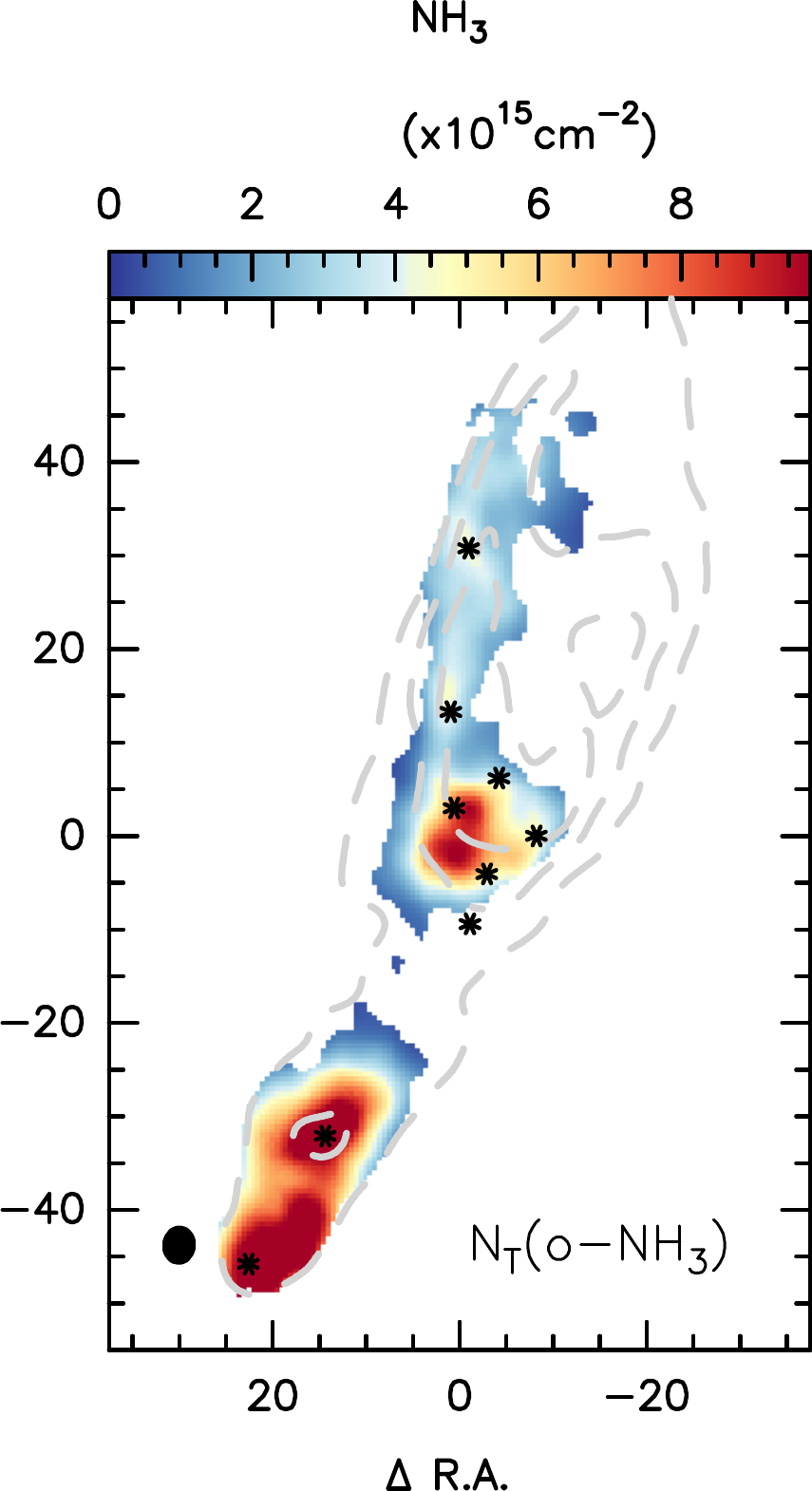}
&\includegraphics[clip, trim=1.cm 1.4cm 0.0cm 1.0cm, height=7.75cm]{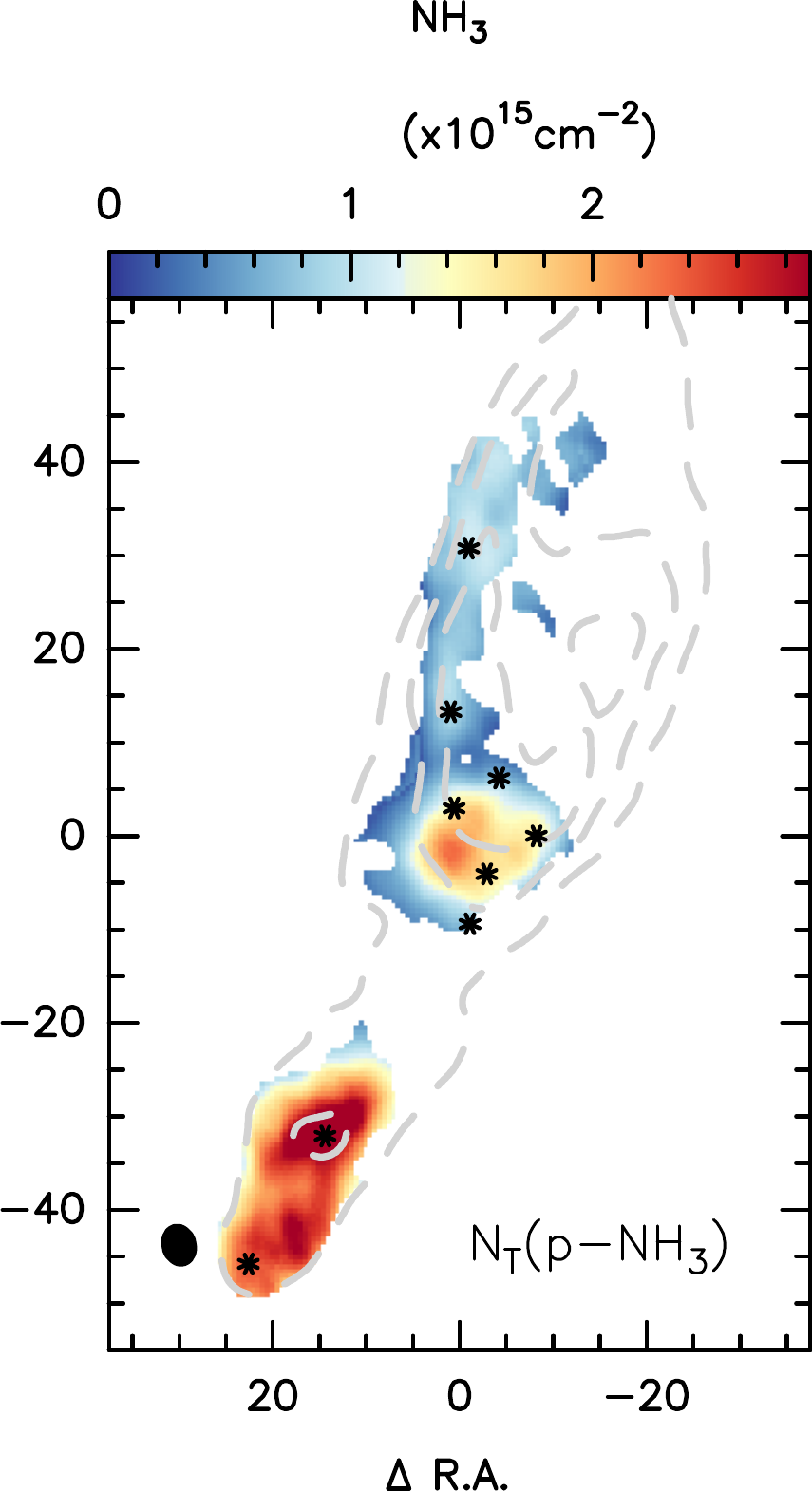}\\

&$\rm \Delta V=9\,km\,s^{-1}$, ff=1\\
&\includegraphics[clip, trim=0cm 0.5cm 0.0cm 1.0cm, height=7.95cm]{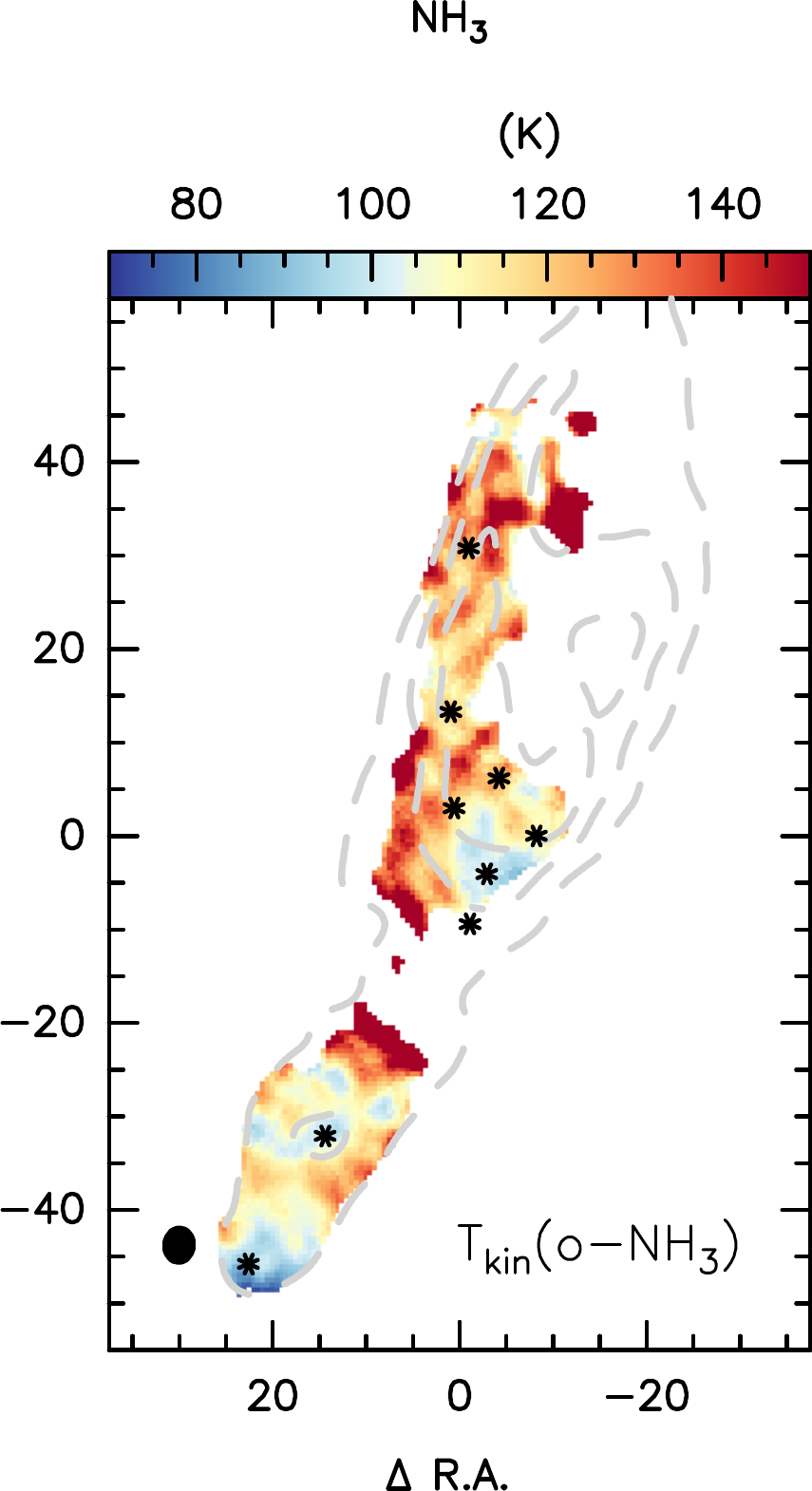}
&\includegraphics[clip, trim=1.cm 0.5cm 0.0cm 1.0cm, height=7.95cm]{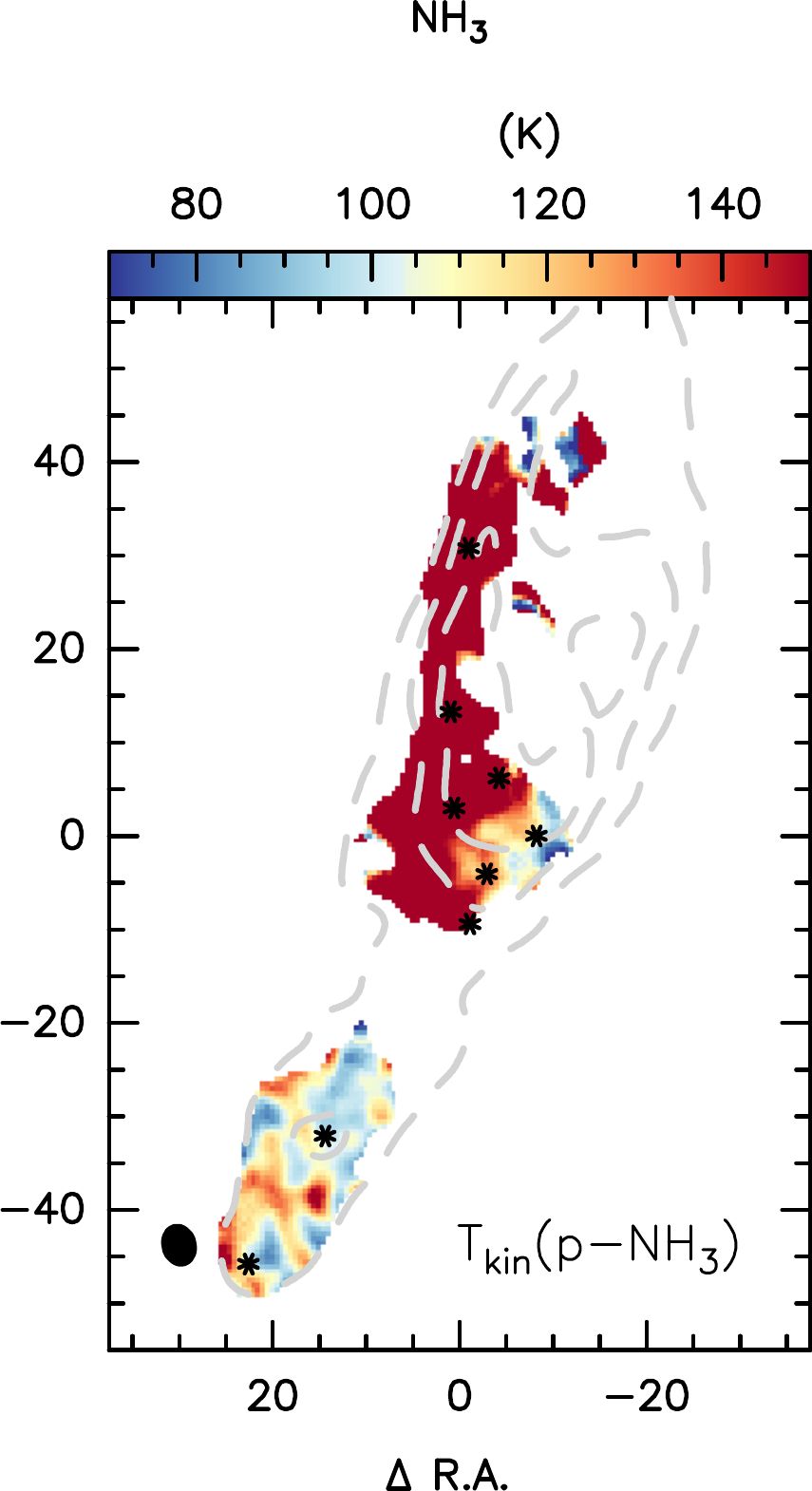}
&\includegraphics[clip, trim=1.cm 0.5cm 0.0cm 1.0cm, height=7.95cm]{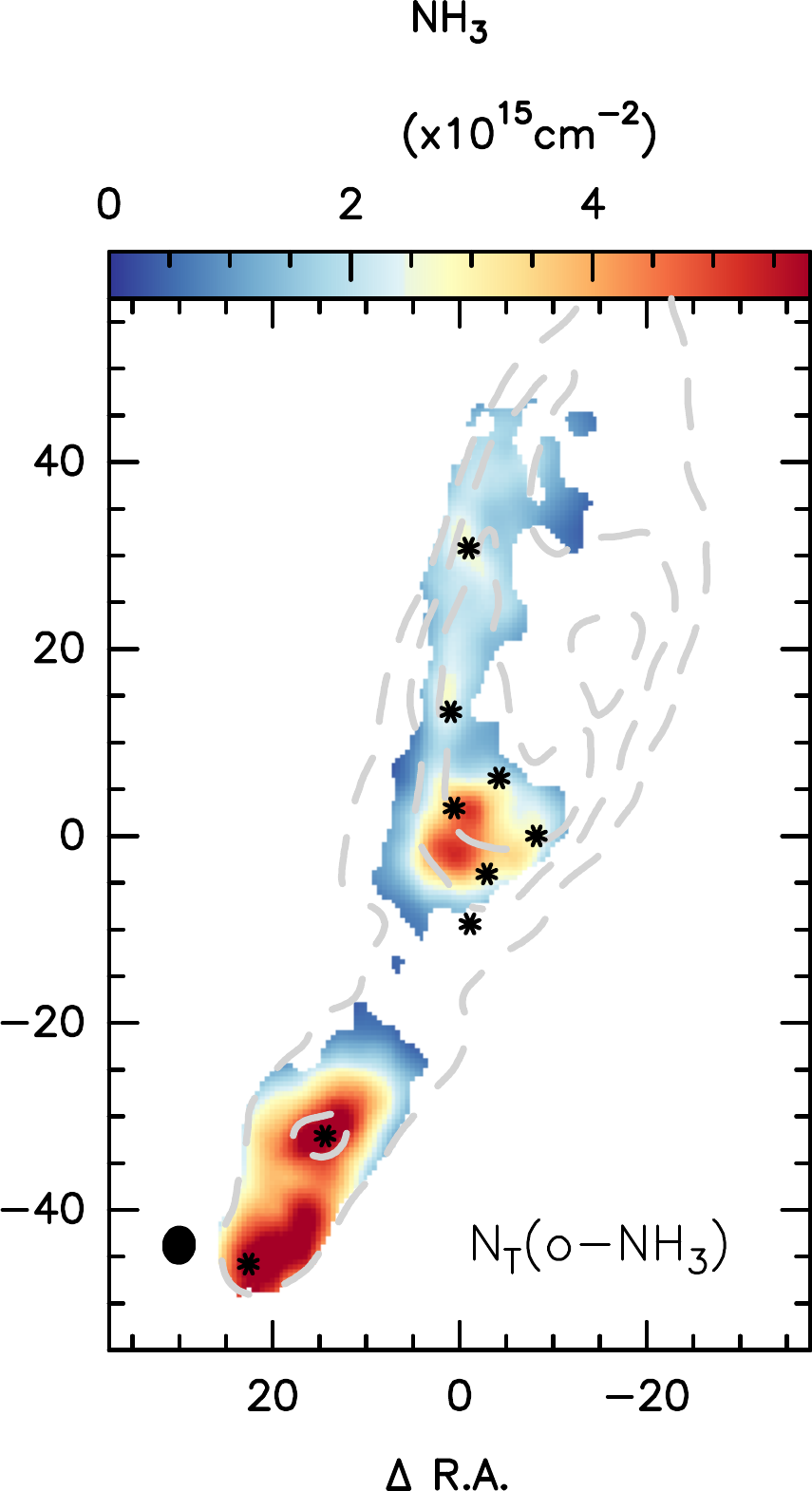}
&\includegraphics[clip, trim=1.cm 0.5cm 0.0cm 1.0cm, height=7.95cm]{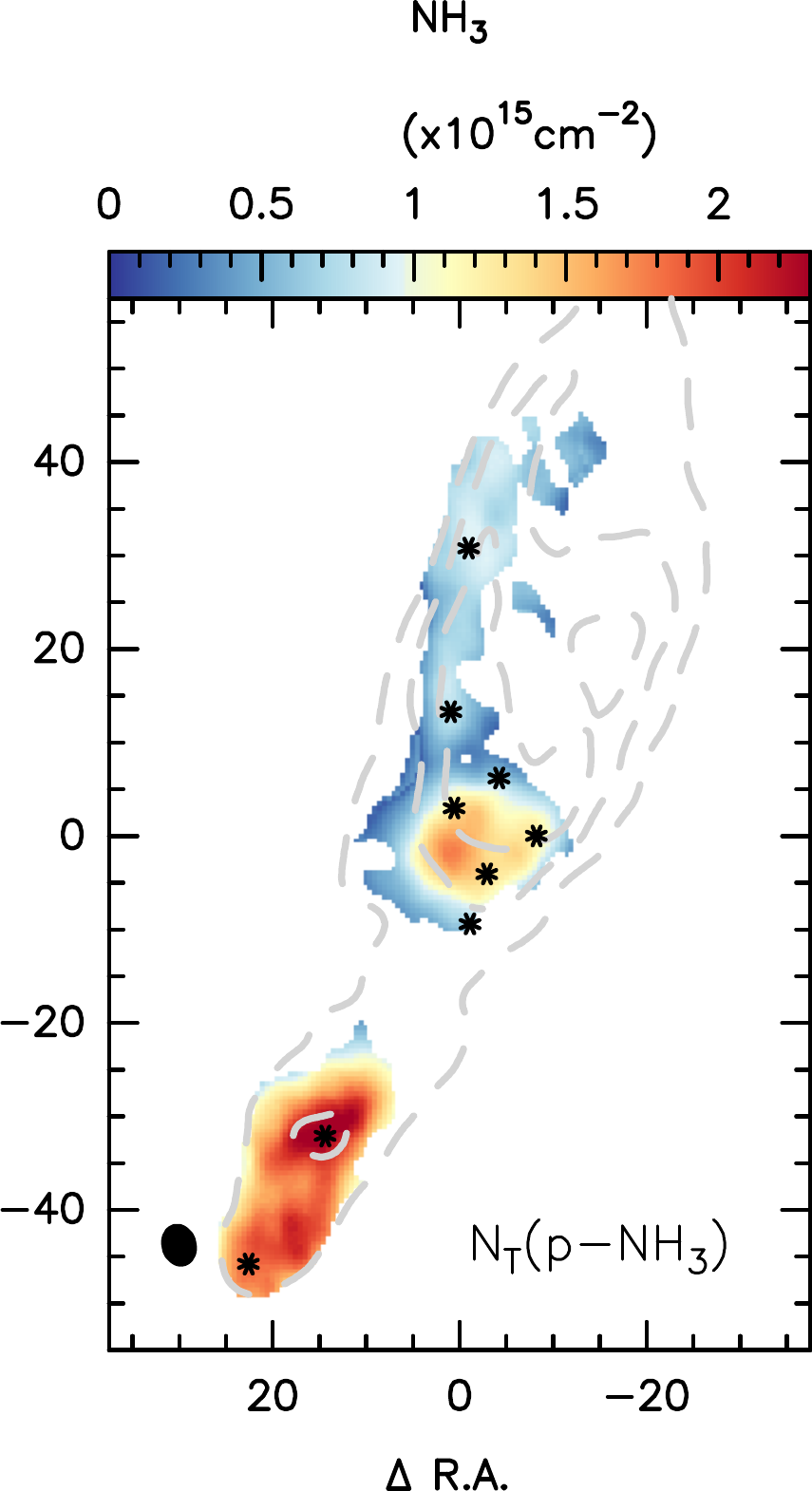}\\

&\multicolumn{4}{c}{R.A. offset (\arcsec)}
\end{tabular}
\caption{The best fit of the kinetic temperature maps and column density maps for $o$- and $p$-$\rm NH_3$, derived from LVG MultiNest, by assuming a single component and {unity filling factor for all lines} toward each position. The results by assuming the line width as $\rm 4\,km\,s^{-1}$ and $\rm 9\,km\,s^{-1}$ are shown in the upper and lower rows, respectively.
The pixels where the $\rm NH_3$ (5,5) and (6,6) lines show $\rm <5\sigma$ emission and CO\,(1--0) show $\rm <8\sigma$ emission are blank for  $p$- and $o$-$\rm NH_3$, respectively.  
The gray dashed contours and the labeled clumpy substructures are the same as shown in Figure~\ref{fig:jet}.
These maps are derived by smoothing all lines to the same pixel size and angular resolution, i.e., $\rm 4.90\arcsec\times3.81\arcsec$,  
shown as the synthesized beam in the bottom left of each panel.
\label{fig:lvgtestmap}}
\end{figure*}

\begin{figure*}[tbh]
\centering
\begin{tabular}{p{8cm}p{8cm}}
\includegraphics[clip, trim=0.cm 0cm 0.0cm 0.0cm, height=5cm]{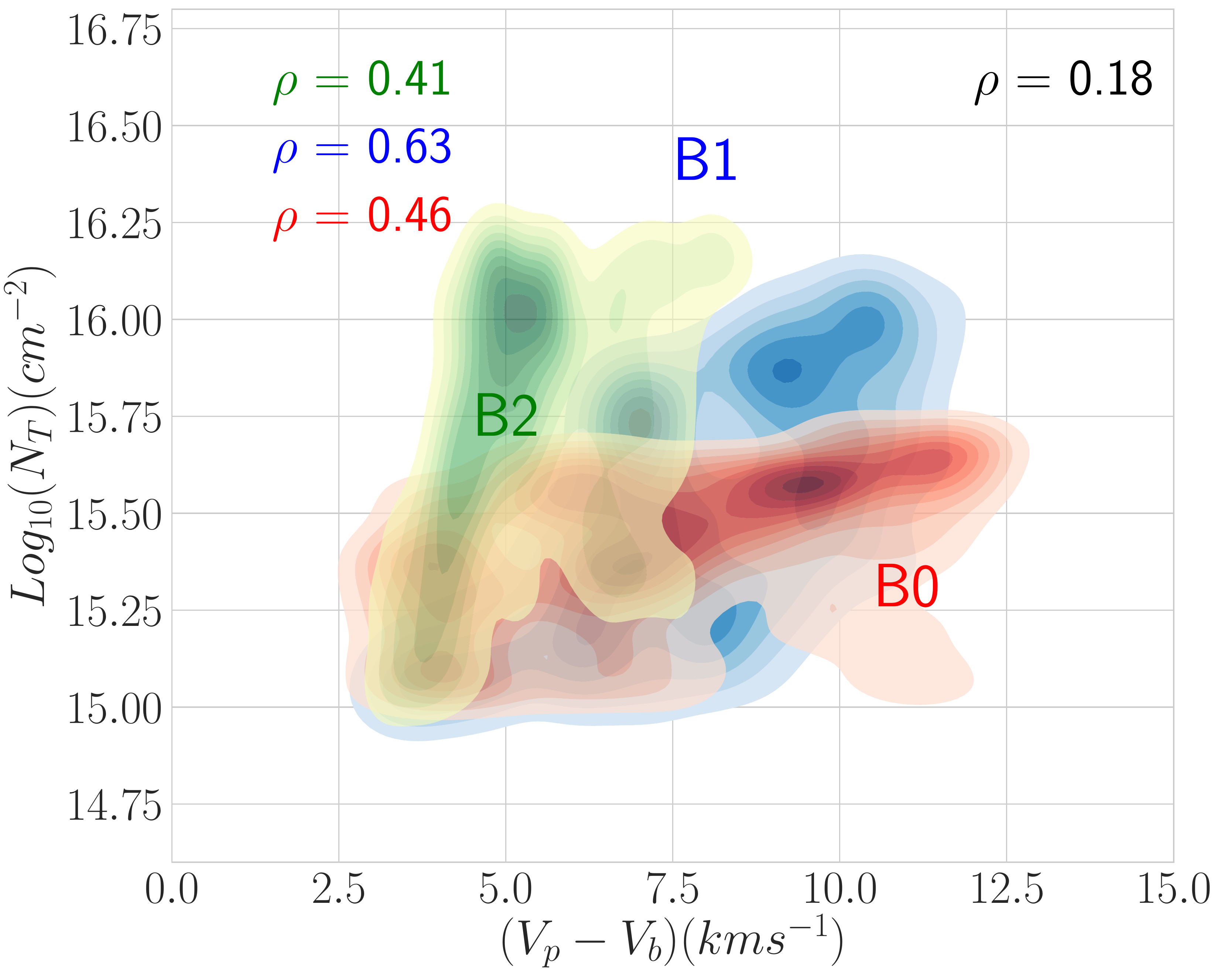}
&\includegraphics[clip, trim=0.cm 0cm 0.0cm 0.0cm, height=5cm]{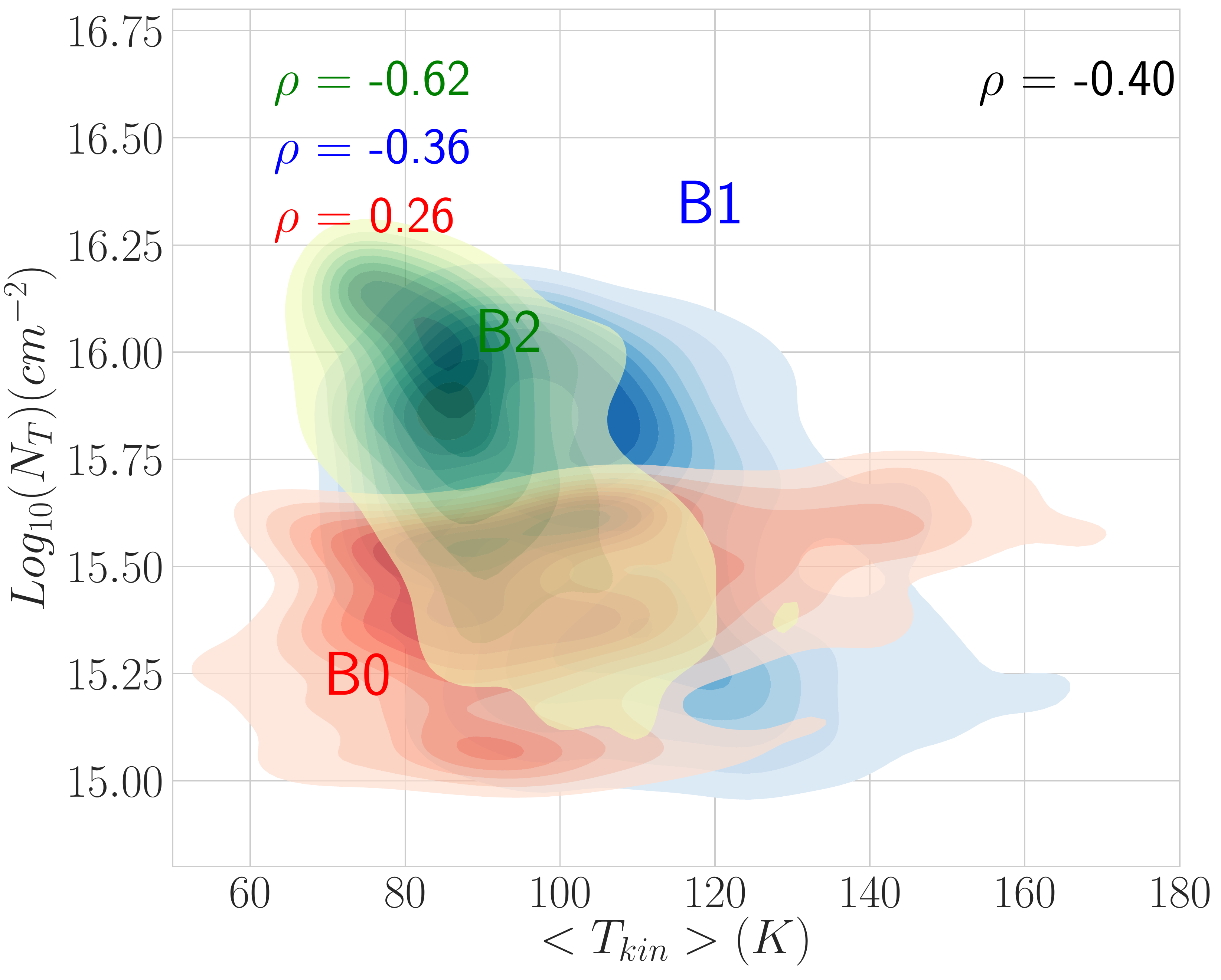}\\
\includegraphics[clip, trim=0cm 0cm 0.0cm 0.0cm, height=5cm]{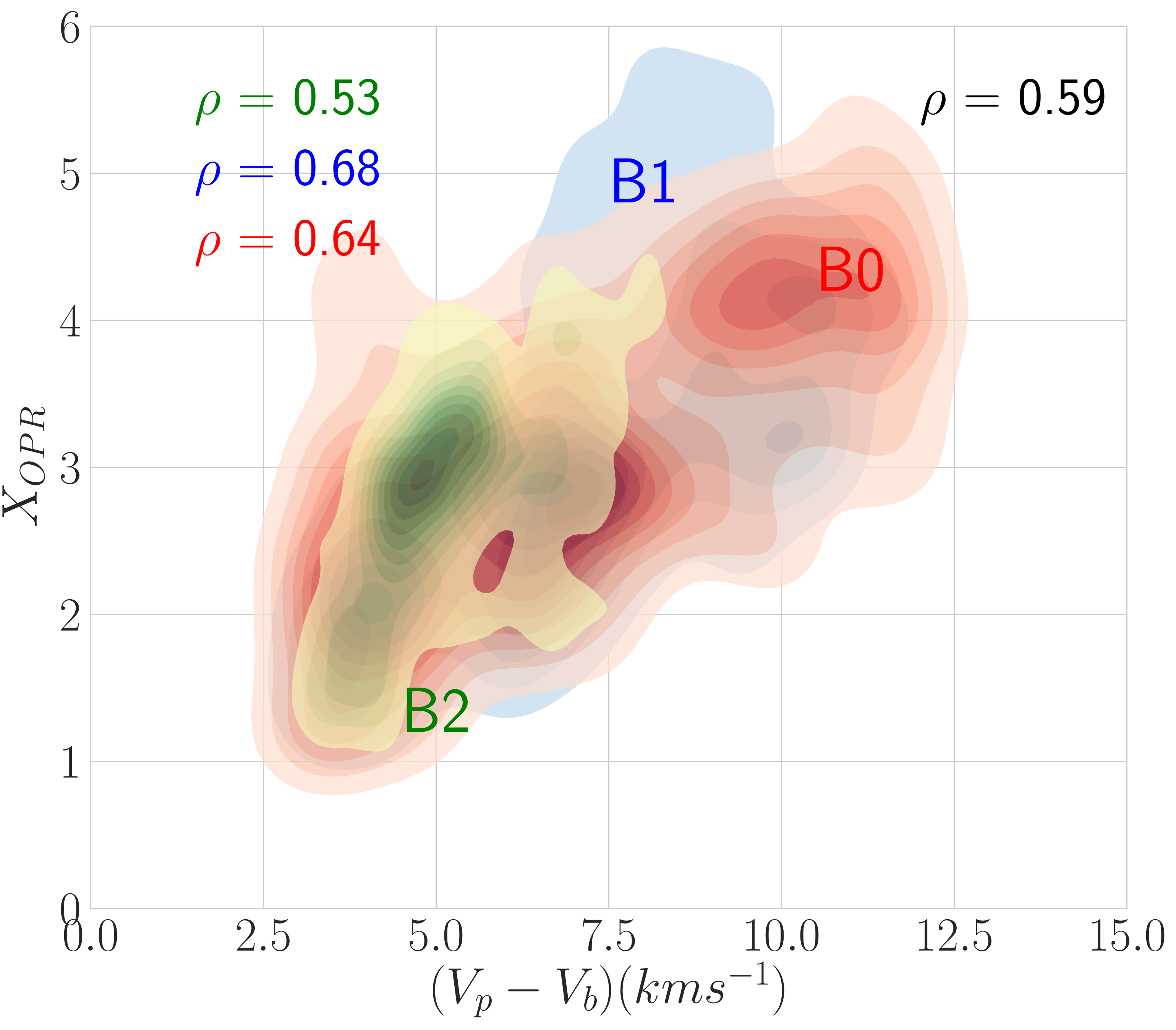}
&\includegraphics[clip, trim=0.cm 0cm 0.0cm 0.0cm, height=5cm]{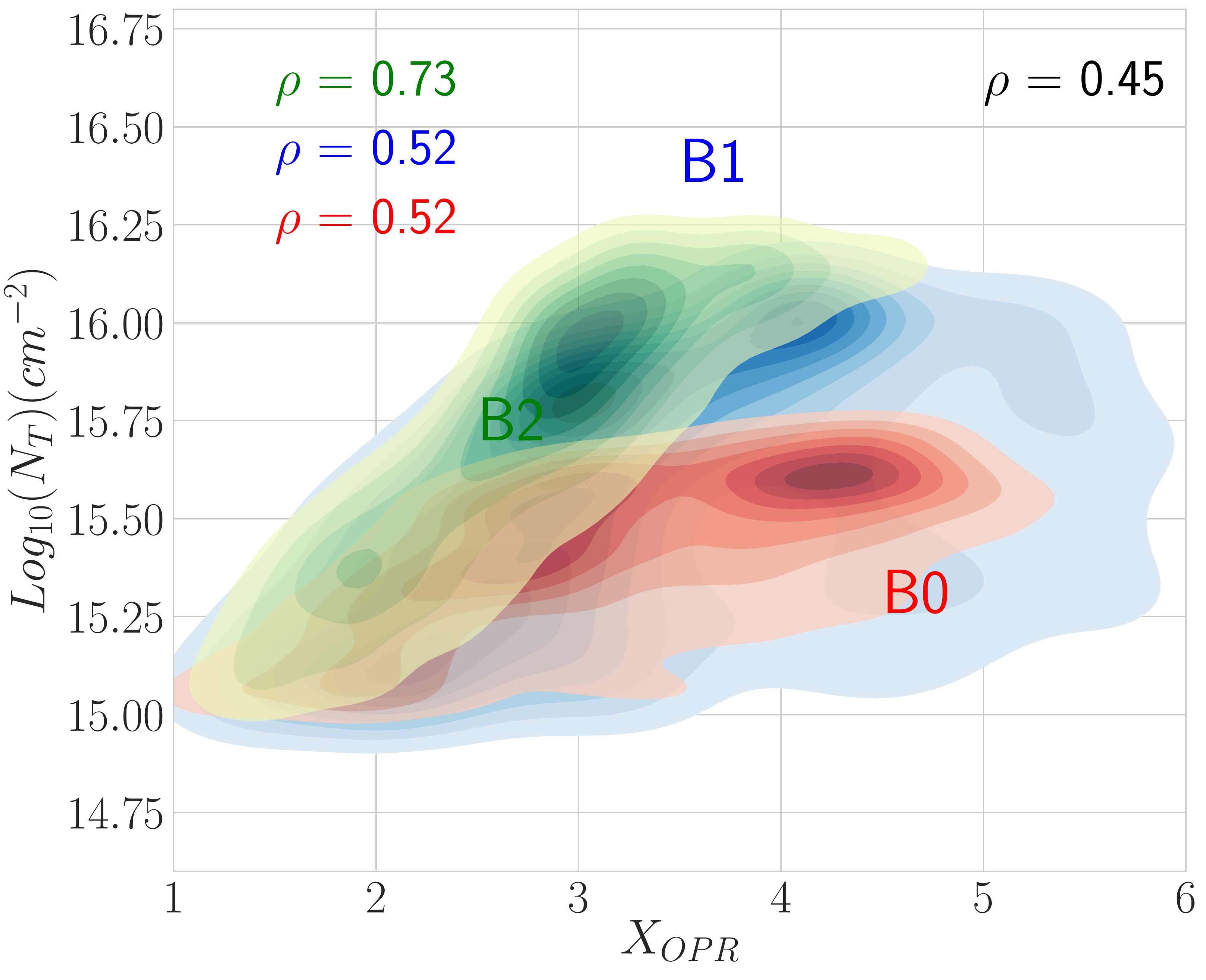}
\end{tabular}
\caption{Possible correlation between variables. Values are extracted from pixels after smoothing the parameter maps to the same angular resolution and plotted with a bivariate Gaussian kernel density estimate as contours. B0, B1, and B2 are plotted, with the SpearmanÕs rank correlation coefficient $\rho$ given in red, blue, and yellowish green, respectively, while the  correlation coefficient  of the entire target is given in black. The pixels where the $\rm NH_3$ (5,5) and (6,6) lines show $\rm <5\sigma$ emission are blank.  
\label{fig:correlation}}
\end{figure*}

Figure~\ref{fig:lvgeg} gives a PDF example of the $T_{kin}$, $N_{T}$,  and $n$  from LVG MultiNest fitting toward  B1a, by treating the $o$- and $p$-$\rm NH_3$ as distinct species.
Table~\ref{tab:lvgfit} lists the mean, standard deviation, and the best fit of $T_{kin}$ and $N_T$ for $o$/$p$-$\rm NH_3$ toward nine clumps.
Figure~\ref{fig:lvgtestmap} gives a combination of $T_{kin}$ and $N_T$ maps for $o$/$p$-$\rm NH_3$,  by testing the cases of assuming the beam-filling factor to be unity for all lines and the line width to be $\rm 4\,km\,s^{-1}$ or  $\rm 9\,km\,s^{-1}$.
Figure~\ref{fig:correlation} gives possible spatial correlations between the variables derived from LVG fittings. 

\begin{table*}[tbh]
\small
\centering
\caption{Mean and standard deviation of the PDF and best-fit results using the  LVG method \label{tab:lvgfit}}
 \scalebox{0.9}{
\begin{tabular}{cc|cccccccccc}
\hline\hline
\multirow{2}{*}{Location} &\multirow{2}{*}{Species$^a$}      &ff$^b$   &\multicolumn{2}{c}{$N_{T,n}$$^c$ ($\rm 10^{15}\,cm^{-2}$)}                                                  &\multicolumn{2}{c}{$T_{kin,n}$$^c$ (K)}       &\multicolumn{2}{c}{$N_{T,b}$$^d$ ($\rm 10^{15}\,cm^{-2}$)}                                                   &\multicolumn{2}{c}{$T_{kin,b}$$^d$ (K)}   &$n$$^e$ ($\rm 10^5\,cm^{-3}$)        \\
               &             &Case                     &mean$\rm \pm$std     &best fit   &mean$\rm \pm$std     &best fit                &mean$\rm \pm$std     &best fit           &mean$\rm \pm$std     &best fit            &best fit       \\

\hline
\multirow{4}{*}{B0a}    &\multirow{2}{*}{$p$-$\rm NH_3$ }        &I             &$\rm 1.0\pm0.0$  &1.1  &$\rm436\pm158$   &327   &$\rm 0.8\pm0.1$ &0.9  &$\rm525\pm174$   &802    &0.1--8.1     \\  
                                  &&II             &$\rm 1.0\pm0.0$  &1.0  &$\rm305\pm124$   &204   &$\rm 0.9\pm0.1$ &0.9  &$\rm 369\pm145$   &271    &0.03-33.7     \\  
                                   
                                  &\multirow{2}{*}{$o$-$\rm NH_3$}    	&I   	   &$\rm 4.2\pm0.8$ &4.4  &$\rm102\pm8$   &100  &$\rm 2.6\pm0.4$  &2.6  &$\rm122\pm11$   &121   &15.2--73.4   \\    
                                  &&II             &$\rm 4.2\pm0.8$ &4.4  &$\rm102\pm8$   &100    &$\rm 2.6\pm0.4$  &2.6  &$\rm122\pm11$   &121   &0.5--5.0   \\        
         
\hline

\multirow{4}{*}{B0e}    &\multirow{2}{*}{$p$-$\rm NH_3$ }          &I                &$\rm 0.8\pm0.0$  &0.9 &$\rm278\pm91$    &233   &$\rm 0.8\pm0.1$  &0.8 &$\rm386\pm132$  &292    &4.8--61.4    \\  
                                           &&II             &$\rm 1.0\pm0.0$  &0.9  &$\rm294\pm78$   &196   &$0.8\rm \pm0.1$ &0.8  &$439\rm \pm181$   &198    &0.03--0.1    \\    
                                   &\multirow{2}{*}{$o$-$\rm NH_3$}      &I  		 &$\rm 4.0\pm0.7$   &4.2 &$\rm103\pm9$     &102  &$\rm 2.5\pm0.4$  &2.6  &$\rm124\pm11$    &122     &0.3--4.9    \\  
                                 &&I  		 &$\rm 4.1\pm0.7$   &4.2 &$\rm103\pm9$     &102     &$\rm 2.5\pm0.4$  &2.6  &$\rm124\pm11$    &123     &0.8--92   \\  
 \hline

\multirow{4}{*}{B1f}   &\multirow{2}{*}{$p$-$\rm NH_3$ }       &I                 &$\rm 0.8\pm0.1$  &0.8   &$\rm175\pm41$   &154  &$\rm0.6\pm0.1$  &0.7 &$\rm206\pm57$   &178     &5.8--46.0     \\  
				&&II             &$\rm 0.8\pm0.1$  &0.8  &$\rm122\pm18$   &117   &$\rm 0.6\pm0.1$ &0.6  &$\rm 139\pm24$   &130    &6.4--29    \\ 
                                 &\multirow{2}{*}{$o$-$\rm NH_3$}    		&I  	 &$\rm 3.7\pm0.7$  &3.9  &$\rm106\pm9$    &104   &$\rm2.3\pm0.3$  &2.4  &$\rm127\pm11$     &126       &0.1--0.2   \\    
                                  &&II  	 &$\rm 3.7\pm0.7$  &3.9  &$\rm106\pm9$    &105          &$\rm2.3\pm0.3$  &2.4  &$\rm128\pm12$     &126       &11--84   \\

\hline

\multirow{4}{*}{B1a}   &\multirow{2}{*}{$p$-$\rm NH_3$ }         &I                &$\rm 1.6\pm0.1$  &1.6  &$\rm275\pm98$   &211   &$\rm1.3\pm0.0$  &1.3  &$\rm400\pm144$   &319    &1.1--32.2       \\  
				&&II             &$\rm 1.6\pm0.2$  &1.6  &$\rm182\pm61$   &146   &$\rm 1.3\pm0.1$ &1.3  &$\rm 243\pm90$   &184    &5.2--6.4   \\ 
                                  &\multirow{2}{*}{$o$-$\rm NH_3$}          &I                &$\rm 7.7\pm1.5$  &8.1  &$\rm98\pm8$  &97   &$\rm4.3\pm0.7$  &4.4  &$\rm119\pm11$   &117    &10.3--14.1  \\   
                                  &&II               &$\rm 7.8\pm1.5$  &8.1  &$\rm98\pm8$  &97         &$\rm4.3\pm0.7$  &4.4  &$\rm119\pm11$   &117    &1.7--2.8  \\

\hline
\multirow{4}{*}{B1b}    &\multirow{2}{*}{$p$-$\rm NH_3$ }       &I                &$\rm 1.5\pm0.1$  &1.6  &$\rm103\pm11$  &100     &$\rm1.3\pm0.0$  &1.3 &$\rm118\pm13$   &115    &1.9--2.9       \\  
				&&II             &$\rm 1.6\pm0.2$  &1.7  &$\rm83\pm8$   &82   &$\rm 1.3\pm0.1$ &1.3  &$\rm 96\pm11$   &96    &3--98    \\ 
                                  &\multirow{2}{*}{$o$-$\rm NH_3$}     &I  			  &$\rm 4.7\pm0.9$  &4.9   &$\rm94\pm7$  &93   &$\rm2.8\pm0.4$  &2.9  &$\rm112\pm9$    &110      &1.4--4.7  \\         
                                   &&II  			  &$\rm 4.7\pm0.9$  &4.9   &$\rm94\pm7$  &93        &$\rm2.8\pm0.4$  &2.9  &$\rm112\pm9$    &110      &0.01--13  \\ 
                                                                     
\hline
\multirow{4}{*}{B1c}    &\multirow{2}{*}{$p$-$\rm NH_3$ }     &I                    &$\rm 1.7\pm0.2$  &1.7   &$\rm116\pm16$   &112  &$\rm1.4\pm0.1$ &1.4  &$\rm136\pm20$   &131    &63.1--70.4    \\  
 				&&II             &$\rm 1.8\pm0.2$  &1.9  &$\rm87\pm9$   &85   &$\rm 1.4\pm0.1$ &1.4  &$\rm 102\pm12$   &100    &0.4--2.1    \\ 
                                  &\multirow{2}{*}{$o$-$\rm NH_3$}     &I  			 &$\rm 5.4\pm1.0$   &5.6  &$\rm86\pm6$  &86    &$\rm3.1\pm0.5$  &3.1  &$\rm102\pm8$     &102        &0.2--27.6  \\                                                                                                                
             &&II  			 &$\rm 5.4\pm1.0$   &5.6  &$\rm86\pm6$  &86    
             &$\rm3.1\pm0.5$  &3.2  &$\rm102\pm8$     &101        &0.1--0.2  \\

\hline
\multirow{4}{*}{B1i}    &\multirow{2}{*}{$p$-$\rm NH_3$ }     &I                 &$\rm0.6\pm0.1$  &0.7  &$\rm329\pm117$    &253   &$\rm0.6\pm0.1$  &0.6  &$\rm433\pm152$     &319       &1.8--2.4      \\  
 				&&II             &$\rm 0.7\pm0.1$  &0.7  &$\rm476\pm181$   &92   &$\rm 0.6\pm0.1$ &0.6  &$\rm 523\pm189$   &92    &0.01--11.2    \\   
                                   &\multirow{2}{*}{$o$-$\rm NH_3$}      &I  		 &$\rm 1.5\pm0.2$  &1.5   &$\rm98\pm7$    &98  &$\rm1.1\pm0.1$ &1.1  &$\rm109\pm7$    &108    &38.2--42.7    \\      
             &&II  		 &$\rm 1.5\pm0.2$  &1.5   &$\rm98\pm7$    &98    &$\rm1.1\pm0.1$ &1.1  &$\rm109\pm7$    &108    &9--34    \\          
\hline

\multirow{4}{*}{B2a}    &\multirow{2}{*}{$p$-$\rm NH_3$ }     &I                &$\rm 3.2\pm0.4$  &3.3  &$\rm86\pm8$   &84   &$\rm2.3\pm0.2$  &2.3  &$\rm101\pm10$    &99   &1.7--23.2\\  
 				&&II             &$\rm 3.7\pm0.4$  &3.8  &$\rm68\pm6$   &67   &$\rm 2.5\pm0.2$ &2.5  &$\rm 81\pm7$   &80    &0.01--0.02    \\                                
                                &\multirow{2}{*}{$o$-$\rm NH_3$}    &I  			&$\rm 11.7\pm2.3$  &12.1  &$\rm83\pm6$  &83  &$\rm6.0\pm1.0$ &6.2  &$\rm100\pm8$     &99       &1.3--13.6      \\   
                                   &&II  			&$\rm 11.7\pm2.3$  &12.1  &$\rm84\pm6$  &83       &$\rm6.0\pm1.0$ &6.2  &$\rm99\pm7$     &98       &0.2--10      \\    
\hline
\multirow{4}{*}{B2b}    &\multirow{2}{*}{$p$-$\rm NH_3$ }      &I                 &$\rm 2.4\pm0.2$  &2.4   &$\rm95\pm11$    &93   &$\rm1.8\pm0.1$  &1.8  &$\rm111\pm13$   &109   &6.7--11.2\\  
 				&&II             &$\rm 2.7\pm0.3$  &2.7  &$\rm71\pm6$   &71   &$\rm 1.9\pm0.2$ &2.0  &$\rm 84\pm8$   &83    &0.01--6    \\           
                                  &\multirow{2}{*}{$o$-$\rm NH_3$}    	&I  		 &$\rm 11.9\pm2.4$  &12.4   &$\rm75\pm4$    &75  &$\rm6.1\pm1.1$ &6.4   &$\rm89\pm6$      &88       &1.1--51.0\\   
                                   	&&II  		 &$\rm 12.6\pm2.4$  &12.4   &$\rm75\pm4$    &75  	&$\rm6.1\pm1.1$ &6.4   &$\rm89\pm6$      &88       &0.5--3\\

\hline
\hline
\multicolumn{11}{l}{{\bf Note.} {\it a}. Estimated by excluding the (7,7) line because no collision rate is given in LAMDA.}\\
\multicolumn{11}{l}{~~~~~~~~~~~{\it b}.  Case I corresponds to a unity filling factor for all lines at any pixel; }\\
\multicolumn{11}{l}{~~~~~~~~~~~~~~ Case II corresponds to a unity  filling factor for (1,1)--(3,3) and as 0.1 for (4,4)--(7,7) at any pixel;}\\ 
\multicolumn{11}{l}{~~~~~~~~~~~{\it c}.    Assuming a narrow (``$n$") FWHM line width of $\rm 4\,km\,s^{-1}$ (the median from the fittings to the observed spectrum in Table~\ref{tab:line}).}\\ 
\multicolumn{11}{l}{~~~~~~~~~~~{\it d}.  Assuming a broad  (``$b$") FWHM line width of $\rm 9\,km\,s^{-1}$ adopted from \citet[][]{bachiller93,lefloch12}.}\\  
\multicolumn{11}{l}{~~~~~~~~~~~{\it e}.  The PDF is not obvious, so the range of the best fits based on different line width assumptions is listed.}\\  
\end{tabular}

}

\end{table*}

One major caveat in the above LVG fitting is that LAMDA only provides the collision rates between the $p$-/$o$-$\rm NH_3$ and $p$-$\rm H_2$ \citep{danby88}.  Although we note that some $o$-$\rm H_2$ cross sections were computed at a few selected energies \citep[e.g., ][]{offer89,rist93,bouhafs17},  no collision rates between $p$-/$o$-$\rm NH_3$ and $o$-$\rm H_2$ is available in LAMDA. 
Unlike the case in cold dense cores, the abundance of $o$-$\rm H_2$ can become significant in warm and shocked regions \citep[e.g., ][]{flower06,neufeld19}.

 A second caveat is the beam-filling factor, which we take as unity for all lines in the fitting, given that the line emissions show more extended spatial distributions toward all the clumpy substructures  than the synthesized beam. However, considering that {shock discontinuities} cannot be spatially resolved by observations,  substructures with high column density and high temperature might be beam smeared. This is likely the case after checking the peak brightness temperatures  of all transitions toward the representative positions (Figure~\ref{fig:velpro}) and their optical depth maps (Figure~\ref{fig:tauline}). {For testing, we set it as unity for (1,1)--(3,3)  and 0.5, or 0.1, or  0.01 for (4,4)--(7,7), and then compare the testing results with the results by assuming the beam-filling factor of unity for all lines. When the filling factor for higher transition levels is adopted in the range of 0.1--1, a smaller filling factor decreases the kinetic temperature significantly, by 30\% for $p$-$\rm NH_3$  but by $\rm <10\%$ for $o$-$\rm NH_3$. In contrast, the column density of $o/p$-$\rm NH_3$ only shows imperceptible changes ($\rm <5\%$) when a different beam-filling factor is adopted.}

 A third caveat is the FWMH line width  we adopted, i.e., the median from all spectrum fitting  as $\rm 4\,km\,s^{-1}$.  This seems to be consistent with the FWHM line width of (1,1)--(3,3) at an angular resolution of 40\arcsec, which was reported to be 8--9\,$\rm \,km\,s^{-1}$  by \citet[][]{bachiller93}. However, {the peak-to-bluest velocity offset varies pixel by pixel  (Figure~\ref{fig:gradient}), being broader along the jet path than within the cavity by a factor of 2}.  As a compromise to the $\rm NH_3$ hyperfine multiplets fitting, we assume one velocity component. In the case of multivelocity components being smeared within the VLA's synthesized beam, an FWHM line width  of 9\,$\rm km\,s^{-1}$ is adopted for the test. We find that such a  line width  leads to a 15\% and 20\% increase in the best-fit $T_{kin}$  for $p$-$\rm NH_3$ and $o$-$\rm NH_3$, as well as a 15\% and 45\% decrease in their  $N_T$, respectively (Table~\ref{tab:lvgfit}).  Nevertheless,  these increase and decrease factors are globally consistent throughout the entire map pixels, so the contrast (local gradient) of each map does not change.

A minor fourth caveat is {the velocity range over which we integrate the line intensity. 
In Table~\ref{tab:hfsline}, we list the uncertainty of the  line-integrated intensity $I$, which is given by comparing the integration (i) where the line emission has $\rm S/N>3$ (-25 to $\rm +25\,km\,s^{-1}$); (ii) where the line emission is  above zero (-27 to $\rm +33\,km\,s^{-1}$), and (iii) includes only the main component of the hyperfine multiplets (-10 to $\rm +5km\,s^{-1}$). 
Due to  strong satellite emissions, the uncertainty for the (1,1) line is  $\rm >20\%$. Nevertheless, it does not change our best-fit result significantly because  the uncertainty for the (2,2) line is $\rm <30\%$ and for the rest lines is $\rm <10\%$  (less than the systematic uncertainty).}


\end{document}